\documentclass[preprint,12pt]{elsarticle}

\usepackage{amssymb}
\usepackage{amsmath}
\usepackage{gensymb}

\usepackage{makecell}
\usepackage{threeparttable} % to use table notes

\usepackage{rotating}

\newcommand{\neut}{$\rm n_{eq}/cm^2$}

\journal{Physics Reports}

\begin{document}

\begin{frontmatter}



%\bibliographystyle{elsarticle-num}

\title{Precision timing detectors}

\author[INFN]{Martina Malberti} %% Author name
\affiliation[INFN]{organization={INFN Sezione di Milano Bicocca},%Department and Organization
            addressline={Piazza della Scienza 3}, 
            city={Milan},
            postcode={20126}, 
            %state={},
            country={Italy}}
\author[Peking]{Xiaohu Sun\corref{cor1}} %% Author name
\cortext[cor1]{Corresponding author}
\ead{Xiaohu.Sun@pku.edu.cn}
\affiliation[Peking]{organization={Department of Physics and State Key Laboratory of Nuclear Physics and Technology, Peking University},%Department and Organization
            addressline={No.5 Yiheyuan Road}, 
            city={Beijing},
            postcode={100871}, 
            %state={},
            country={China}}

%% Abstract
\begin{abstract}

Precision timing has played a critical role in high-energy physics experiments, particularly for particle identification and the suppression of pileup under the challenging conditions expected at future colliders like the High-Luminosity Large Hadron Collider (HL-LHC). Over the past decades, significant advancements in timing measurement technologies have been made to meet the demands of increasingly complex collider environments. After introducing the motivation for precision timing in collider experiments, the underlying physical principles of timing measurements and the most important factors influencing the time resolution of a detector, this review presents a survey of key detector technologies developed in recent years, including scintillators read out by silicon photo-multipliers (SiPMs), low-gain avalanche diodes (LGADs), multi-gap resistive plate chambers (MRPCs). The integration of precision timing into large-scale systems is discussed with examples from detectors at current collider experiments. Finally, we explore emerging technologies and future directions in the field, highlighting their potential impact on the next generation of high-energy physics experiments.
%Precision timing has played a critical role in high-energy physics, serving as a key element in particle identification and the suppression of pileup vertices within the same bunch crossing, such as in the conditions of the high luminosity LHC (HL-LHC) experiments. Over the past decades, timing measurement technologies have been advanced to address emerging challenges. In this article, we will outline the motivation behind timing measurements, the fundamental principles of timing detection, and several representative detector sensors developed in recent years for the current generation of experiments at colliders (like MRPCs, scintillators read out by SiPMs, Low Gain Avalanche Diodes). We will examine the primary factors influencing timing resolution, including the characteristics of sensors and front-end electronics. Subsequently, the integration of timing detectors into modern large-scale high-energy particle detection systems will be discussed, to conclude with an outlook discussing several novel technologies for timing measurements.

\end{abstract}

\begin{graphicalabstract}
\end{graphicalabstract}

\begin{highlights}
%\item Precision timing at colliders: pileup mitigation, 4D tracking, time-of flight particle identification, impact on BSM searches
%\item Advanced technologies enabling picosecond timing: SiPMs+fast scintillators, LGADs, MRPC 
%\items Implementation in major experiments: ATLAS High Granularity Timing, CMS MTD
%\items 

\item Timing challenges in high energy collider experiments: TOF, pileup and BSM searches.
\item The frontier of timing at 20-30 ps: scintillation, gaseous and silicon detectors.
\item Applications in large-scale experiments: CMS MTD BTL\&ETL, ATLAS HGTD, ALICE TOF etc.
\item To sub-20 ps: scint., Cherenkov, quantum, hybrid gaseous, advanced silicon techs etc.

%\item Challenges in time measurements of high energy particles are driven by the high-precision particle identification through time-of-flight, the overwhelming pileup collisions in future colliders and the exhilarating pursuit of physics beyond the standard model.
%\item Developments of various technologies ranging from scintillation coupled with photo-detector, gaseous detectors to silicon-based sensors have pushed the frontier of time resolution to 20-30 ps in large-scale collider experiments.
%\item Novel technologies including but not limited to scintillator engineering, exploration of Cherenkov light, quantum confinement, hybrid gaseous chambers, resistive silicon sensors, columnar 3D sensors and monolithic sensors keep thriving and drive the time resolution to a few ps and even lower.
\end{highlights}

%% Keywords

\end{frontmatter}


\tableofcontents


\section{Introduction}
\label{sec:introduction}

Precision timing detectors have become an essential tool in modern high energy physics (HEP) experiments, with an increasing interest and advancements in sensor technologies in recent years driven by the need to address the experimental challenges of future collider facilities, like the upcoming High Luminosity Large Hadron Collider (HL-LHC)~\cite{Apollinari:2120673}, the Future Circular Collider (FCC)~\cite{FCC-ee, FCC-hh} and the Muon Collider~\cite{Long2021}.

% Examples of TOF detectors in past and present hep experiments
% 1960's : ADONE e+e- MAE for rejection of bkg from cosmic rays. 
% Mark-II at SLC TOF  scintillators 180-250 ps - https://www.sciencedirect.com/science/article/pii/0168900289912175
%1980s : NA35 TOF 
%1990s : NA49 TOF - scintillator, PesTOF counters -  between 60 and 85 ps https://www.sciencedirect.com/science/article/pii/S0168900299002399
% NA56:  scintillators, 74-100 ps intrinsic time resolution https://link.springer.com/article/10.1007/s100529900145
% PHOENIX @ BNL - plastic scintillators -  better than 90 ps: https://www.sciencedirect.com/science/article/pii/S0168900299002612?via=ihub 
% STAR TOF - scintillator based - average time resolution 87 ps
% ALICE TOF - MRPCs - 56 ps 

%Historically, timing information has been used mainly to perform time-of-flight (TOF) measurements for particle identification (PID) or suppression of backgrounds from non-beam events~\cite{Ash1974, ABRAMS198955, AFANASIEV1999210, Ambrosinietal.1999,  CARLEN1999123}. These TOF systems, typically based on scintillators, offered timing resolutions of the order of hundreds picoseconds, sufficient for distinguishing cosmic rays from collisions events or separating pions, kaons, and protons at low to intermediate momenta over large distances. 
Historically, timing detectors have been used mainly to perform time-of-flight (TOF) measurements for particle identification (PID) or suppression of backgrounds from non-beam events. 
One of the first examples of TOF detector application in a collider experiment is from the ADONE e$^+$e$^-$ ring at LNF in the 1960s, where plastic scintillator counters placed at opposite sides to the collision region were used to distinguish collision events from cosmic rays by means of a TOF measurement with $\sim$350~ps resolution~\cite{Ash1974}. TOF systems for PID were employed in several 
early collider and fixed target experiments starting from the 1970s.
%to distinguish between particle species based on differences in their velocities over known path lengths. 
These systems, typically based on scintillators, offered timing resolutions of the order of hundreds picoseconds, sufficient for separating pions, kaons, and protons at low to intermediate momenta. Over time, TOF detectors became standard components in a variety of experiments (like NA49~\cite{AFANASIEV1999210}, NA56/Spy~\cite{Ambrosinietal.1999} at the CERN SpS, CDF TOF~\cite{CABRERA2002416} at Fermilab, STAR~\cite{LLOPE2004252} and PHENIX~\cite{CARLEN1999123} at Brookhaven National Laboratory, to cite a few examples), reaching time resolutions of $\sim$100~ps or better.
%The state-of-the-art time-of-flight detector installed presently in a HEP experiment is the ALICE TOF detector~\cite{CERN-LHCC-2000-012,Cortese:545834}, dedicated to the identification of charged particles in the intermediate momentum region (up to $\sim$5~GeV) in heavy ion collisions at the LHC. The ALICE TOF has achieved an in-situ time resolution of 56~ps using Multi Gap Resistive Plate Chambers~\cite{Carnesecchi:2018oss}.
The state-of-the-art time-of-flight detector installed presently in a HEP experiment is the ALICE TOF~\cite{CERN-LHCC-2000-012,Cortese:545834}, a large area detector dedicated to the identification of charged particles in the intermediate momentum region (up to $\sim$5~GeV) in heavy ion collisions at the CERN Large Hadron Collider. Based on Multi Gap Resistive Plate Chambers, it has achieved an in-situ time resolution of 56~ps~\cite{Carnesecchi:2018oss}.

With the increasing complexity of collider environments and the prospects of high-luminosity running conditions, the scope of timing detectors has expanded beyond traditional TOF-based PID. The focus and technology developments have shifted toward high-precision timing layers with resolutions in the tens of picoseconds, used not only for particle identification but also to mitigate pileup effects. In addition, precision timing offers the opportunity to expand the physics reach in searches for Beyond the Standard Model (BSM) long-lived unstable particles.

\subsection{Challenges and opportunities from precision timing measurements}

\paragraph{Particle identification through time-of-flight} 
The principle of particle identification through time-of-flight is based on the measurement of the particle flight time $t$ across a known distance $L$ to obtain the particle velocity, $\beta = L/ct$, combined with a measurement of its momentum $p$. The time-of-flight difference $\Delta t_{TOF}$ of two particles with equal momentum but different masses, and thus velocity, is: 

\begin{equation}
\Delta t_{TOF} =  \frac{L}{c} \left(\frac{1}{\beta_1} - \frac{1}{\beta_2}\right)
\end{equation}

\begin{figure}[!h]
\centering
\includegraphics[width=0.80\linewidth]{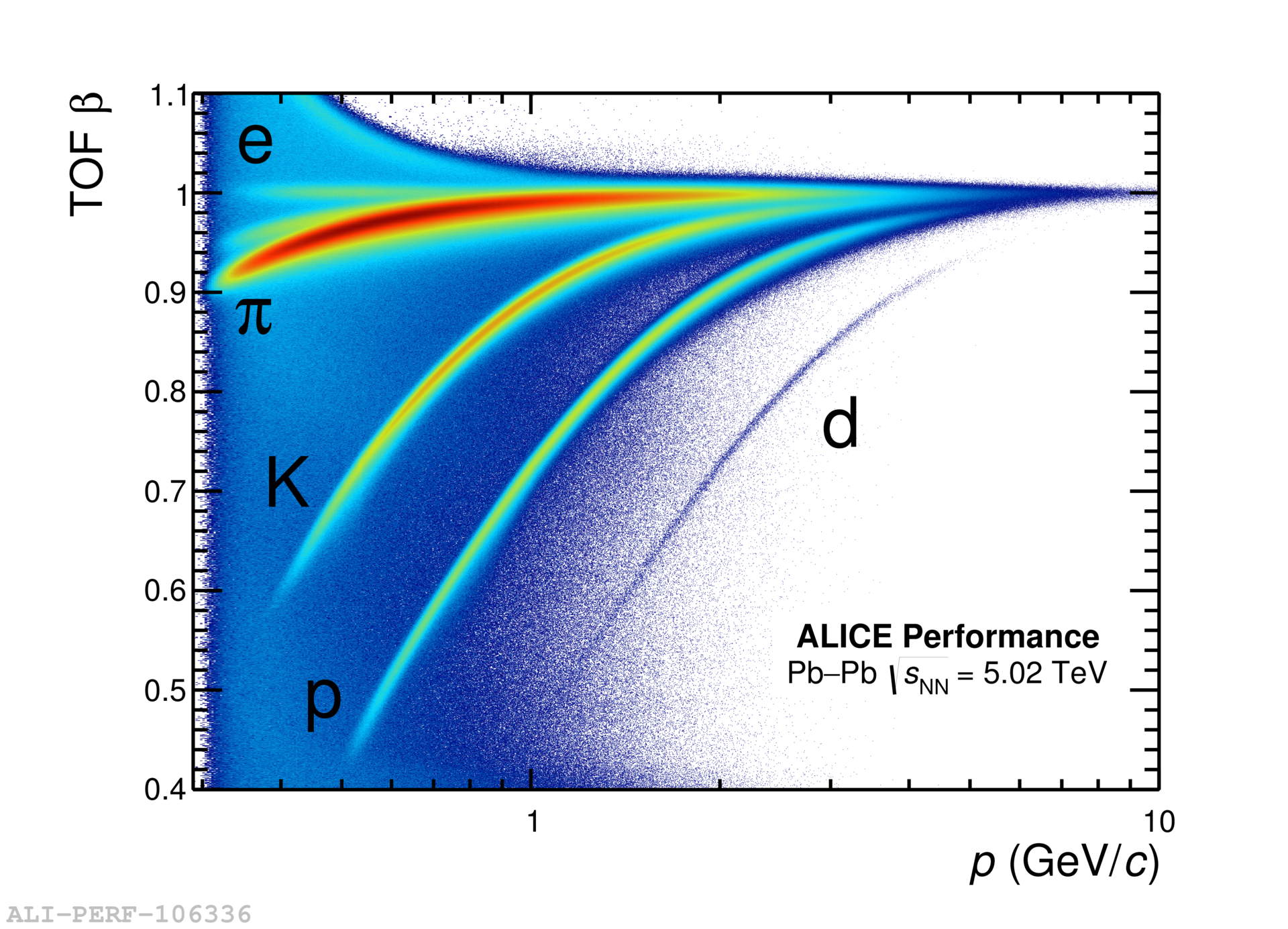}
\caption{ Particle velocity $\beta = v/c$ measured by the ALICE TOF as a function of momentum in Pb–Pb collisions at $\sqrt{s_{NN}}$ = 5.02~TeV~\cite{Carnesecchi:2018oss}.}
\label{fig:tofpid}
\end{figure}

As an example, Fig.~\ref{fig:tofpid} reports the particle velocity $\beta$ measured by the ALICE TOF as a function of the particle momentum, showing a good separation between different particles species up to $\sim$5~GeV. 
TOF particle identification is an important element for heavy ions and heavy flavour physics. The integration of new dedicated timing layers with a precision of a few tens of picoseconds in detectors at future colliders will add or further enhance TOF particle identification capabilities in the experiments.
%TOF particle identification capabilities will be further expanded in the future ALICE3 upgrade (see Sec.~\ref{sec:noveltechnologies}). Precision timing with the MIP Timing detector of the CMS HL-LHC upgrade will also enable new particle identification capabilities to the experiment, to be exploited in the context of charmed hadrons reconstruction in heavy-ion collisions measurements and in the study of exclusive B-hadrons decays~\cite{CERN-LHCC-2019-003}.

\paragraph{Pileup mitigation}
The upcoming high luminosity phase of the LHC, currently scheduled to start in 2030, will see an increase of the instantaneous luminosity up to 5.0-7.5~$\times$~10$^{34}$~cm$^{-2}$s$^{-1}$ aiming to collect over 3~ab$^{-1}$ of data in ten years of operation. This large amount of data, ten times larger than the current one, will allow experiments to perform precision measurements of the Higgs boson and of the fundamental interactions of the Standard Model (SM), and to explore the potential existence of new rare phenomena.

\begin{figure}[!h]
\centering
\includegraphics[width=0.47\linewidth,trim=0 -1.4cm 0 0, clip]{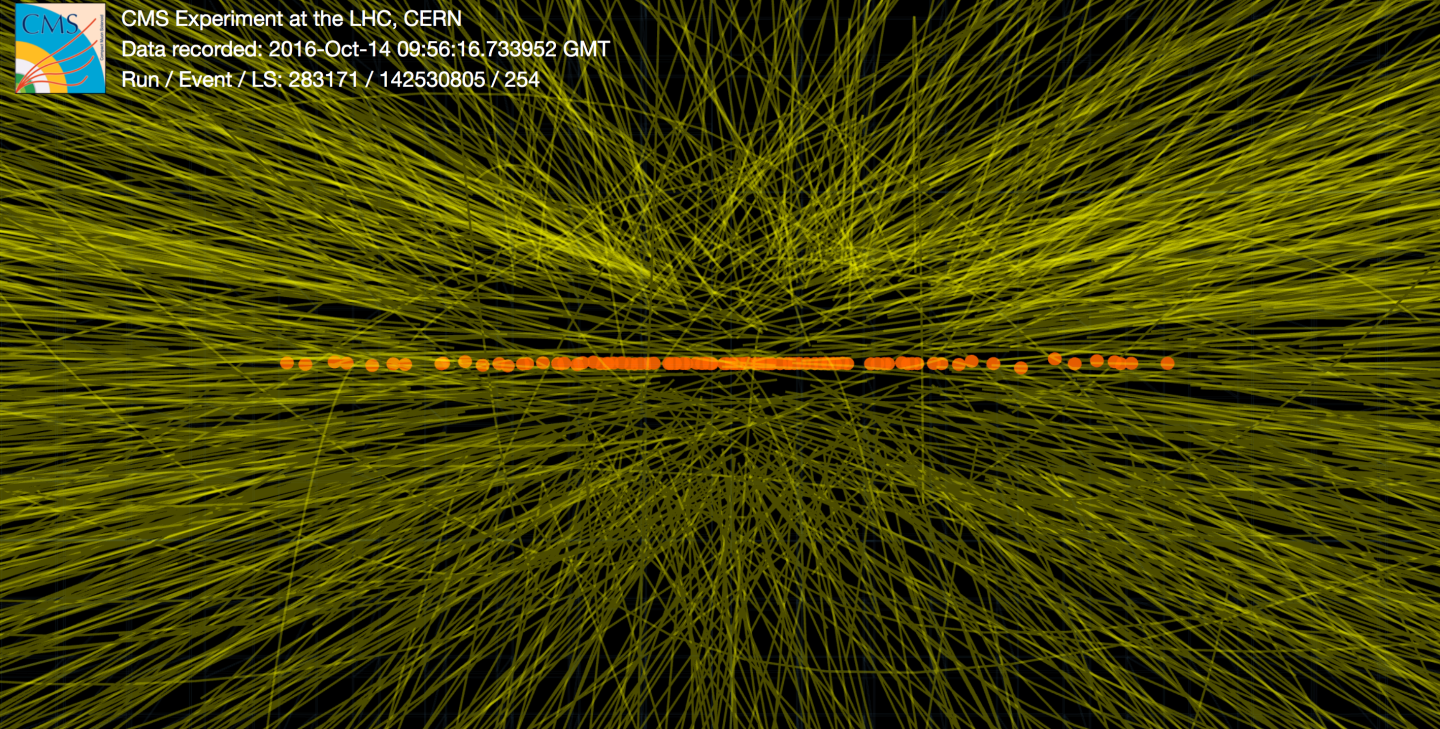}
\includegraphics[width=0.51\linewidth]{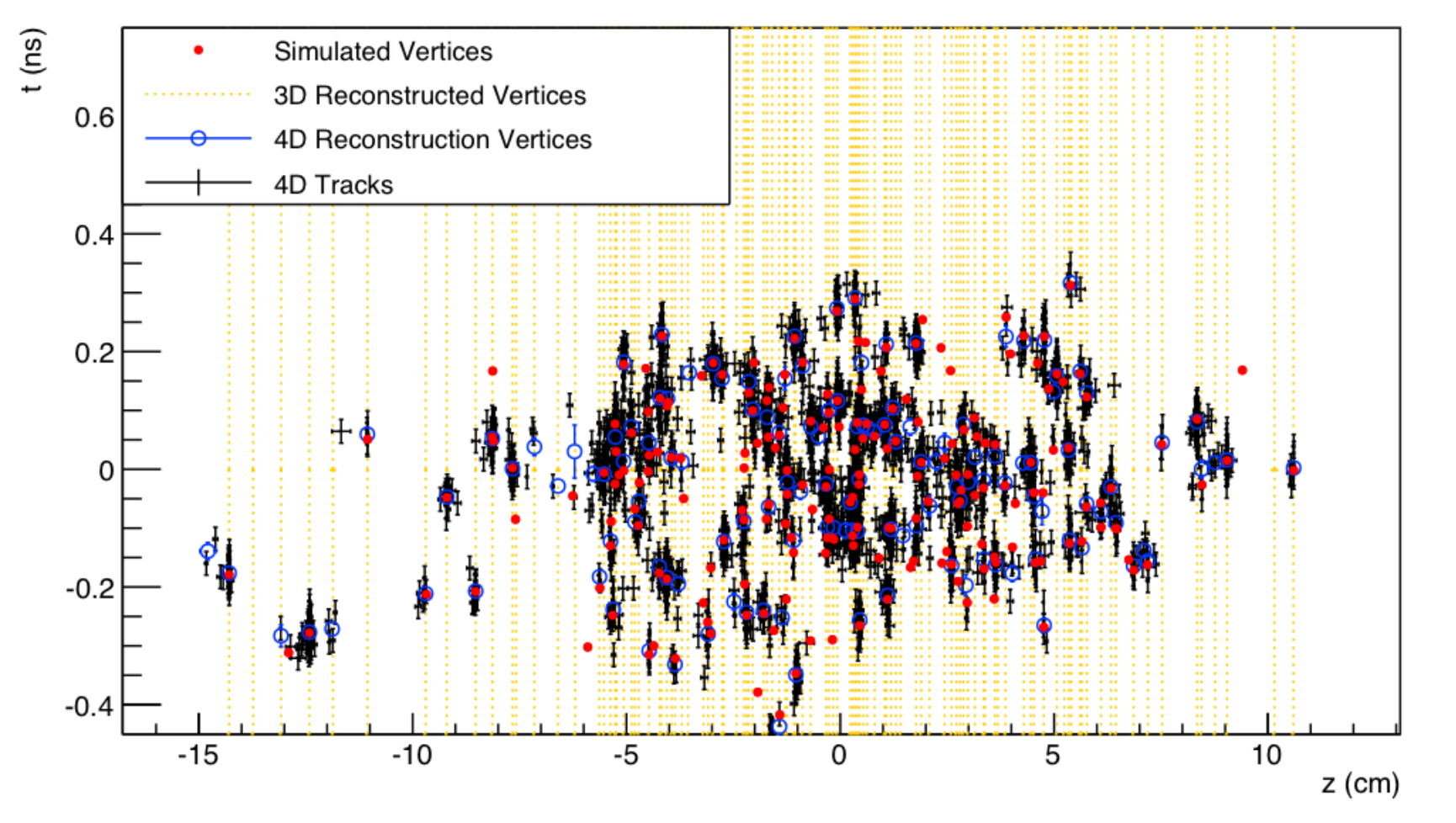}
\caption{Left: event display showing reconstructed tracks and vertices in a real collision event from a high pileup run in 2016 with 130 vertices~\cite{Collaboration:2231915}. Right: simulated (red-dots) and reconstructed vertices (blue) and tracks (black) in 200 pileup collisions using 4D tracking; vertices merged in 3D (yellow lines) are separated in 4D~\cite{CERN-LHCC-2019-003}.}
\label{fig:pileup}
\end{figure}

One main challenge arising from high intensities is the large increase in the number of concurrent interactions that occur in one proton-proton bunch crossing, referred to as pileup. At the HL-LHC, there will be 140–200 pileup events per bunch crossing, to be compared to the current $\sim$60. In this condition, the vertex and track density will be so large that events will overlap in space (as shown in Fig.~\ref{fig:pileup}), challenging the precision of the tracking systems. The spatial overlap of tracks and energy deposits can degrade the identification and reconstruction of the hard interaction and increase the rate of false triggers.
Precision timing is a powerful handle to cope with these demanding experimental conditions. 
%To cope with these demanding experimental conditions, experiments at the LHC are undergoing an extensive upgrade program which includes the installation of new dedicated timing layers, like the ATLAS High Granularity Calorimeter (HGTD)~\cite{CERN-LHCC-2020-007} and the CMS MIP Timing Detector (MTD)~\cite{CERN-LHCC-2019-003}, to precisely measure the time of arrival of charged particles as a powerful handle for pileup mitigation. 
Due to the longitudinal spread of the luminous region along the beam direction, interactions are spread in time with an RMS of about 180~ps. A precise measurement of the time of arrival of charged particles with a time resolution of the order of a few tens of picoseconds allows resolving in the time dimension interaction vertices which are overlapping in space, as illustrated in Fig.~\ref{fig:pileup}-right, recovering the effective level of pileup of the current LHC.
Both the ATLAS and CMS experiments have opted
to include in their upgrades for the HL-LHC phase new dedicated timing layers: the High Granularity Timing Detector (HGTD)~\cite{CERN-LHCC-2020-007} and the MIP Timing Detector (MTD)~\cite{CERN-LHCC-2019-003}, respectively.

\begin{figure}[!h]
\centering
\includegraphics[width=0.56\linewidth,trim=0 -1cm 0 0, clip]{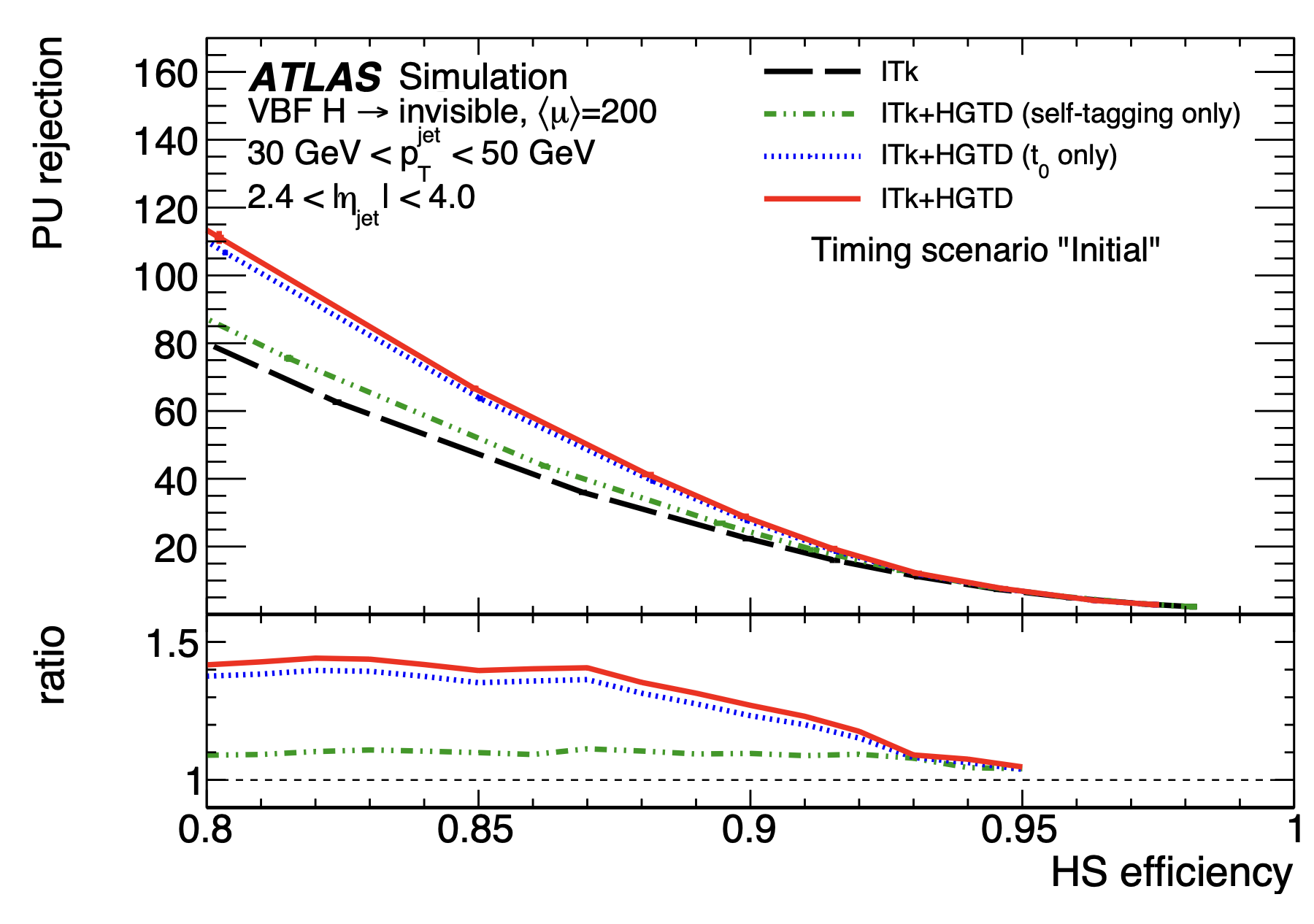}
\includegraphics[width=0.42\linewidth]{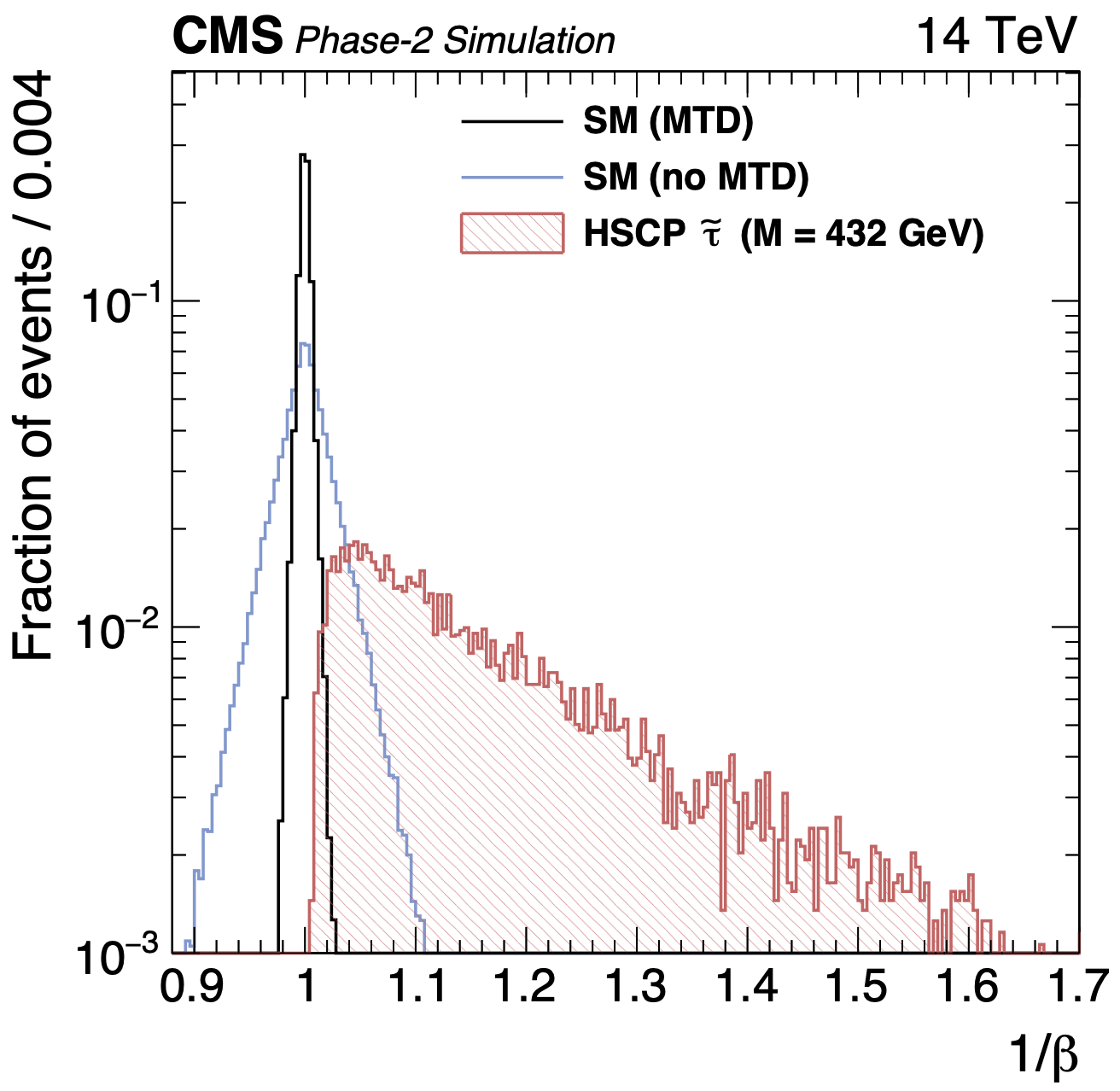}
\caption{Left: pileup jet rejection as a function of hard-scatter jet efficiency with and without using timing information from the ATLAS High Granularity Timing Detector~\cite{CERN-LHCC-2020-007}. Right: distribution of $1/\beta$ for Heavy Stable Charged Particle signal events and background events from the SM Drell Yan + jets production,  with and without using the precision track time information provided by the CMS MIP Timing Detector~\cite{CERN-LHCC-2019-003}.}
\label{fig:physicsPerformance}
\end{figure}

Removing pileup tracks that are inconsistent with the primary hard interaction significantly enhances the quality of most event observables, leading to improved sensitivity across a broad spectrum of physics measurements. For example, the suppression of pileup jets (Fig.~\ref{fig:physicsPerformance}-left) and improved lepton isolation efficiency in the forward detector region contribute to increased sensitivity in processes such as vector boson fusion, vector boson scattering, and lepton-based forward-backward asymmetry measurements, as documented in the ATLAS HGTD Technical Design Report (TDR)~\cite{CERN-LHCC-2020-007}.
The study of di-Higgs production, one of the major goals of the HL-LHC physics program, will also largely benefit from improvements in jet reconstruction, tagging of jets originating from b-quarks, and enhanced identification and isolation of leptons and photons. Additional performance studies and projections can be found in the CMS MTD TDR~\cite{CERN-LHCC-2019-003}.

\paragraph{Precision timing for BSM searches}
Precision timing can also allow expanding the reach and strategies for searches of new particles, like long-lived particles (LLPs) or heavy stable charged particles, which are predicted in several extensions of the Standard Model. LLPs may have lifetimes long enough to travel significant distances in the detector before decaying, often resulting in final-state signatures with delayed objects such as leptons, photons, or jets. As reported in~\cite{CERN-LHCC-2019-003}, precision track timing with 30~ps resolution can be exploited to extend the sensitivity to LLPs in different ways. For neutrals LLPs with delayed photons signatures, the precise measurement of the time of the primary vertex is a key aspect to measure accurately the photon time-of-flight, which would be otherwise dominated by the time spread of the luminous region, and infer the LLP lifetime. For charged LLPs, the possibility to reconstruct the time of primary and secondary vertices provide kinematic constraints enabling a direct measurement of the LLP mass. In the case of Heavy Stable Charged Particles (HSCP), the measurement the particle velocity, $\beta$, from the path length and the time difference between the primary vertex and the particle hits in the timing layer, is useful to discriminate between signal and background SM processes, where particles typically travel at velocities close to the speed of light (Fig.~\ref{fig:physicsPerformance}-right).

\vspace{1cm}
This review aims to provide an overview of timing detectors in collider experiments, focusing on the main underlying technologies, current implementations, and future directions. We begin by introducing the physical principles governing timing measurements and the main factors influencing the time resolution of a detector (Sec.~\ref{sec:principles}). We then examine in details a selection of sensors technologies, namely scintillators coupled to photo-detectors (Sec.~\ref{sec:sensor-scintillator}), silicon-based (Sec.~\ref{sec:sensor-lgad}) and gas-based detectors (Sec.~\ref{sec:sensor-mrpc}), focusing on those applied either in operational HEP experiments or in detectors which are under construction. The integration of timing detectors into major collider experiments, with particular attention to the upgrades planned for the HL-LHC, is presented in Sec.~\ref{sec:integration}. Finally, we discuss the emerging R\&D efforts aimed at pushing timing performance even further in the context of next-generation collider facilities.
\section{Basic principles in timing measurements}
\label{sec:principles}
%A timing detector is a device designed to measure accurately the time of arrival of the particle hitting the sensor.

\subsection{Interaction of particles passing through matter}
%- Charged particles: ionization/excitation
%   - energy e-h pair
%   - average energy losso dE/dx Bethe Block
%   - minimum ionizing particles
% Photons
    %- interaction via photo-electric, Compton, pair prod, with probabilities depending on the energy and material (Z)
    %- photo-electric important for example for photo-multipliers where the photon in converted to an electron....

The basic principle of particle detectors, including those designed for timing measurements, is transferring part or all the particle's energy to the sensor and its conversion to a measurable signal. 

Heavy charged particles primarily interact with matter through Coulomb interactions with atomic electrons. This interaction can lead to excitation, where an electron is promoted to a higher energy level within the absorber atom, or ionization, which occurs when the energy transferred is sufficient to completely remove the electron from the atom. 

The average energy loss per unit length, denoted as $dE/dx$ and referred to as stopping power, is well described by the Bethe-Block formula~\cite{Bethe:1934za}. The value of $dE/dx$ approaches a near-constant broad minimum of about 2~MeV g$^{-1}$cm$^2$ for particles with $\beta\gamma\approx 3-4$, where $\beta=v/c$ is the particle's speed. In practice, most relativistic particles have mean energy loss rates close to the minimum and are known as {\it minimum ionizing particles} (MIPs). In addition to ionization and excitation, charged particles can produce Cherenkov radiation if they travel faster than the phase velocity of light in the medium.

Photons can interact with matter mainly through photo-electric effect, Compton scattering, and pair production, with probabilities depending on their energy and  the material properties   

The way the converted energy transforms into a detectable signal depends on the type and design of the detector. In silicon or gas-based detectors, the motion of charge carriers produced by ionization generates a current signal. In scintillator materials, the energy lost by particles is converted into optical photons, which are then transformed into an electrical signal using a photo-detector.

\subsection{Typical chain for timing measurements}
\begin{figure}[h]
\centering
\includegraphics[width=\linewidth]{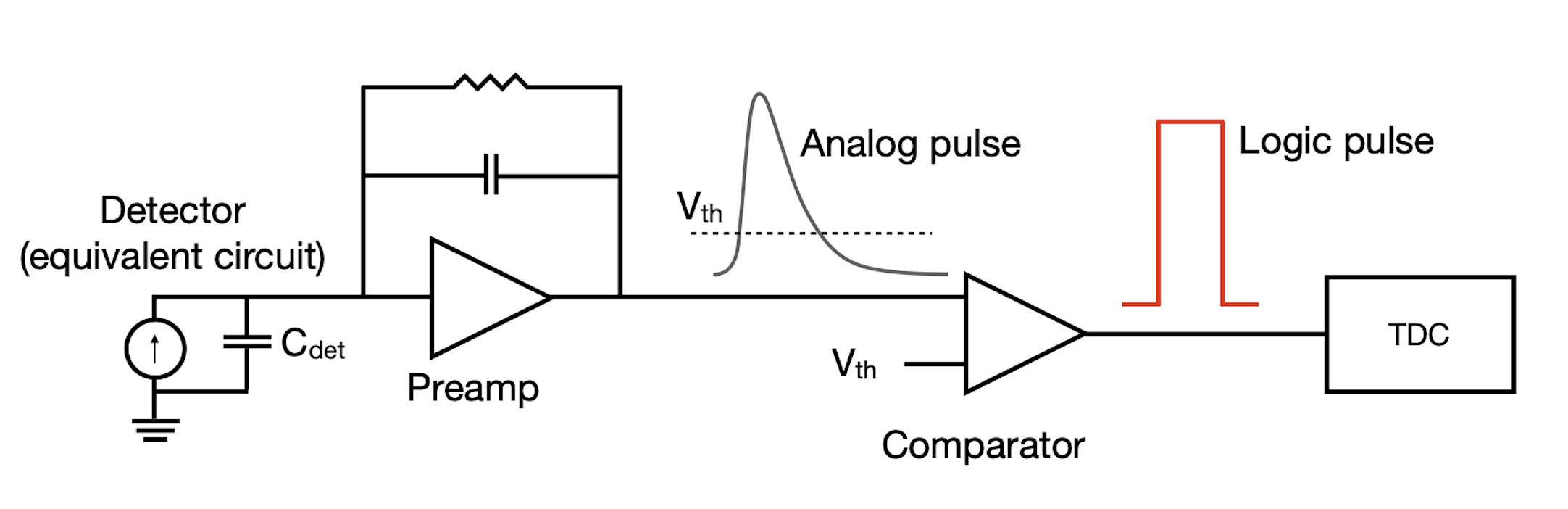}
\caption{Schematic representation of the typical chain for timing measurements. Adapted from \cite{Rivetti}.}
\label{fig:timing_chain}
\end{figure}

A simplified representation of the typical chain for timing measurements is reported in Fig.~\ref{fig:timing_chain}. The current signal from the sensor is usually amplified and shaped and then fed to a discriminator. If the signal is above a given threshold (V$_{th}$), a comparator generates a logic pulse whose leading edge indicates the time of occurrence of the input analog pulse. The time of the logic pulse, relative to a clock, is eventually digitized by a time-to-digital converter (TDC).

\subsection{Factors influencing time resolution}
From Fig.~\ref{fig:timing_chain}, one can realize that the time resolution of a detector depends on several factors related to both the sensor and electronics. In general, it can be expressed as the sum in quadrature of different contributions:

\begin{equation}
\sigma^2_{t} = \sigma^2_{jitter} + \sigma^2_{S} +  \sigma^2_{time-walk} + \sigma^2_{TDC} + \sigma^2_{clock}
\label{eq:timeResolution}
\end{equation}

where $\sigma_{jitter}$ is the term due to noise, $\sigma_{S}$ is related to stochastic fluctuations in the signal formation, $\sigma_{time-walk}$ is due to the time-walk effect, $\sigma_{TDC}$ is the uncertainty from the TDC and $\sigma_{clock}$ is the contribution due to the jitter of the reference clock.

\subsubsection{Jitter}
\begin{figure}[h]
\centering
\includegraphics[width=0.8\linewidth]{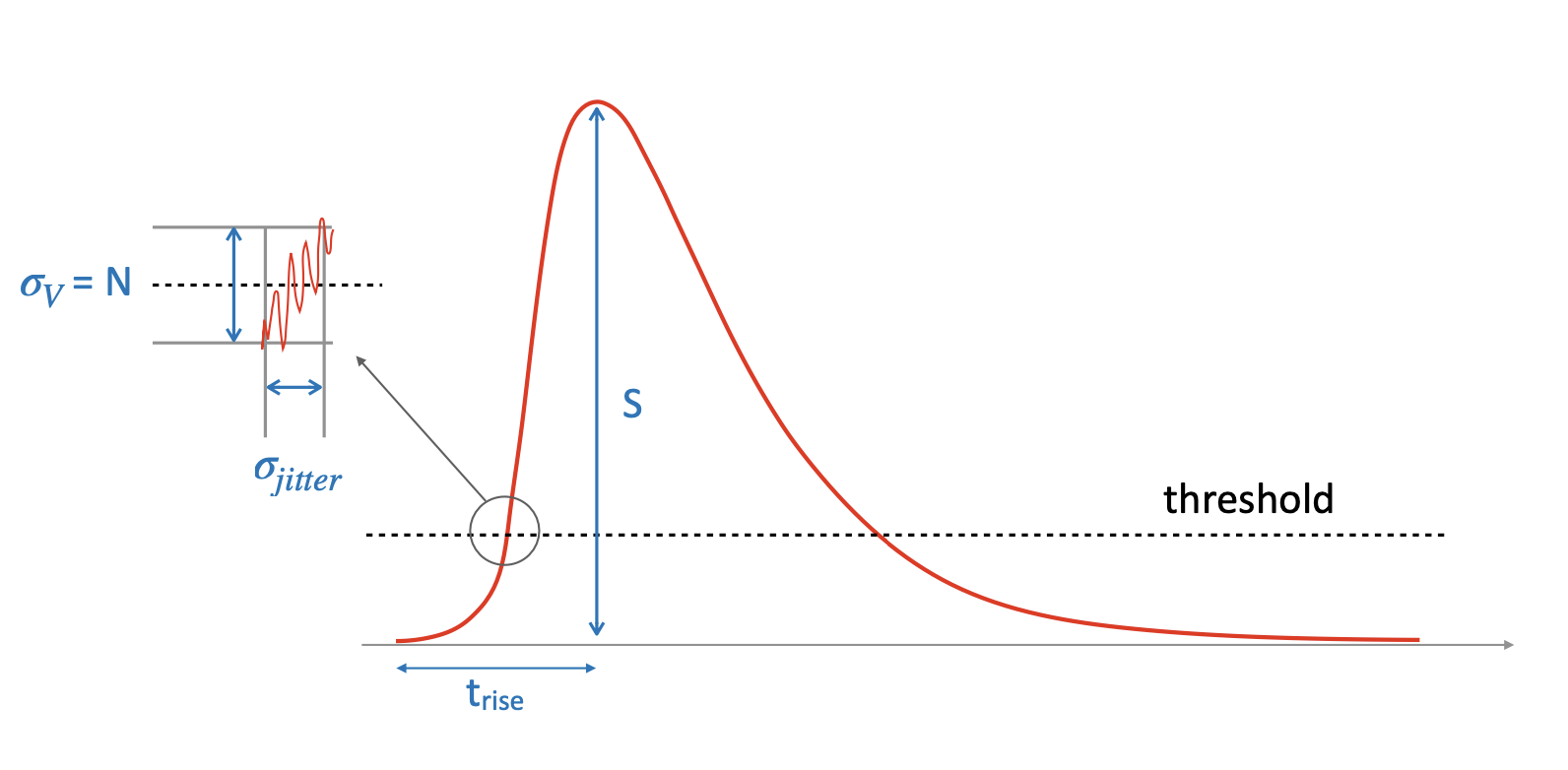}
\caption{Effect of the noise on time measurements.}
\label{fig:jitter}
\end{figure}

The jitter or noise term, $\sigma_{jitter}$, is related to the presence of noise on the signal, either from the electronics or the sensor itself. The noise causes fluctuations in the time at which the discriminator threshold is crossed, as represented in Fig.~\ref{fig:jitter}. The jitter term depends on the noise N and the inverse of the slope dV/dt of the signal at the timing threshold V$_{th}$, as 

\begin{equation}
\sigma_{jitter} = \frac{N}{dV/dt} \approx \frac{t_{r}}{S/N}
\end{equation}

\noindent where, assuming a constant slope, dV/dt is approximated by the ratio between the signal amplitude S and the signal rise time t$_r$.
Fast rise time and a high signal-over-noise ratio S/N are fundamental to achieving a good time resolution. For instance, for a typical rise time of the order of 200 ps, a S/N$\sim$10 is needed to keep this contribution to the time resolution at the level of 20~ps. 
Since the rise time is determined by both the sensor rise time and the amplifier rise time, the minimization of the jitter term requires an accurate optimization of the readout electronics. 

\subsubsection{Stochastic fluctuations in the signal formation}
The signal from the detector originates from the energy deposition of a particle in the sensitive volume. One intrinsic limit to time precision comes thus from statistical fluctuations of the deposited energy, which result in variations of the signal amplitude and shape event by event.
Non-uniformities in the energy deposition or dependencies of the response on the impact point of the particle can also affect the signal shape and consequently spoil the time resolution. 

\subsubsection{Time-walk}
\label{sec:timewalk}
\begin{figure}[h]
\centering
\includegraphics[width=0.7\linewidth]{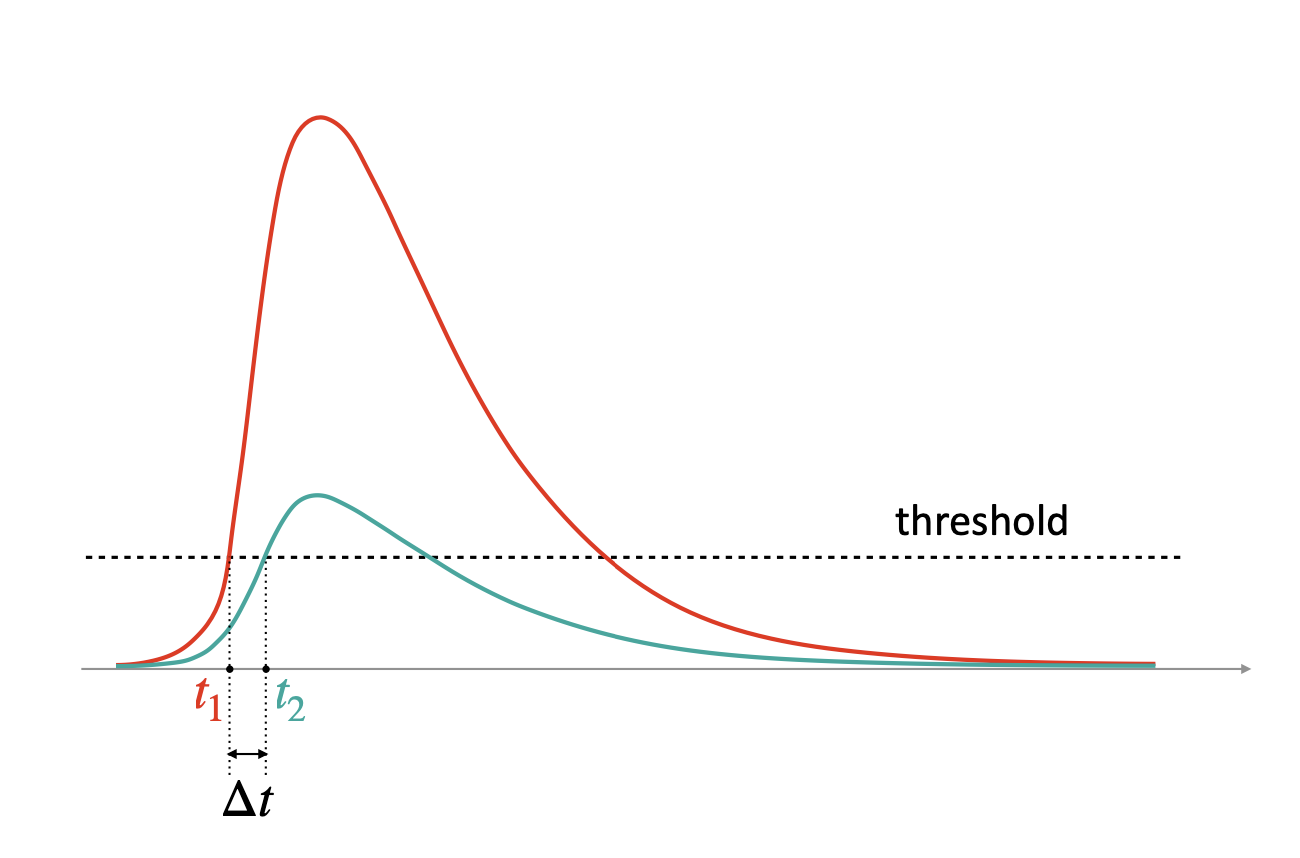}
\caption{Time-walk effect: signals of different amplitudes, starting at the same time, are detected at two different times t$_1$, t$_2$.}
\label{fig:time-walk}
\end{figure}

An important source of inaccuracy in timing measurements is the {\it time-walk} effect, which arises from the interplay between signal amplitude variations and the time pick-off method. This effect is illustrated in Fig.~\ref{fig:time-walk}. Signals of different amplitudes, starting at the same time, cross the fixed leading edge discrimination threshold at different times. The resulting delay $\Delta t$ can be approximated by the linear relation $\Delta t \sim t_r V_{th}/S$. The contribution to the time resolution due to the time-walk effect is the RMS of the $\Delta t$ distribution
\begin{equation}
\sigma_{time-walk} = [\Delta t]_{RMS} = \left[\frac{t_r V_{th}}{S}\right]_{RMS} \propto   \left[\frac{N}{dV/dt}\right]_{RMS}
\label{eq:twalk}
\end{equation}

The time-walk effect cannot be avoided in systems based on a fixed discrimination threshold. Following Eq.~\ref{eq:twalk}, its impact on the time resolution can be reduced by increasing the steepness of the signals and lowering the timing threshold, provided that the noise level permits it. There are also several methods to correct the time-walk effect, either through hardware or software.
Constant Fraction Discriminators (CFD) measure the time when a given fraction of the amplitude is reached. The principle of operation is based on detecting the zero-crossing of the bipolar pulse obtained by subtracting a fraction of the input signal from a delayed copy of it~\cite{GEDCKE1967377, GEDCKE1968253}. The time of zero crossing of the resulting pulse does not depend on the signal amplitude. 
Time-walk can be corrected offline by exploiting the additional measurement of a quantity proportional to the signal amplitude, like the integrated charge of the pulse or the time-over-threshold (ToT), which corresponds to the time interval during which the signal stays above the fixed threshold.
Another possibility is to use a system with fine waveform sampling and process the digitized waveforms offline. While this approach can potentially offer the best performance, it is usually impractical in high-energy physics particle detectors with a large number of read-out channels because it is very demanding in terms of computing power and bandwidth.

\subsubsection{TDC}
The term $\sigma_{TDC}$ is due to the finite TDC bin size ($\Delta t_{TDC}$) and corresponds to

\begin{equation}
\sigma_{TDC} = \frac{\Delta t_{TDC}}{\sqrt{12}}
\end{equation}

This contribution is usually made small by design thanks
to the fine binning of TDCs commonly used in high energy physics experiments: for instance, the CMS Barrel Timing Layer ASIC (TOFHIR2~\cite{Albuquerque_2024}) has a TDC with 10~ps binning, the CMS Endcap Timing Layer ETROC and HGTD ALTIROC feature 20~ps binning, which make the TDC bin size contribution to the time resolution negligible.

\subsubsection{Clock}
Timing measurements are usually performed using a reference clock.
The clock distribution, especially in large systems with many channels, must be carefully designed. Any jitter or instability in the reference clock directly results in a jitter of the timing measurement, degrading the resolution. 

At the LHC, the precision clock is synchronized to the 40~MHz LHC bunch frequency and transmitted to the detectors via high-speed data links. To retain the excellent resolution of precision timing detectors such as the CMS MTD and the ATLAS HGTD, which will be installed for the HL-LHC phase, the clock jitter must be kept below 15~ps RMS. This clock jitter includes both the contribution from the radiofrequency (RF) delivered by the LHC, which for the current system is at the level of 9~ps~\cite{SBaron_2012}, and the contribution from the clock distribution to the readout of the different subdetectors. Various effects, like differences in clock paths (fibre/cable lengths) or temperature variations which can affect the time propagation along the fibres, can generate relative time offsets across the systems. The monitoring of the channel synchronization and calibration of possible offsets are therefore essential aspects in large timing systems.
% replace sensor with 3 standalone chapters 20250207
%\input{Sensors}
\section{Scintillators coupled to photo-detectors}
\label{sec:sensor-scintillator}

In recent years, scintillators have demonstrated excellent time resolution at the order of $O(10)$ ps in detecting MIPs and have been adopted in modern large-scale experiments at colliders. Both inorganic scintillators featured with high light yield and organic scintillators with relatively shorter decay time can find their applications suitable for high-performance timing measurements. The scintillators are coupled to photo-detectors to convert the scintillation light into electrons. The resulting electrical signal is then amplified and processed for precise timing or energy measurements.

\subsection{Formation of scintillation light}
\label{sec:sci-light}
%{\bf - Principle of detection: dE/dx ionizaion/excitation --$>$ converted to visible or near to visible photons}

The incident charged and neutral particles interact with the scintillation material that causes an amount of the energy loss (-dE/dx)~\cite{Rossi:99081,Fano:1963xu,Fleck:2021njs,Fabjan:2020wnt}. A fraction of this energy leads to excited states of the atoms or molecules or dedicated luminescent centers that eventually generate light in the scintillator. The dominant form of the energy loss depends on the energy and the type of the incident particle. Charged particles mainly interact with the electrons of the scintillator and lose energy via ionization at lower energy, while the radiative energy loss via bremsstrahlung becomes dominant when the energy rises. Photons interact with the electrons at lower energy, with energy loss dominantly from photoelectric absorption mainly below about 500~keV and from Compton scattering roughly between 800~keV and 3~MeV, while they are stimulated by atomic nucleus in high energy resulting in energy loss mainly via pair production dominating above roughly 10~MeV.

%%% inorganic scintillator

\begin{figure}[!h]
\centering
\includegraphics[width=1.0\linewidth]{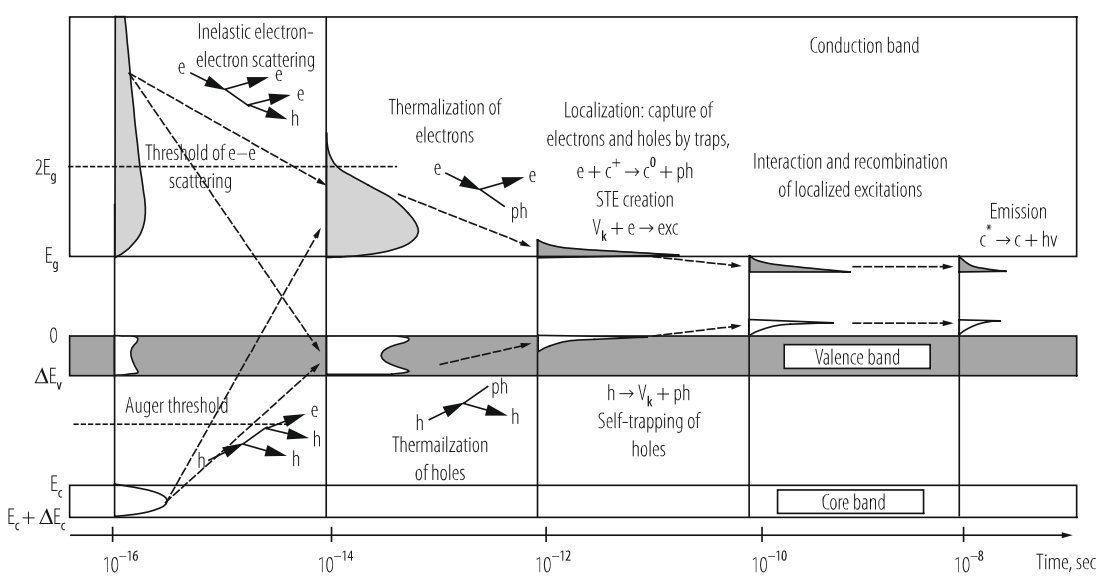}
\caption{Relaxation scheme for electronic excitations in an insulator: e, electrons; h, holes; ph, phonons; h$\nu$, photons; V$_k$, self-trapped holes; STE, self-trapped excitons; c$^n$, ionic centers with charge n. Density of states is represented by grey and white areas for electrons and holes respectively (courtesy A. Vasiliev)~\cite{vasiliev2000,Fabjan:2020wnt,Lecoq:2006tzy}.}
\label{fig:relaxation}
\end{figure}

In the inorganic scintillator, the electrons primarily produced in the above processes will undergo a chain of complex steps before the generation of the scintillation light, which is referred as the relaxation of electric excitations~\cite{Rodnyi95,BARTRAM1996225,VASILEV1996165}. A simplified diagram is proposed by A. Vasiliev provides a pedagogical description shown in Fig.~\ref{fig:relaxation}~\cite{vasiliev2000,Fabjan:2020wnt,Lecoq:2006tzy} in which the typical time of each step is indicated.
\begin{itemize}
\item The first step is the production of secondary and more particles from the primary particles ionized by the incident particle. These primary particles include the hot electrons e in the conduction band and the corresponding deep holes h in the inner core hands. They swiftly spread out and interact with the scintillator materials through inelastic electron-electron scattering and Auger processes, generating secondary and more electrons in the conduction band and holes in core and valence bands. The cascade keeps evolving till the energy of electrons in the conduction band is no longer higher than 2E$_g$ (the bandgap width) to kick more electrons to the conduction band and all the hole reach the valence band making Auger process impossible. The typical time of this step is $10^{-16}$ to $10^{-14}$ s.
\item The second step is the thermalization of the electrons and holes from the last step via phonons with the crystal lattice. Electrons eventually reach to the bottom of the conduction band and holes to the top of the valence band. This step takes time of sub-picosecond, $10^{-14}$ to $10^{-12}$ s.
\item The third step is the localization of the electrons and holes by capture with self-trapping, stable defects and impurities of the material. This takes $10^{-12}$ to $10^{-10}$ s, entering sub-nanosecond era.
\item The fourth step is the energy transfer from excitations to the luminescent centers taking $10^{-10}$ to $10^{-8}$ s.
\item The last step is the production of the scintillation light from the luminescent centers which could range from sub-nanosecond to hundreds of nanosecond depending on the structure of the scintillator.
\end{itemize}
\noindent After the relaxation, the scintillation light will be transported to the photo-detector. The energy of the emitted photons should be smaller than the forbidden energy bandgap of the scintillation material to avoid absorption.

The luminescent centers of the inorganic scintillator can be the crystal lattice itself or a dopant such as thallium (Tl) or cerium (Ce), for scintillation photon production. In general, inorganic crystals have a much higher density of 5-10 g/cm$^{-3}$ with respect to organic materials, leading to strong stopping power. This results in more energy deposit per distance, making a compact detector design possible and providing signals with high significance for energy and time measurements. The light yield ranges from about 10,000 to 80,000 per MeV. Pages 433-434 of Ref.~\cite{Fleck:2021njs} list the characteristics of a variety of crystals. BaF$_2$, the fastest crystal scintillator known so far, has a fast component (about 10\% of the total light yield) with a less than 0.6 ns decay time and a slow component with 630 ns decay time that is shown to be reduced by doping with yttrium in recent studies~\cite{Zhu:2018jpp,8720279}. The fast component is due to a special configuration of the energy bands in which the forbidden energy bandgap is larger than the total value of the energy gap between the upper core and valence bands and the width of the valence band. Holes that emerge in this core band could not provide enough energy for an Auger electron from the valence to the conduction band when recombining with an electron from the valence band, resulting in only radiative emission of the energy in core-valence transition in the UV range that is very fast. This mechanism is referred as core-valence luminescence (CVL), cross-luminescence or Auger-free luminescence~\cite{rodnyi1992core,VANEIJK1994936,itoh1997temperature,KHANIN2023113399}. Lutetium oxyorthosilicate doped with Cerium (Lu$_2$SiO$_5$:Ce, or LSO:Ce)~\cite{MELCHER1992212} and lutetium-yttrium oxyorthosilicate doped with Cerium (Lu$_{2(1-x)}$Y$_{2x}$SiO$_5$, LYSO:Ce)~\cite{10.1063/1.1328775} with a high density have a high light yield of above 30,000 photons per MeV and a fast decay time of 40 ns with a feature of excellent radiation hardness~\cite{1462080,4291695}. Cerium doped lanthanum tri-halides, such as LaBr$_3$ and CeBr$_3$, have even higher light yield and faster decay time than LYSO, but they are lower in density and need special treatment in practice due to high hygroscopic nature. Properties of typical inorganic crystals are listed in Tab~\ref{tab:crystal}.

\begin{table}[!ht]
    \centering
    \begin{threeparttable}[b]
    \begin{scriptsize}
    \renewcommand{\arraystretch}{1.2}
    \begin{tabular}{cccccccc}
    \hline
        Scintillator & $\rho$ & $\tau_d$ & $\lambda_{max}$ & $n$ & Light yield & Hygro- & Price \\
        ~ & [g/cm$^3$] & [ns] & [nm] & ~ & [ph/MeV] & scopic & [USD/cm$^3$] \\  \hline
        Nal:Tl & 3.67 & 245 & 410 & 1.85 & 37700 & yes & 4-8  \\
        BGO & 7.13 & 300 & 480 & 2.15 & 8500 & no & 13-30  \\
        (Bi$_4$Ge$_3$O$_{12}$) & & & & & & &\\
        GAGG:Ce & 6.7 & 53 & 520 & 1.9 & 42217 & no & 200   \\
        (Gd$_3$Al$_2$Ga$_3$O$_{12}$) & & & & & & & \\
        BaF$_2$ & 4.89 & 650$^s$ & 300$^s$ & 1.5 & 10000$^s$ & no & 45   \\
        ~ & ~ & $<$0.6$^f$ & 220$^f$ & ~ & 1400$^f$ & ~ &  \\
        Csl:Tl & 4.51 & 1220 & 550 & 1.79 & 64800 & slight & 7-12  \\
        Csl:Na & 4.51 & 690 & 420 & 1.84 & 38500 & yes & 7-10  \\
        Csl & 4.51 & 30$^s$ & 310 & 1.95 & 1357$^s$ & slight & -  \\
        ~ & ~ & 6$^f$ & ~ & ~ & 415$^f$ & ~ &  \\
        PbWO$_4$ & 8.3 & 30$^s$ & 425$^s$ & 2.2 & 113$^s$ & no & 20-40 \\ 
        ~ & ~ & 10$^f$ & 420$^f$ & ~ & 29$^f$ & ~ &   \\ 
        LYSO:Ce & 7.1 & 40 & 420 & 1.81 & 32000 & no & 70   \\ 
        LuAG:Pr & 6.7 & 21 & 310 & 1.85 & 17000 & no & 400  \\ 
        (Lu$_3$Al$_5$O$_{12}$:Pr) & & & & & & &\\
        YAP:Ce & 5.55 & 27 & 350 & 1.94 & 18000 & no & 125 \\ 
        (YAlO$_3$:Ce) & ~ & ~ & ~ & ~ & ~ & ~ &  \\ 
        CeF$_3$ & 6.16 & 30 & 340 & 1.62 & 2752 & no & -   \\ 
        LaBr$_3$:Ce & 5.29 & 20 & 356 & 1.9 & 63000 & yes & 135-175  \\ 
        CeBr$_3$ & 5.23 & 17 & 371 & 1.9 & 45000 & yes & 140-170  \\ \hline
    \end{tabular}
    \caption{Properties of typical inorganic crystals. $\rho$ is the density, $\tau_d$ the decay time of the scintillation light, $\lambda_{max}$ the wavelength of the scintillation light at maximum and $n$ the index of refraction at $\lambda_{max}$. For the scintillator that process multiply components in time, $^s$ refers to the slow component and $^f$ for the fast. Data come from Refs.~\cite{pdg:2024cfk,Fleck:2021njs,Fabjan:2020wnt}.}
    \end{scriptsize}
    \end{threeparttable}
    \label{tab:crystal}
\end{table}

%%% organic scintillator

\begin{figure}[!h]
\centering
\includegraphics[width=0.7\linewidth]{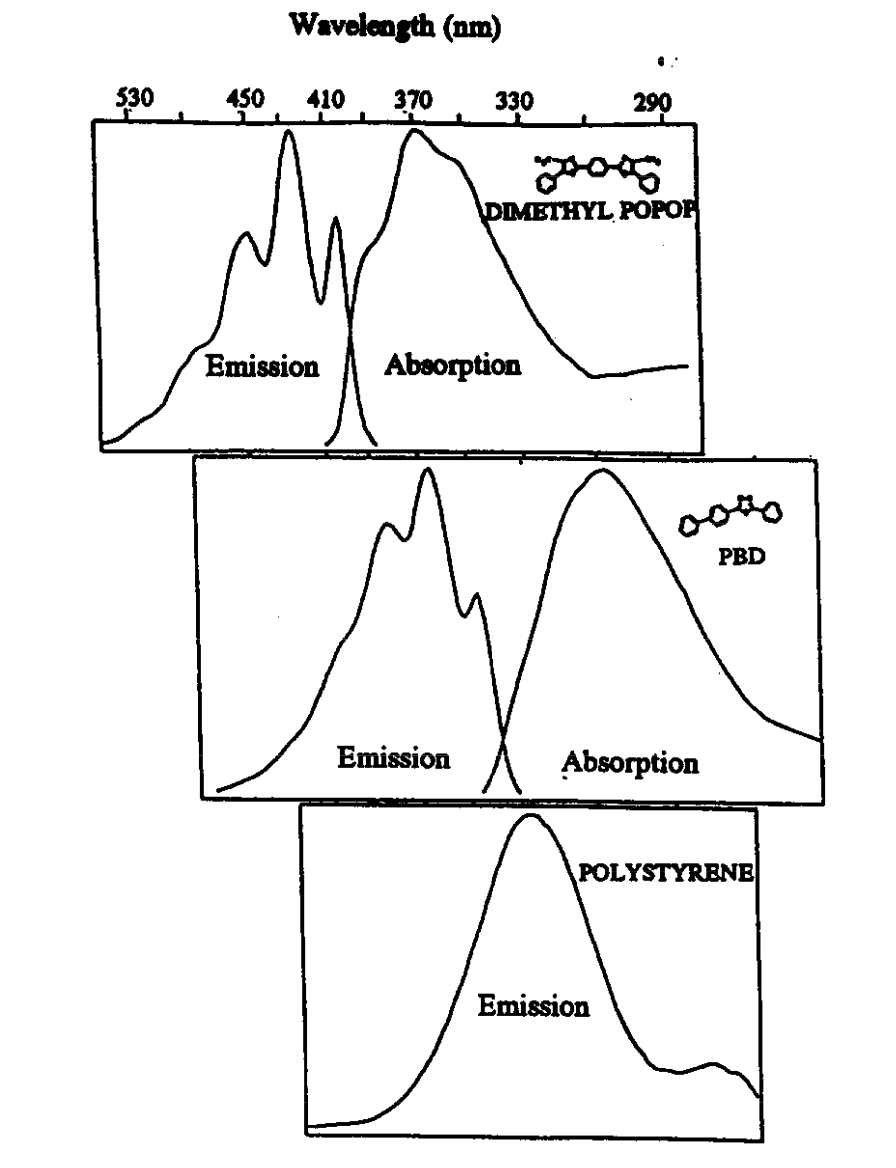}
\caption{Overlap of the emission and absorption spectra of a typical blue emitting plastic scintillator from Ref.~\cite{MOSER199331}}
\label{fig:plastic_spectra}
\end{figure}

Among organic scintillating materials, plastic scintillators that contain aromatic rings, such as polystyrene (PS) and polyvinyltoluene (PVT), have fast timing performance and are most widely used in collider experiments. Liquids that are mostly used in non-collider experiments also have good timing and are not discussed here. MIPs pass through the plastic scintillator leaving a wake of excited molecules. The de-excitation of these molecules in the plastic base can produce photons, but the quantum efficiency of the output energy is low, at a few percent (for example, toluene with a considerably high efficiency can only emit 9\% of input energy as optical outputs). The emission spectrum tupically peaks at 300-350 nm which reaches the limits of input ranges of most photo-detectors. The attenuation length is too short ($<$ 1 cm) making the plastic base not transparent to its own emission at the typical size of detector granularity (O(1) cm to O(1) m). To get over these issues, fluorescent materials are often introduced to the plastic base. A primary fluor is added to absorb the energy emitted by the plastic base and emit photons with a high quantum yield (usually close to 1.0), resulting in a longer wavelength typically of 350 - 400 nm and a longer attenuation length at O(10) cm. The primary fluor is dissolved with high concentration (1\% or larger) resulting in proximity to the molecules of the plastic base at a distance of 10$^{-8}$ m that is less than the wavelength of spectrum peak of the plastic base. At such a distance, a non-radiative dipole-dipole exchange known as Forster transfer becomes the predominant way of energy transfer and strongly shortens the decays time of the plastic base (for example, 16 ns of polystyrene) by an order of magnitude to a few nanoseconds. Furthermore, a secondary fluor can be added to shift the emission to a longer wavelength ($>$ 400 nm) and increase attenuation length up to 2 m and longer. Sometimes a third fluor is added to further extend the attenuation length. The transfer of the energy is radiative since the primary fluor. The overlap of the emission and absorption spectra of a typical blue emitting plastic scintillator is shown in Fig.~\ref{fig:plastic_spectra} from Ref.~\cite{MOSER199331}. A variety of fluors can be found in Appendix II.A of Ref.~\cite{Sauli:1992bu}. Eventually, the density ranges from 1.03 to 1.20 g/cm$^{-3}$ and the light yield ranges from 5,000 to 10,000 photons per MeV energy left by MIPs. Pages 441-442 of Ref.~\cite{Fleck:2021njs} list the properties of common plastic scintillators. The time resolution of plastic scintillators reaches a sub-nanosecond level~\cite{BIALKOWSKI1974221,BENGTSON1974227,MOSZYNSKI19791}.

\subsection{Conversion to electrons}

%%% photo-detector
The scintillation photons generated in the scintillating materials are then passed to the photo-detectors for the conversion to electrical signals. PMT (photomultiplier tube), MCP-PMT (micro-channel-plate photomultiplier tube) and SiPM (silicon photomultiplier, also called MPPC, multi-pixel photon counter) are often coupled to scintillators for timing measurement in high energy physics experiments. PMTs have a wide range of input spectrum from 115 to 1700 nm with a combined quantum efficiency ($\epsilon_Q$) and light collection efficiency ($\epsilon_C$) of 0.15-0.25. The gain is up to $10^{7}$ and rise time is down to 0.7 ns. The single photon time resolution is about 200 ps. MCP-PMT has a narrower range for the inputs from 115 to 650 nm with a lower $\epsilon_Q \epsilon_C$ of 0.01-0.10, but the rise time can be as low as 0.15 ns, leading to an excellent of single photon time resolution of about 20 ps. PMTs are in general vulnerable to magnetic field, while SiPMs are not. SiPMs also has a wide range of inputs from 125 to 1000 nm with $\epsilon_Q \epsilon_C$ of 0.15-0.40. The rise time is about 1 ns and the single photon time resolution can reach 50 ps~\cite{pdg:2024cfk}. Features of high photon detection efficiency, fast response, insensitivity to magnetic field, compactness in size, lower bias voltage (tens of volts) and so on, made SiPMs a popular choice for timing detection since 2010s~\cite{Gundacker_2020}.

\subsection{Time resolution}

%{\bf - time resolution of the scintillator determined by the density of photo-electrons at the detection threshold, which depends on the time distribution of photons being converted in the photo-detector. 
%
%- most important scintillator parameters influencing the time resolution are: the intrinsic scintillator light yield, the scintillator rise and decay time, the efficiency of light transport from the production point to the photo-detector:
%\begin{equation}
%    \sigma_S \propto \sqrt{\frac{\tau_r\tau_d}{LY}} 
%\end{equation}
%
%- best time resolution with scintillators with high light yield and fast decay and rise time ~\cite{Lecoq}. 
%}

In scintillators coupled to photo-detectors, the time resolution ($\sigma_{ph}$) is directly driven by factors of the rise ($\tau_r$) and decay time ($\tau_d$) of the scintillation light, and the light output in number of photoelectrons converted from the scintillation light at the photo-detector ($N_{pe}$). It is expressed in the following form derived from Hyman theory~\cite{10.1063/1.1718829,10.1063/1.1719516}.

\begin{equation}
  \sigma_{ph} \propto \sqrt{\frac{\tau_r \cdot \tau_d}{N_{pe}} }
  \label{eq:reso-sci}
\end{equation}

The rise time $\tau_r$ and the decay time $\tau_d$, determining the timing distribution of photoelectrons at the photo-detectors, play a key role in the time resolution. These are driven by the intrinsic properties of the scintillators. The origin of the rise time $\tau_r$ mainly come from the localization of electrons and holes at the end of the thermalization process and the energy transfer to the luminescent centers~\cite{6303850}, ranging up to tens of nanoseconds as described in the relaxation scheme in Sec.~\ref{sec:sci-light}. Ref~\cite{Fabjan:2020wnt} summarizes several techniques to improve the performance of the energy transfer. The spin-allowed dipole-dipole transitions observed between $5d$ and $4f$ energy levels in Ce$^{3+}$ and Pr$^{3+}$ can provide fast timing~\cite{Lecoq}. Cross-luminescence like BaF$_2$ introduced in Sec~\ref{sec:sci-light} is also intrinsically fast and provides temperature independent performance, although they emit deep-UV light that reaches the limit of photo-detectors. The strongly quenched intrinsic luminescence such as PWO can also generate fast light. The intrinsic and self-activated scintillators CeF$_3$, CdWO$_4$, Bi$_4$G$_3$O$_{12}$ and CsI have $\tau_r$ less than 30 ps, while the activated scintillators CaF$_2$:Eu, ZnO:Ga and Lu$_2$SiO$_5$:Ce have $\tau_r$ less than 40 ps~\cite{842466}. 

The decay time $\tau_d$ is mostly driven by the time of optical transition in the scintillator. The probability $\Gamma$ of radiative transitions in the medium with index of refraction $n$ is written as the following form according to Ref.~\cite{Henderson06}.
\begin{equation}
\Gamma = {1 \over \tau_{\rm d}} \propto {n\over \lambda_{\rm em}^{3}}\left({n^{2} + 2\over 3}\right)^{2}\sum_{f} \vert \langle {f} \vert \mu \vert {i} \rangle \vert^{2}
\end{equation}
\noindent where $\lambda_{\rm em}$ is the emission wavelength, and $\mu$ the oscillator strength of the
transition between the initial state $i$ and the final state $f$. From this, a high index of refraction, a short emission wavelength and a stronger oscillator strength would help to suppress $\tau_d$. However, the acceptance of light could be reduced by the difference of the indices of refraction between the scintillator and the grease or glue used to couple to the photo-detector. The decay time is in general longer than the rise time by a few orders of magnitude, ranging from roughly 1 ns to tens or hundreds of nanoseconds as shown in Tab~\ref{tab:crystal}.

The denominator in Eq.~\ref{eq:reso-sci}, the light output in number of photoelectrons converted from the scintillation light at the photo-detector ($N_{pe}$) is the product of the energy deposit ($E_{dep}$) in MeV of the particle passing through and the light output ($LO$) in number of photoelectrons per MeV.
\begin{equation}
    N_{pe} = E_{dep} \cdot LO
\end{equation}

The energy deposit $E_{dep}$ depends on the interaction of the particle passing through the scintillation medium. It is quantified by the stopping power (-dE/dx) of the medium that is proportional to the atomic number $Z$. Accordingly, materials with heavy elements in general give rise to larger energy deposit and yield more scintillation light. The energy deposit is also scaled with the size of the scintillator. However, the size is limited in reality. The timing detector is often placed in front of calorimetry which requires as few matters as possible in its upstream to reduce the energy loss before particle reaching it. Besides that, the impact of the light absorption in the transition should also be considered in large-size scintillators.

The light output ($LO$) can be broken down into the original light yield in number of photons ($LY$) per MeV of the scintillation, the light collection efficiency ($LCE$) accounting for all factors on the way from where the light is emitted to the photo-detector, and the photon detection efficiency ($PDE$) of the photo-detector:

\begin{equation}
  LO = LY \cdot LCE \cdot PDE
\end{equation}
\label{eq:lo}

$LY$ is driven by $N_{eh}$, the number of thermalized electron-hole pairs originated from the ionization of the incident particle and the secondary ionization. $N_{eh}$ is derived from the energy deposit of the incident particle $E_{dep}$ divided by the mean energy required to generate one electron-hole pair after thermalization which is proportional to the bandgap $E_g$. Ref.~\cite{PhysRev.139.A1702,Robbins_1980} pointed out the minimum mean energy needed for one pair that is about $2.3 E_g$. The factor $\beta$ represents the efficiency of $N_{eh}$ production with respect to the case using the minimum energy $2.3 E_g$.

\begin{equation}
  N_{eh} = \frac{E_{dep}}{2.3 E_g} \cdot \beta
\end{equation}

Two more efficiencies should be considered before the scintillation light is generated: $S$, the efficiency of energy transfer to the luminescent center, and $Q$, the quantum efficiency of the luminescent center. Then, $LY$ reads~\cite{Robbins_1980}

\begin{equation}
LY = \frac{N_{eh}}{E_{dep}} \cdot S \cdot Q = \frac{10^6}{2.3 E_g} \cdot \beta \cdot S \cdot Q 
\end{equation}

\noindent where $LY$ is defined per MeV and $E_g$ is in eV. In principle, lower $E_g$ leads to higher $LY$. However, vanishing $E_g$ also means potentially more non-radiation transitions resulting in a low $Q$, and the scintillator light falling beyond the sensitive range of photo-detectors in wavelength. The limit of $E_g$ is around 3 eV corresponding to $LY$ of about 140,000 photons/MeV assuming $\beta \cdot S \cdot Q$ as unity. Tab. 2 in Ref.~\cite{LEMPICKI1993304} provides a good summary of these efficiencies.

The light collection efficiency $LCE$ in Eq.~\ref{eq:lo} is affected by many factors along the way to the photo-detector, including the size and shape of the scintillator, the optical transparency, the interface between the scintillator and the photo-detector and reflection materials used for wrapping the scintillator and so on. Given the fraction of light reflected (R) and transmitted through (T) at normal incidence,

\begin{equation}
R = \frac{(n_1-n_2)^2}{(n_1+n_2)^2}, \,\, T = \frac{4n_1n_2}{(n_1+n_2)^2}
\end{equation}

\noindent it is important to match the index of refraction between the scintillator ($n_1$) and the coupling materials ($n_2$) to the photo-detector (usually glass, grease or glue). A CsI scintillator with $n_{CsI} = 1.79$ coupled to a photomultiplier with a glass entrance window with $n_{glass} = 1.52$ has $R = 0.0067$ at normal incidence, which does not impact much~\cite{Fleck:2021njs}. However, light reach the scintillator surface isotropically, and can go beyond the cone of acceptance that is limited by the critical angle $\theta_c = sin^{-1} (n_2 / n_1)$. The CsI scintillator coupled to the glass entrance window would have more than half of the light totally reflected internally. In general, scintillators with high densities have $LCE$ about 10-60\%.

The photon detection efficiency ($PDE$) at photo-detectors in Eq.~\ref{eq:lo} affects the final light output. It is the ratio of the number of photons detected over the number of photons that arrive at the detector. Taking SiPMs as example, $PDE$ is a product of the quantum efficiency ($QE$), the avalanche trigger probability ($P_{trig}$) and the effective geometrical fill-factor ($FF$).

\begin{equation}
  PDE = QE \cdot P_{trig} \cdot FF
\end{equation}

\noindent $QE$ represents the probability of a photon entering the detector and absorbed in the sensitive area, instead of being reflected, with a dependence on the wavelength. $P_{trig}$ depends on the electric field that is driven by the overvoltage and highly affected by the position where the carriers are generated. Light of UV to blue tends to be absorbed close to the surface of the junction leading to a higher $P_{trig}$ in a p-on-n junction as the electrons transverse more distance in the electric field, while red light are more sensitive in a n-on-p junction~\cite{6519337,Zappalà_2016,OTTE2017106}. $FF$ is the active area of all SPADs divided by the total area of the SiPM. Typically, SiPMs with larger SPAD of a size of $50 \times 50 ~\mu\text{m}^2$ has $FF$ as large as 80\%, while 30\% with smaller SPAD of $10 \times 10 ~\mu\text{m}^2$.

\subsection{Radiation effects}
\label{sec:sci_rad}

Inorganic scintillators in general performs with excellent radiation hardness benefiting from the strong electrostatic field of the crystal lattice that shields the luminescent centers. However, the transportation of the light in the scintillator can be affected and the corresponding light output would decrease impacting the time resolution. This is driven by the degraded transparency of the scintillator due to the formation of color centers originated from the impurities or point defects in the crystal most commonly caused by the ionization dose induced radiation damage~\cite{Fleck:2021njs}. LYSO maintains 75\% of the light output, while BGO and BaF$_2$ crystals also maintain 45\% light output after 120 Mrad induced by gamma-ray~\cite{Yang:2016six}. Damages induced by hadrons and neutrons cause additional displacement damage and nuclear breakup. More studies can be found in Ref.~\cite{7762196,9075296,DISSERTORI20141}.

Plastic scintillators are less radiative hard~\cite{Kharzheev:2019tfk}. Broken atomic bonds (radicals) are formed during irradiation and absorbs light preferentially in UV, which turns the scintillator yellow or even brown depending with higher dose. Radicals can be recovered by “annealing” afterwards with the bonds reformed, while they can also polymerize via cross linking resulting in a permanent reduction in light yield~\cite{Sauli:1992bu}. It is worth noting that oxygen can accelerate the annealing process which originally could take months in a inert environment~\cite{Sirunyan_2020,GILLEN19924358,SEGUCHI1981195}.

Photo-detectors, such as SiPMs, are more affected by irradiation damage that adds defects in sensitive area, leading to an increase in the leakage current~\cite{Moll:1999kv}. It can be expressed as the dark count rate DCR = $I_d / (G \cdot q)$ where $I_d$ is the leakage current, $G$ the gain and $q$ the elementary charge $1.602\times 10^{-19}$~C. DCR is in general below 100 $kHz/mm^2$ from the SiPMs on the market. With radiation, DCR increases rapidly with about 1 $kHz/mm^2$ for every $10^9 n_{eq} / cm^{-2}$ (1 MeV equivalent neutron)~\cite{Gundacker:2020cnv}. An increased DCR introduces more fake pulses in SiPMs and increases the self-heating. To mitigate these, lowering the temperature is usually effective before the plateau of DCR at low cryogenic temperatures due to trap-assisted tunneling. DCR is reduced by a factor of 2 every 10~$\degree$C drop in temperature. However, the range of working temperature is usually limited, as other detectors in the same volume require stable temperatures. Thus, an adjustment of the operating voltage is usually performed to further reduce DCR and the corresponding power consumption. This weakens the electric field and leads to a lower $P_{trig}$ and $PDE$ that degrade the final time resolution.

\section{Timing with low gain avalanche detectors}
\label{sec:sensor-lgad}
%{\bf
%- A silicon detector is a semiconductor device which basically is a solid state ionisation chamber.
%- Principle of detection:  external bias voltage polarizes the p-n junction inversely, creating a large depleted volume. When a particle crosses the sensors the dE/dx along its path produces electron-hole pairs. Motion of the charge carriers due to the application of an external electric field induces a current on the readout electrodes.
%- average energy to produce e-h pair in silicon is 3.6 eV, MIP creates about 75 e/h pairs per um 
%- Time resolution in silicon detectors is mainly driven by fluctuations in the charge distribution created by the ionizing particle
%(Landau fluctuations) and by the jitter term. Essential to achieve good time resolution are therefore fast and uniform charge collection, steep signal rise time, low noise.
%}

\subsection{Principle of operation and signal formation in silicon sensors}
The principle of operation of silicon sensors relies on the $pn$ junction.
A external bias voltage applied to the electrodes polarizes the $pn$ junction inversely, creating a large depleted volume. When a particle crosses the sensors, the energy loss  along its path produces electron-hole pairs. In silicon, the average energy needed to create an e-h pair is 3.6~eV and a MIP produces about 75 e-h pairs (most probable value) per $\mu m$. 
Under the influence of the electric field of the depleted diode, charge carriers drift towards the electrodes where they induce a current signal. The signal ends when the last charge carrier reaches the electrodes. When a sufficiently high electric field ($>$10–20~kV/cm) is applied, charge carriers move with a saturated drift velocity $v_{sat} \sim 100~\mu m$/ns. For typical sensor thicknesses of about 200-300~$\mu m$, the drift time is of the order of a few nanoseconds.

The signal induced on a given electrode can be determined using the Ramo-Shockely's theorem~\cite{10.1063/1.1710367, 1686997}:

\begin{equation}
i(t) =  -q \vec{v} \cdot \vec{E_w}
\label{eq:Ramo}
\end{equation}

\noindent where q is the electric charge of the charge carrier, $\vec{v}$ is the drift velocity and $\vec{E_w}$ is the weighting field, which describes how the moving charge couples to a specific electrode and depends uniquely on the geometry of the sensor.

In planar silicon sensors, the peak current reaches a value of about 1-2~$\mu A$~\cite{CARTIGLIA2015141}, independent of the sensor thickness: thick sensors produce a larger number of e-h pairs; on the other hand, each pair produces a lower initial current as the weighting field is inversely proportional to the detector thickness. The interplay of these two effects results in a small signal amplitude, which cannot be increased by increasing the sensor thickness, and is at the core of the limited time resolution of standard planar silicon sensors.  
The best example of timing performance with thin planar silicon sensors in a working experiment is presently represented by the NA62 GigaTracKer, where a single hit time resolution of $\sim$115~ps has been achieved with $300 \times 300~\mu m^2$, 200~$\mu m$ thick sensors~\cite{CortinaGil:20245q}.

As discussed in Sec.~\ref{sec:principles}, essential to achieve good time resolution are fast and uniform charge collection, steep signal rise time, and low noise.
Hybrid silicon sensors with precision timing capabilities ($\sigma_t <$~100~ps) comprise 3D~\cite{PARKER1997328} and Low Gain Avalanche Diodes (LGADs)~\cite{CARTIGLIA2015141, PELLEGRINI201412}. 
3D silicon sensors consist of arrays of columnar electrodes oriented perpendicular to the wafer surface, resulting in a structure where the inter-electrode distance and active substrate thickness are decoupled. Thanks to their short inter-electrode drift distance (tens of $\mu m$) giving rise to very fast signals~\cite{PARKER1997328}, they are attractive candidates for precision timing. While 3D sensors for particle tracking are already in operation in HEP experiments like the ATLAS IBL~\cite{Capeans:1291633}, R\&D efforts to achieve O(10)~ps time resolution are ongoing.
LGADs are thin silicon sensors with low internal gain specifically optimized for precision timing measurements. This sensor technology is the one chosen for the forward timing layers of the ATLAS and CMS experiments at the HL-LHC and is presented in details in the next sections. A comprehensive introduction to LGAD design can be found in~\cite{UltrafastSiliconDetectors}.
Ongoing trends in silicon sensors for timing applications will be further discussed in Sec.~\ref{sec:noveltechnologies}.

%3D silicon sensors with columnar geometry are suitable candidates for precision timing thanks to their short inter-electrode drift distance (tens of $\mu m$) giving rise to very fast signals~\cite{PARKER1997328}. 
%{\bf add a comment about negligible contribution to the time resolution from Landau fluctuations due to drift direction being perpendicular to ionization direction}
%A time resolution of about 30~ps has been achieved as reported in \cite{KRAMBERGER201926}. Albeit this good result, the performance is ultimately limited by the electric field and weighting field spatial non uniformities. These effects can be partially mitigated by employing trench-like geometries, for which a time resolution of O(10)~ps has been reached in prototypes~\cite{Lampis_2023}.

\subsection{LGAD design}
\begin{figure}[!h]
\centering
\includegraphics[width=0.9\linewidth]{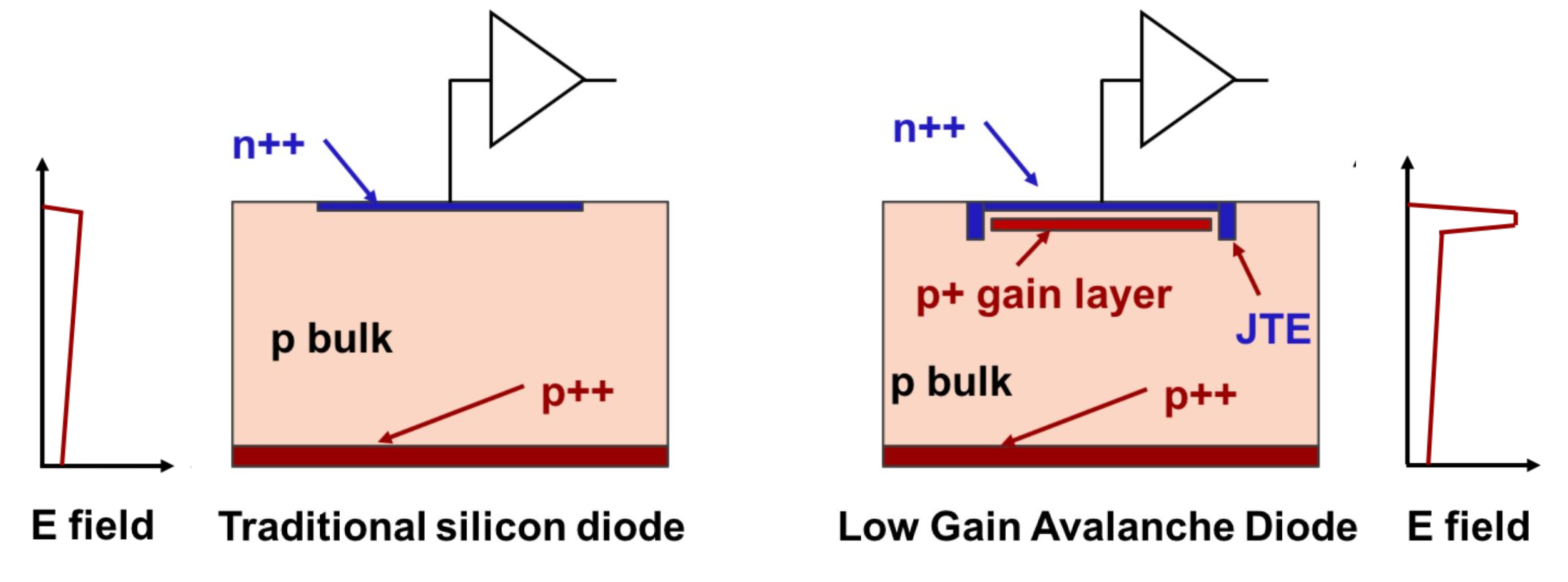}
\caption{Simplified representation of a standard silicon diode (left) and of a Low-Gain Avalanche Diode (right). The additional p$^+$ implant in the LGAD design provides charge multiplication. Adapted from~\cite{CERN-LHCC-2019-003}.}
\label{fig:LGAD}
\end{figure}

In silicon sensors, charge multiplication occurs when charge carriers drift in a region with a high electric field ($>$300~kV/cm), thanks to which electrons (and to a lesser extent the holes) are accelerated enough to produce further e-h pairs.
To achieve this condition, the core feature of the LGAD design is the addition of a p$^{+}$ layer (boron or gallium, with an acceptor density $N_A \sim 10^{16}~atoms~/cm^3$) implanted near to the $pn$ junction, which generates, in a localized region, an electric field high enough to trigger charge multiplication. A simplified representation of the LGAD structure is shown in Fig.~\ref{fig:LGAD}, compared to the one of a traditional silicon diode. Two regions characterize therefore the LGAD sensor volume: the drift volume, corresponding to the largest part of the bulk, with a lower electric field (E $\sim$ 30~kV/cm) and a thin multiplication region (0.5–1~$\mu$m) located within a depth of a few micrometres with very high field (E $\sim$ 300~kV/cm).
A termination structure called Junction Termination Extension (JTE) is present to control the electric field at the junction edges.

Thanks to their internal gain, LGADs provide larger signals and much higher slew rates with respect to standard silicon diodes.

\begin{figure}[!h]
\centering
\includegraphics[width=0.7\linewidth]{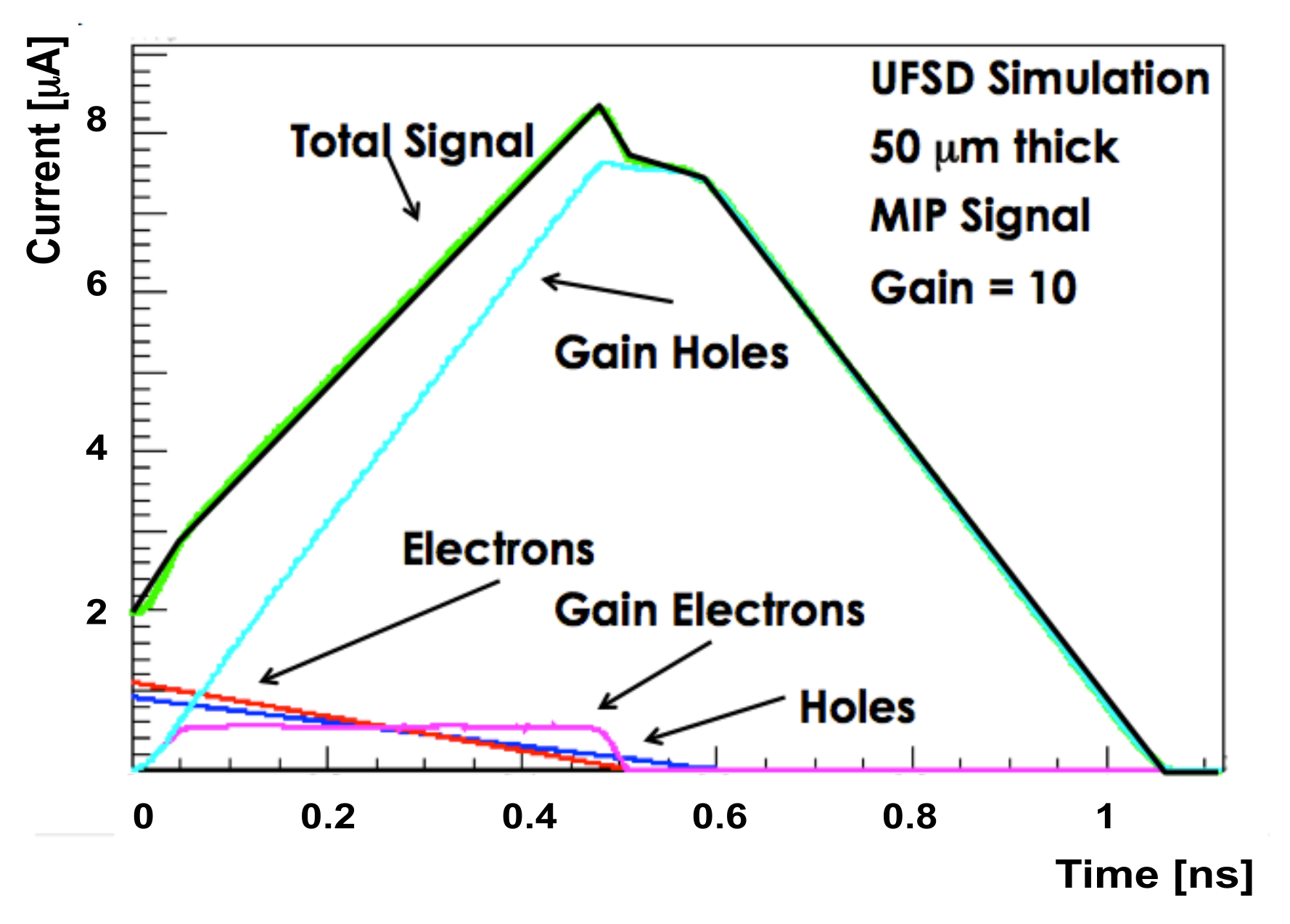}
\caption{Simulated current signal for a 50~$\mu$m thick LGAD: initial electrons (red) and holes (blue), gain electrons (violet), gain holes (light blue). From~\cite{CARTIGLIA2015141}}
\label{fig:LGAD_signalComponents}
\end{figure}

The effect of the gain mechanism on signal formation is illustrated in Fig.~\ref{fig:LGAD_signalComponents}, which reports the simulated current, and its separate components, for a 50~$\mu$m thick LGAD: the initial
electrons, drifting towards the n$^{++}$ electrode, go through the gain layer and generate additional e/h pairs. Since the multiplication happens very near the cathode, the gain electrons are immediately absorbed, while the gain holes drift almost the full bulk thickness before being collected by the anode, yielding a large signal.

The current generated by the multiplication mechanism can be estimated from the number of electrons entering the gain layer in a time interval $dt$ as $dN_{Gain}=G N_{e-h} v_{sat} dt$, where $G$ is the gain, $N_{e-h}$ the number of e-h pairs produces per unit length ($N_{e-h}\sim$ 75/$\mu$m) and $v_{sat}$ is the saturated drift velocity.
Using Ramo’s theorem and under the simple assumption of a parallel plate geometry ($E_w \propto 1/d$), the current induced by these secondary charges can be calculated as: 

\begin{equation}
%dN_{Gain} = G N_{e-h} v_{sat} dt
dI_{Gain} = dN_{Gain} q v_{sat} \frac{1}{d} \propto \frac{G}{d} dt  
\label{eq:gain_current}
\end{equation}

The signal slew rate in sensors with gain is therefore proportional to the gain over the sensor thickness ratio $G/d$:  

\begin{equation}
\frac{dI_{Gain}}{dt} \sim \frac{dV}{dt} \propto \frac{G}{d}
\label{eq:gain_slewRate}
\end{equation}

\begin{figure}[!h]
\centering
\includegraphics[width=0.47\linewidth]{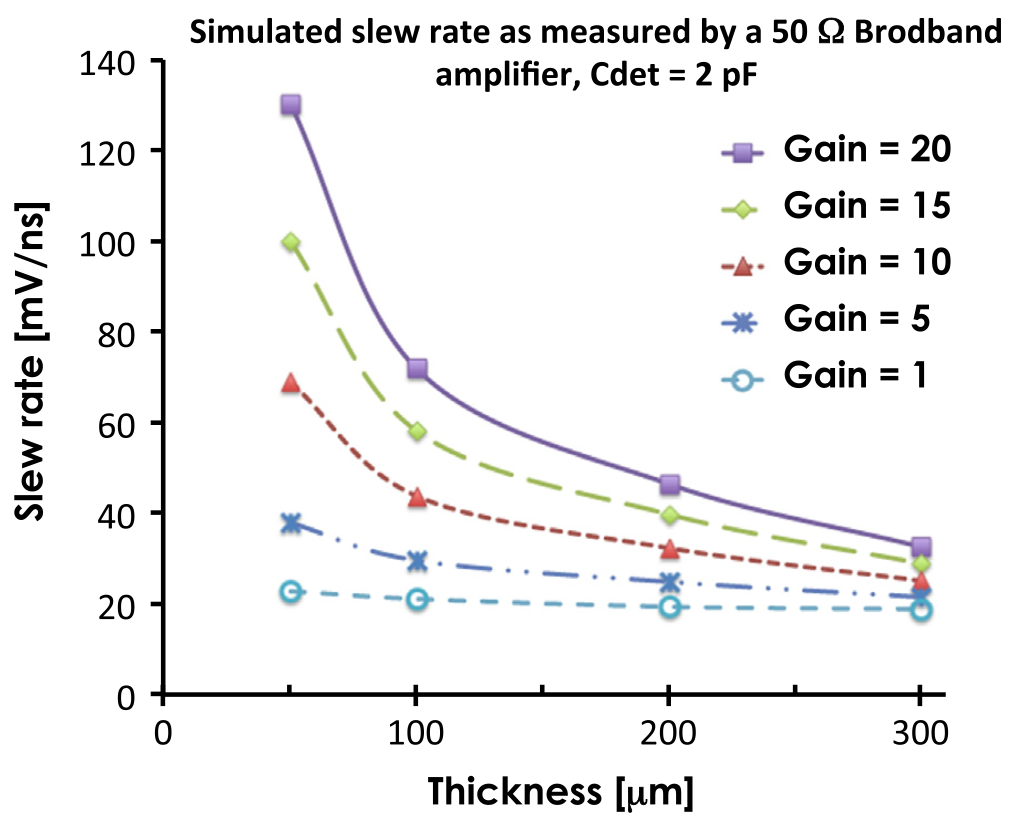}
\includegraphics[width=0.51\linewidth]{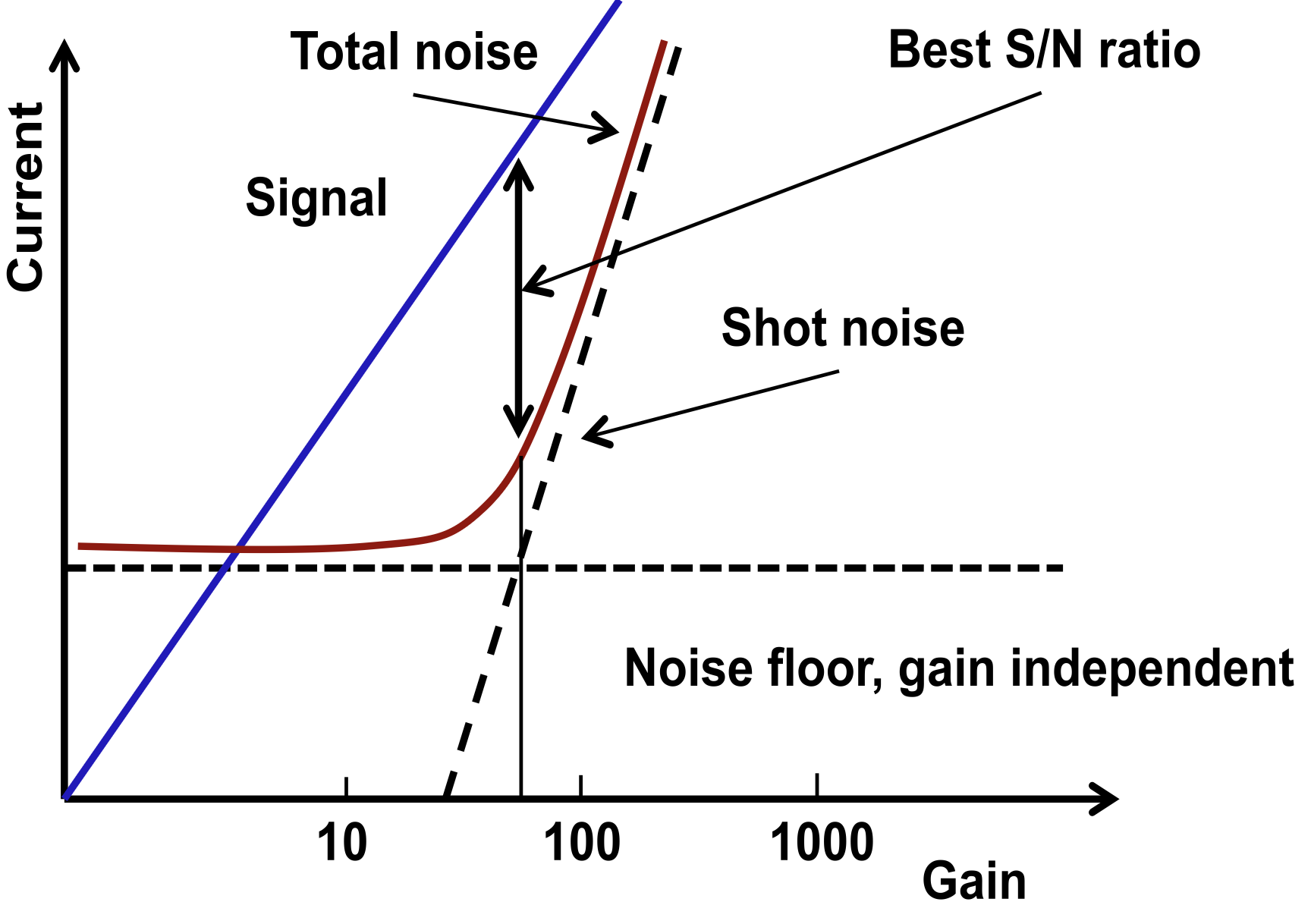}
\caption{Left: simulated slew rate as a function of gain and sensor thickness~\cite{CARTIGLIA2015141}. Right: signal and noise current as a function of the gain~\cite{Sadrozinski_2018}.}
\label{fig:slewRateLGADs}
\end{figure}

Eq.~\ref{eq:gain_slewRate} shows that thin sensors with high gain can provide large signals with high slew rate suitable for precision timing measurements. A simulation of the slew rate for sensors with different thickness and gain is reported in Fig.~\ref{fig:slewRateLGADs}.

It is important to note that the internal gain mechanism increases the signal but also the sensor noise. Shot noise in silicon sensors arises from fluctuations in the leakage current. When carriers undergo multiplication, shot noise is further amplified by the so-called {\it excess noise factor}, resulting in $\sigma_{shot} \sim \sqrt{2 e I_{bulk} G^2 F \tau_{int}}$, where $I_{bulk}$ is the bulk leakage current, $G$ the gain, $F\sim G^x$ the excess noise factor, $\tau_{int}$ is the electronics integration time~\cite{UltrafastSiliconDetectors}. 
To achieve a beneficial effect, the gain has to be therefore sufficiently large to increase the signal but not but not so large that the sensor noise becomes dominant over the electronics noise floor, as sketched in Fig.~\ref{fig:slewRateLGADs} (right).
LGAD sensors are designed to operate with a moderate internal gain (10-30) to minimize the excess noise contribution and allow low-power operation also after irradiation.

\subsection{Time resolution of LGADs}
As discussed in Sec.~\ref{sec:principles}, several effects contribute to the time resolution of a detector.
Expanding Eq.~\ref{eq:timeResolution}, the time resolution of LGADs can be expressed as:

\begin{equation}
\sigma^2_{t} = \sigma^2_{jitter} + \sigma^2_{LandauNoise} + \sigma^2_{distortion} + \sigma^2_{time-walk} + \sigma^2_{TDC}
\label{eq:timeResolutionLGADs}
\end{equation}

\begin{figure}[!h]
\centering
\includegraphics[width=0.8\linewidth]{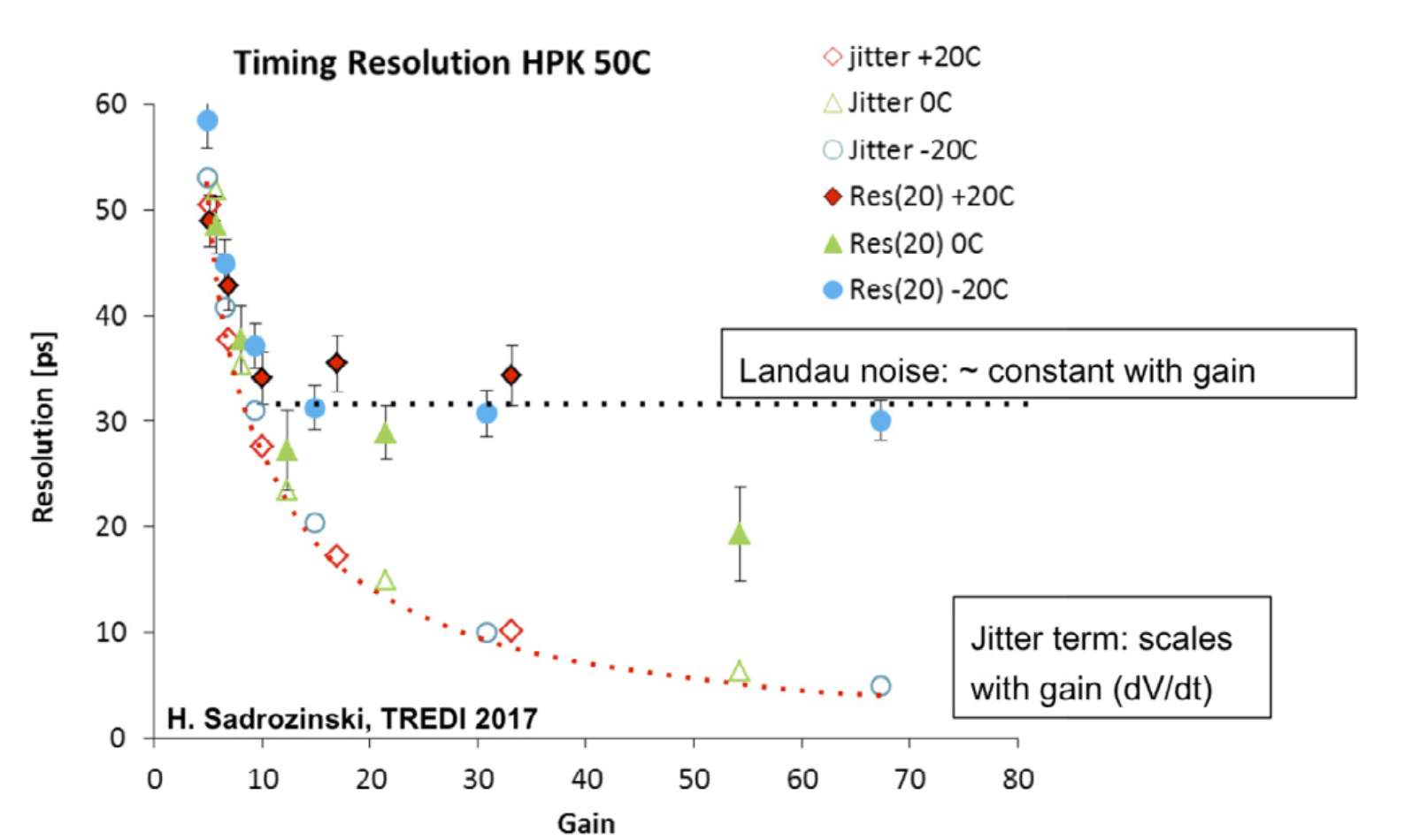}
\caption{Total time resolution and jitter contribution as a function of the sensor gain, as measured on 50-micron thick sensor manifactured by Hamamatsu. The jitter term (open markers) decreases with gain, while the total time resolution (full markers) flattens around 30~ps due due to the Landau noise. From~\cite{CARTIGLIA2019350}.}
\label{fig:timeResolutionLGADs}
\end{figure}

The last two terms, $\sigma_{time-walk}$ and $\sigma_{TDC}$, can be considered non-leading contributions, as $\sigma_{time-walk}$ can be compensated by means of an appropriate circuit (see Sec.~\ref{sec:timewalk}) and $\sigma_{TDC}$ can be kept typically below 10~ps thanks to the fine TDC binning.  
The distortion term, $\sigma_{distortion}$, arises from signal shape variations due to non-uniform weighting field and non-uniform drift velocities (following from eq.~\ref{eq:Ramo}). In LGADs, this is a sub-leading contribution as well: signal shape variations are minimized by the geometry, close to a parallel plate capacitor, and by the high electric field, which saturates the drift velocity.
The jitter term $\sigma_{jitter} = N/(dV/dt)$ is reduced thanks to the large slew rate and is driven by the noise of the electronics.
The last term $\sigma_{LandauNoise}$, referred to as {\it Landau noise}, is related to variations on an event-by-event
basis caused by the random nature of electron–hole pairs creation along the particle path. The Landau noise represents an intrinsic limit to the time resolution of both LGADs and no-gain silicon sensors. In Fig.~\ref{fig:timeResolutionLGADs} the total time resolution and the jitter term measured for 50~$\mu m$ thick LGADs as a function of the sensor gain are shown. While the jitter term decreases with gain due to the slew rate increase, the total resolution approaches a constant term around 30~ps for gain values above $\sim$20 as a consequence of the Landau noise.

\subsection{Radiation hardness of LGADs}
\label{sec:RadiationHardnessLGADs}
Radiation damage in semiconductors arises from two mechanisms: surface damage and bulk damage. For a detailed review of radiation damage in silicon sensors see~\cite{MMoll}.
Bulk damage is due to displacement of atoms from their lattice sites resulting in defect energy levels inside the silicon band gap and leads to three main macroscopic effects on sensors:
\begin{itemize}
\item  {\it Leakage current increase} - The increase leakage current due to bulk damage is proportional to the number of defects in the sensor volume V after a fluence $\Phi$ : $\Delta I= \alpha \Phi V$, where $\alpha$ is known as current-related damage constant ($\alpha \sim 4 \times 10^{-17} A/cm$ for neutrons). The leakage current has strong temperature dependence as $I(T) = I_0 T^2 e^{\frac{-E_a}{2k_BT}}$, where $E_a$ = 1.2 eV is the activation energy and $k_B$ is the Boltzmann’s constant. It can thus be mitigated by lowering the operating temperature, with a factor approximately two reduction for a decrease of 7$^{\circ}$C in temperature. 
\item {\it Decrease of charge collection efficiency (CCE)} - A fraction of the charge carriers are trapped by deep defects during their drift in a process called trapping. CCE decreases with increasing radiation levels and with increasing drift length. Thin sensors are less sensitive to this effect.
\item {\it Change in the doping concentration} - Three different mechanisms concur in changing the doping concentration $N_{eff}$ of silicon sensors: acceptor creation, donor and acceptor removal. The density of effective dopants $N_{eff}$ varies as a function of the equivalent fluence $\Phi_{eq}$ as

\begin{equation}
N_{eff} = N_{D_0}e^{-c_D\Phi_{eq}} - N_{A_0}e^{-c_A\Phi_{eq}} - g_{eff} \Phi_{eq}
\label{eq:doping}
\end{equation}

where $N_{D_0}$ and $N_{A_0}$ are the initial values of donor/acceptor concentrations, $c_D$ ($c_A$) is the donor (acceptor) removal coefficients and $g_{eff}$ = 0.02~cm$^{-1}$ the coefficient of proportionality between the fluence and the density of new acceptor-like defects~\cite{MMoll}.

\end{itemize}

\begin{figure}[!h]
\centering
\includegraphics[width=0.51\linewidth]{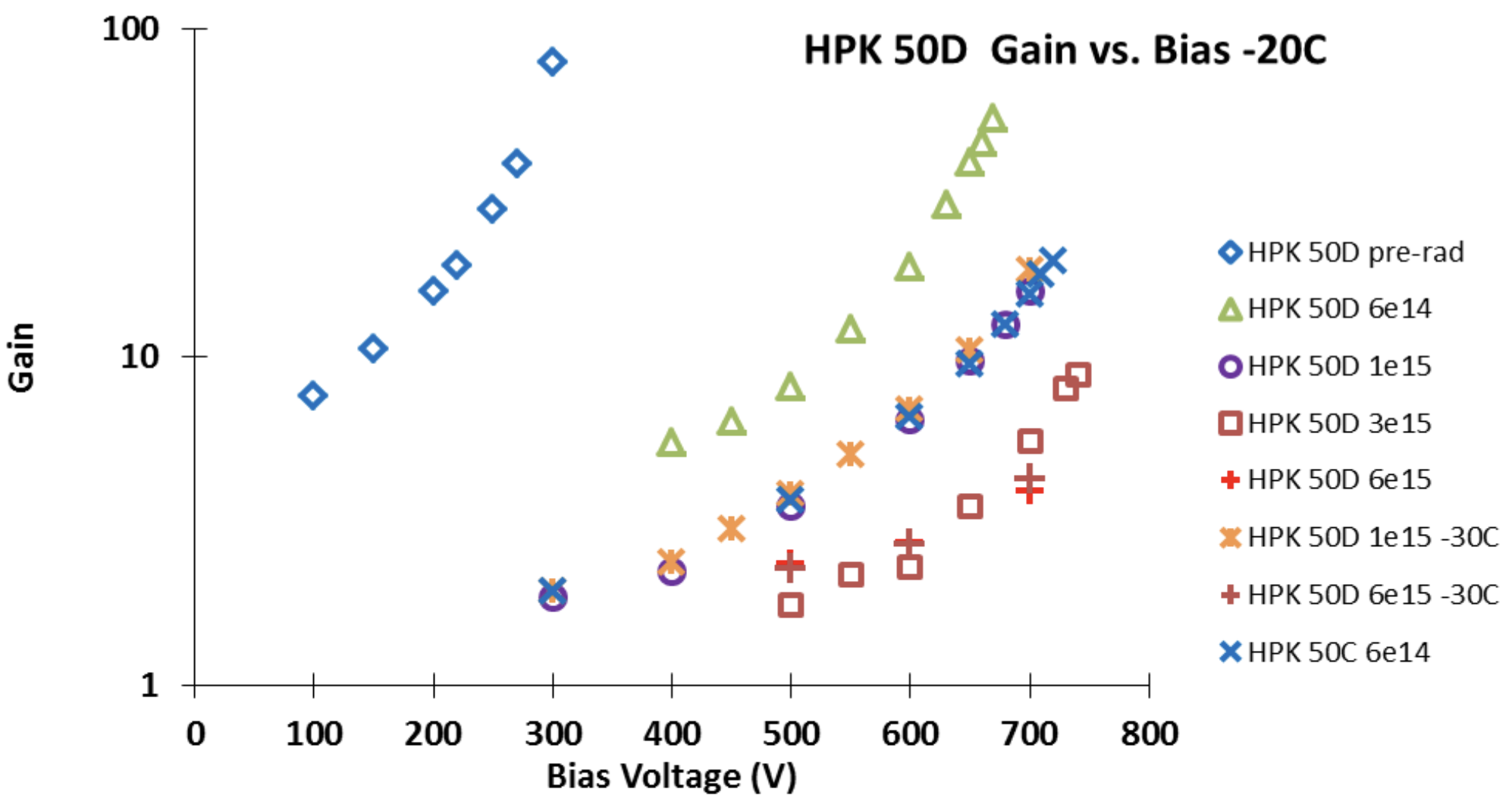}
\includegraphics[width=0.47\linewidth]{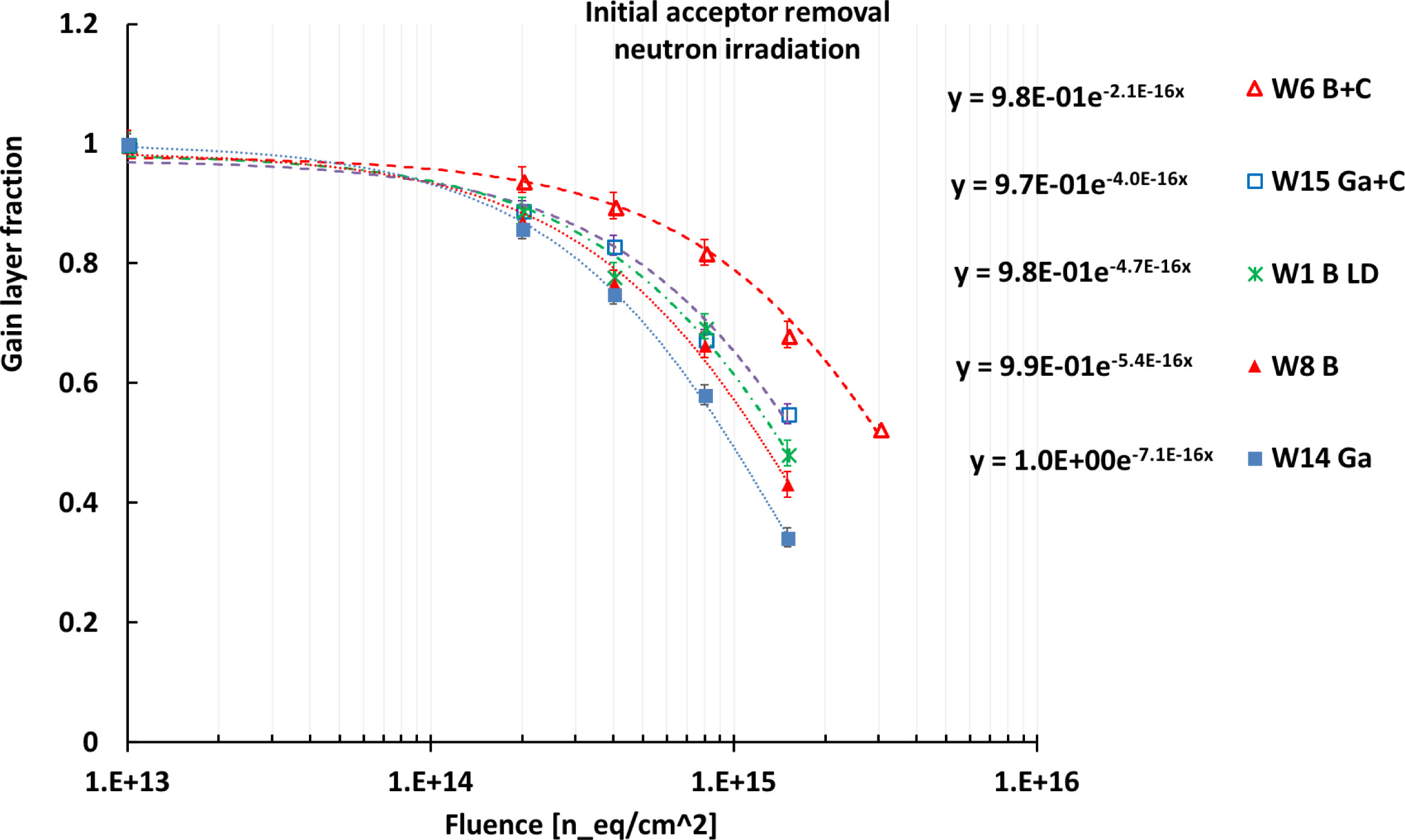}
\caption{Left: gain as a function of the bias voltage applied to the sensors for different levels of irradiation~\cite{Cartiglia:2017O8}. Right: fraction of gain layer still active as a function of the fluence for different gain layer configurations, showing that carbonated gain layers (open markers) are more radiation resistant than the equivalent non-carbonated ones~\cite{FERRERO201916}.}
\label{fig:gainLGADs}
\end{figure}

The change in doping concentration can be particularly harmful to LGADs as it deactivates the p-doping of the gain implant, thus reducing the signal multiplication power. This effect, known as "initial acceptor removal", can be explained by boron atoms from the lattice becoming interstitials thereby turning ineffective as acceptors. 
The gain loss due to acceptor removal can be partly recovered by increasing the bias voltage at which the sensors are operated, as illustrated in Fig.~\ref{fig:gainLGADs} (right), and by optimizing the doping profile. In particular, carbon implantation in the gain layer has been demonstrated to reduce the acceptor removal rate~\cite{FERRERO201916}. Carbon replaces, in fact, boron in the ion-defect complexes formation, mitigating the boron deactivation.
Fig.~\ref{fig:gainLGADs} (left) shows the fraction of gain layer still active as a function of the neutron fluence for different gain layer configurations. State-of-the-art carbonated LGADs can maintain a gain of 10–15 up to fluences of about 2$\times 10^{15}$~\neut. 

It should be noted that the operational voltage of LGAD sensors is limited by the so-called single event burn-out (SEB) phenomenon~\cite{Lastoviska-Medin_2023}\cite{Beresford_2023}: when the electric field in a silicon sensor is above $\sim$11~V/$\mu$m, a single particle crossing the sensor can trigger an avalanche process that makes the sensor irreversibly inoperable. The SEB mechanism therefore reduces the possibility of compensating for the acceptor removal and restoring the gain of irradiated sensors by increasing the bias voltage. R\&D efforts are ongoing to extend the radiation tolerance of silicon sensors up to extreme fluences (1$\times$10$^{17}$~\neut), including for example carbon shield techniques or doping compensation~\cite{SOLA2024169453}.

\section{Gaseous detectors}
\label{sec:sensor-mrpc}

%{\bf
%- Principle of operation: electron-ion pairs produced via ionization along the particle trajectory. In presence of a very large electric
%field, primary electrons will gain a very large kinetic energy leading to further ionization and avalanche formation. Signal is generated by the induced current due to the drift of negative and positive charges towards the electrodes
%
%- Time resolution in gaseous detectors typically limited by fluctuations in the primary ionization, fluctuations due to long drift and in the avalanche formation. 
%%- critical parameters for precision timing with gaseous detectors are: number of electrons contribution to signal leading edge, short drift distances, gases with small longitudinal diffusion, uniform electric field
%%- Factor limiting time resolution in conventional gaseous detectors is due to the spread in drift distances --> for O(mm) drift distances, timing jiter can be of the order of ns
%
%- While the typical time resolution can be of the order of nanoseconds, specific technologies developed for timing applications can reach below 100 ps. Examples: Multi Gap RPCs, Fast Timing Micro Pattern Gaseous Detectors (FTM), Picosec MicroMegas
%
%- A detailed description of MRPCs (adopted for example in ALICE, STAR experiments) will be given in the next chapter.
%}

\subsection{Formation of signals}

When a charged particle passes through the gaseous medium, it transfers a fraction of energy via the electromagnetic interactions with the atomic electrons around, creating electron-ion pairs randomly along its trajectory. The electrons from the primary ionization may have sufficient energy to trigger secondary ionization in the medium, which happens in the vicinity of the primary ionization, forming clusters. The number of pairs of all ionization per cm, $N_T$, is usually several times the number of pairs in primary ionization per cm, $N_P$, as listed in Tab~\ref{tab:gas}, although the clusters with only 1 electron still dominate. For Ar as an example, 65\% of the clusters contain 1 electron, 15\% with 2 electrons, 6\% with 3 electrons and 1.1\% with more than 20 electrons~\cite{FISCHLE1991202}. The locations of these clusters, the number of electrons in each cluster and their spatial extent impact the spatial and time resolution of the gaseous detector~\cite{Fabjan:2020wnt}. 

\begin{table}[!ht]
    \centering
    \begin{threeparttable}[b]
    \begin{scriptsize}
    \renewcommand{\arraystretch}{1.2}
    \begin{tabular}{cccccccc|c}
    \hline
        Gas & Density & $E_X$ & $E_I$ & $W_I$ & $dE/dx|_{min}$ & $N_P$ & $N_T$ & $N_T / N_P$ \\
        & mg cm$^{-3}$ & eV & eV & eV & keV cm$^{-1}$ & cm$^{-1}$ & cm$^{-1}$ & \\
        \hline
        H$_2$ & 0.084 & 10.8 & 13.6 & 37 & 0.34 & 5.2 & 9.2 & 1.8 \\
        He & 0.179 & 19.8 & 24.6 & 41.3 & 0.32 & 3.5 & 8 & 2.3\\ 
        Ne & 0.839 & 16.7 & 21.6 & 37 & 1.45 & 13 & 40 & 3.1 \\ 
        Ar & 1.66 & 11.6 & 15.7 & 26 & 2.53 & 25 & 97 & 3.9 \\ 
        Xe & 5.495 & 8.4 & 12.1 & 22 & 6.87 & 41 & 312 & 7.8 \\ 
        CH$_4$ & 0.667 & 8.8 & 12.6 & 30 & 1.61 & 28 & 54 & 1.9 \\ 
        C$_2$H$_6$ & 1.26 & 8.2 & 11.5 & 26 & 2.91 & 48 & 112 & 2.3 \\ 
        iC$_4$H$_{10}$ & 2.49 & 6.5 & 10.6 & 26 & 5.67 & 90 & 220 & 2.4 \\ 
        CO$_2$ & 1.84 & 7.0 & 13.8 & 34 & 3.35 & 35 & 100 & 2.9 \\ 
        CF$_4$ & 3.78 & 10.0 & 16.0 & 35-52 & 6.38 & 52-63 & 120 & 1.9-2.3 \\
        \hline
    \end{tabular}
    \caption{List of properties of typical gases used at NTP (one atm at 20$\degree C$). $E_X$ and $E_I$ stand for the lowest excitation and ionization energy, $W_I$ for the average energy for creation of ion pair, $dE/dx|_{min}$ for the minimum of the energy loss per unit distance, $N_p$ and $N_T$ for the numbers of primary and all electron-ion pairs, respectively, all of which are measured with MIPs that carry unit charge (remade from ~\cite{pdg:2024cfk}).}
    \end{scriptsize}
    \end{threeparttable}
    \label{tab:gas}
\end{table}

An external electric field is usually applied, under which electrons and ions produced in the ionization drift in opposite directions and diffuse towards the electrodes for signal formation. Taking electrons as an example, they actively interact with molecules. The electron–molecule collision cross-sections depend on the molecular structure and the electron energy that is driven by the electric field. It is pointed out that high values of the total electron scattering cross section reduce the electron diffusion and increase the drift velocity in Ref.~\cite{Sauli_2014}. Noble gases only have inelastic interactions opened up above roughly 10 eV, while molecular gases, such as CH$_4$, open inelastic, rotational and vibrational interactions since about 0.1 eV. The addition of polyatomic gases, such as C$_x$H$_y$, CO$_2$, CF$_4$, can effectively suppress the diffusion at an early stage helping to achieve large drift velocity. The cross sections of typical gases are compared in Ref.~\cite{Sauli_2014} and the open-access database for a wide range of gases can be found on the web of LXCAT.

The electric field can be increased to a level where electrons obtain sufficient energy between collisions to ionize the gaseous medium. The first Townsend coefficient $\alpha$ is defined as $\alpha = 1/\lambda$, where $\lambda$ is the mean free path for ionization, that is the average distance an electron travels before the next ionizing collision. Considering $N_0$ as the number of electrons in a certain location, it increases as the electrons drift by a distance $dx$ in a uniform electric field, $dN = N_0\alpha dx$. Integrating this differential equation yields $N = N_0 e^{\alpha x}$, indicating the exponential growth of electron number and the formation of an electron avalanche.
%Thus, an electron avalanche can be formed by integrating the formula, which leads to $N = N_0 e^{\alpha x}$. 
Additional ways of energy transfer, such as the Penning effect, could exist and provide an increase of the ionization yield~\cite{SAHIN2014104} when the ionization potential of one gaseous component is lower than the excitation potential of the other.

While the avalanche keeps evolving, the multiplication factor continues to rise. Up to an empirical condition of $\alpha x = 20$ (Raether's limit), the high density of the ions and electrons accumulated in the avalanche significantly enhances the electric field in front of and behind the avalanche, and photons emitted in the inelastic collisions also start to contribute to the ionization. All these extend avalanches both forward and backward between the electrodes and eventually form a conductive channel, called a streamer, resulting in a local discharge. Before reaching the geometry limit of the chamber or a reduction of the field, the streamer can develop across the whole gas gap producing a spark breakdown. The signal generated by the stream can range from 50 pC to a few nC~\cite{ALICE:1999osy,BACCI2000342}, usually large enough to be directly detected without pre-amplification, but the time resolution could only be at a level of a few nanoseconds.

\subsection{Time resolution}

There are various gaseous detectors developed for different spatial and timing purposes. In general, wire-based detectors are not competitive for high-precision timing applications, because of the fluctuations in the distance from primary ionization to the wires that is exponentially distributed and the $1/r$ electric field only allowing amplification in the vicinity of the wire after certain drifting time~\cite{Lippmann:2003uaa}. These limit the timing resolution to a few nanoseconds. Thus, suppressing the uncertainty of the localization of the primary ionization, such as by reducing the gas gap and applying a uniform electric field that allows drifting and amplification in the same region, may improve the time resolution.

Detectors with parallel plates can be designed with thin gas gap and an electric field with high uniformity providing amplification with no drift delay, leading to competitive timing performance. One of the earliest examples of such a detector is the Pestov spark counter ~\cite{PARKHOMCHUCK1971269,COOPER200941}. In this design, one of the electrodes uses resistive materials with $\rho = 10^9 - 10^{10} ~\ohm cm$, while the other one is metallic. With an area of 600 cm$^2$ and a gas gap of 1 mm working at atmospheric pressure, the time resolution can reach less than half a nanosecond. Further performance improvements have been demonstrated by reducing the gas gap to 0.1 mm and increasing the working pressure to 12 bar, yielding time resolutions as low as 27 ps.
%Further reduction of the gas gap to 0.1 mm and increasing the working pressure to 12 bar leads to a time resolution down to 27 ps.

\begin{figure}[!h]
\centering
\includegraphics[width=0.9\linewidth]{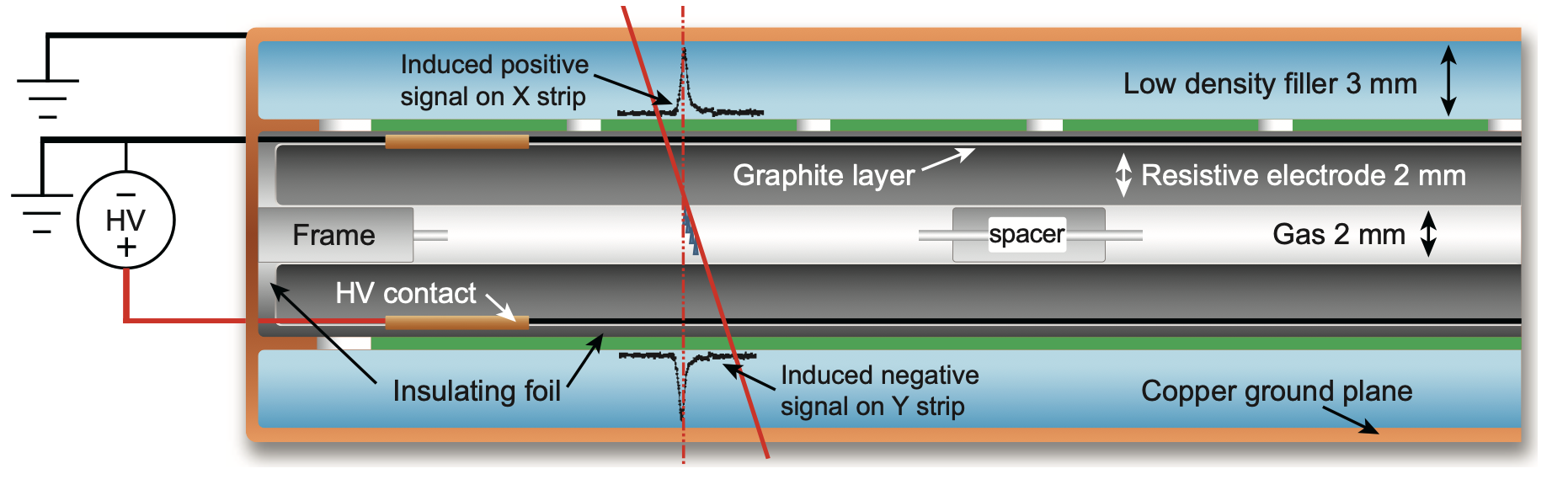}
\caption{Schematic cross section of a generic RPC from Ref.~\cite{pdg:2024cfk}}
\label{fig:rpc}
\end{figure}

Based on a similar structure, Resistive Plate Chambers (RPC) are designed with two parallel high bulk resistivity electrode plates and a thin gap of gas working at 1 atm~\cite{SANTONICO1981377}, as sketched in Fig.~\ref{fig:rpc}. The gas gap thickness limits the range of the ionization locations and drives the time resolution of the RPC. Given the high resistivity, the time to recover from the charge deposited on the resistive layer $\tau = \rho \epsilon$, where $\rho$ is the electrode resistivity and $\epsilon$ its dielectric constant, is typically 1 ms, much longer than the avalanche development time (typically 1 ns). As a result, the electric field is reduced only at the location of the avalanche, avoiding a discharge of the entire counter, providing a stability at very high fields and particle rates.

Large-area RPCs can be built to hundreds of m$^2$ with resistive plates typically made from Bakelite ($\rho = 10^{10} - 10^{12} ~\ohm cm$) or sode-lime glass ($\rho = 10^{12} - 10^{13} ~\ohm cm$). Gases based on Hydrofluorocarbon (HFC) became the standard of RPC gas mixture, such as the widely used tetrafluorethane (C$_2$F$_4$H$_2$) which provides a large number of primary ionization clusters of 80-100/cm and suppresses the formation of streamers with its electro-negativity.

The time resolution for an RPC can be approximated as $\sigma_t \approx 1.28/\alpha v$~\cite{Fabjan:2020wnt}, where $\alpha$ stands for the effective Townsend coefficient of the gases and $v$ the electron drift velocity. A typical RPC is designed with a gas gap of 2 mm at an electric field of 50 kV/mm leading to $\alpha \approx 10$~mm and $v \approx 130~\mu$m/ns. This corresponds to a time resolution of about 1 ns with an efficiency close to 100\%. To further control the range of random location of the primary clusters, improve $\alpha$ and $v$, the gas gap can be reduced to the sub-millimeter scale. However, the full detection efficiency at 1 atm for RPC limits the gas gap to 1 mm. To go beyond the limit, multiple gaps can be combined to effectively recover the efficiency. Thus, multiple gap RPC (MRPC) is introduced~\cite{CERRONZEBALLOS1996132,RPCbook}. In MRPCs, the typical gas gap is reduced to 0.25-0.3 mm and the electric field can be raised to about 100 kV/cm, resulting in $\alpha \approx 113$~mm and $v \approx 210~\mu$m/ns. This eventually can provide a time resolution of 50 ps~\cite{Fabjan:2020wnt}.  Ref.~\cite{RPCbook} derives the time resolution of an MRPC in the same way as single-gap RPC described above
\begin{equation}
    \sigma_t = \frac{1}{\sqrt{N_g \lambda g}} \cdot \frac{U}{\alpha v}
\end{equation}
\noindent where $N_g$ is the number of gas gaps, $g$ their size, $\lambda$ the number of clusters per unit length and $U$ is a factor of order 1 for the avalanche statistics as not all primary ionizations contribute equally~\cite{Gonzalez-Diaz_2017}. It is worth noting that the time resolution is inversely proportional to the square root of the total number of primary charge $N_g \lambda g$, following the same rule as scintillation in Eq.~\ref{eq:reso-sci}. The sub-millimeter gas gap requirement challenges the mechanical structure, leaving only glass suitable for the resistive plates at the moment, and limits large-area detector building. A recent review of RPC detectors can be found in Ref.\cite{FONTE2025170401}.

\subsection{Ageing}

In general, the gaseous contents are less affected by irradiation and ageing and exhibit good long-term stability. However, the internal structures and surfaces of the detector, especially in the case of glass electrodes, can be corroded by the hydrofluoric acid generated from discharges in the environment of fluorocarbons and water. A flow of gas maintained during the operation could mitigate the damage~\cite{AIELLI200486,SAKAI2002153}. Further details on the ageing impact on rate capabilities, performance degradation, and operational longevity can be found in Refs~\cite{pdg:2024cfk,Sauli_2014,RPCbook}.

\section{Integration}
\label{sec:integration}

%A variety of modern large particle experiments adopts the fast sensor technologies introduced above and applied to the higher performance timing detectors. This includes the barrel (BTL) and endcap (ETL) timing layers of MIP Timing Detector (MTD) on CMS, the High Granularity Timing Dectector (HGTD) on ATLAS, the ALICE TOF detector and the BESIII TOF detector, as well as Precision Proton Spectrometer (PPS) and calorimetries on CMS. This section introduces the integration of these timing technologies to the functioning detectors with electronics, cooling and mechanical supports. A summary of the parameters of these timing detectors is listed in Tab.~\ref{tab:comparison}.

Various modern large particle experiments adopt the fast sensor technologies introduced above for high-performance timing detectors. 
In this section, we describe a few examples of detectors which, at the time of writing, are either under construction or in operation: the barrel and endcap timing layers of MIP Timing Detector (MTD) of the CMS experiment and the High Granularity Timing Detector (HGTD) of ATLAS, designed for the high luminosity phase of LHC;  the ALICE TOF and BESIII TOF detectors, MRPC-based and representing the current state-of-the art of TOF operating in collider experiments; the CMS Precision Proton Spectrometer (PPS), using diamond detectors; and finally, the electromagnetic and the High Granularity calorimeters in the CMS upgrade for HL-LHC, as examples of precision timing application in calorimetry. This section introduces the integration of these timing technologies into the detectors with electronics, cooling and mechanical supports. The most relevant design and operational parameters for the CMS BTL, CMS ETL, ATLAS HGTD and ALICE TOF are compared in Tab.~\ref{tab:comparison}.

\begin{table}[h!]
\centering
\begin{threeparttable}[b]
\begin{scriptsize}
\renewcommand{\arraystretch}{1.2}
\begin{tabular}{l | l l l l}
 \hline
        & \textbf{CMS} & \textbf{CMS} & \textbf{ATLAS} & \textbf{ALICE}\\ 
        & \textbf{BTL} & \textbf{ETL} & \textbf{HGTD} & \textbf{TOF}\\ 
 \hline
 Sensor technology & LYSO+SiPM & LGAD &  LGAD &  MRPC \\
 Distance to IP & R=1.148~m & z=$\pm$3~m & z=$\pm$3.5~m & R=3.7~m  \\
 Coverage & $|\eta|<$1.48 & 1.6$<|\eta|<$3 & 2.4$<|\eta|<$4 & $|\eta|<$0.9\\
 Number of layers & 1 & 2 & 2 & 1 \\
 Total area & 38~m$^2$ & 14~m$^2$ & 6.4~m$^2$ & 141~m$^2$\\
 Unit sensor/pad size & 3.75$\times$3.12$\times$54.7~mm$^2$ & 1.3$\times$1.3~mm$^2$  & 1.3$\times$1.3~mm$^2$ & 3.5$\times$2.5~cm$^2$\\
 Number of readout channels & 331776 & 8~$\times$10$^6$ & 3.6~$\times$10$^6$ & 157248\\
 Track time resolution & 30-60~ps & 35~ps\tnote{1} & 30-50~ps & 56~ps\\
 Max fluence (\neut) & 1.9$\times$10$^{14}$ & 1.7$\times$10$^{15}$ & 2.5$\times$10$^{15}$ & 2$\times$10$^{10}$\\
 ASIC & TOFHIR & ETROC & ALTIROC & NINO \\
 Temperature of operation & -45$^{\circ}$C & -30$^{\circ}$C & -30$^{\circ}$C & - \\
 Total power consumption & 1.2~kW/m$^2$ & 3.7~kW/m$^2$ & 6~kW/m$^2$ & 0.8~kW/m$^2$\\
 Ref. & \cite{CERN-LHCC-2019-003} & \cite{CERN-LHCC-2019-003} & \cite{CERN-LHCC-2020-007} & \cite{CERN-LHCC-2000-012,Cortese:545834} \\
 \hline
\end{tabular}
\begin{tablenotes}
       \item [1] with 2 hits per track
\end{tablenotes}   
\end{scriptsize}
\end{threeparttable}
\caption{Table of the most important design and operational parameters of selected timing detectors.}
\label{tab:comparison}
\end{table}

\subsection{The CMS Barrel Timing Layer}

%The MIP Timing Detector (MTD) is designed for CMS during the high-luminosity LHC (HL-LHC) operations when the instantaneous luminosity will be $5 \times 10^{34}$ cm$^{-2}$ s$^{-1}$ in average corresponding to a number of interactions per bunch crossing of 140. MTD will provide a track time resolution of 30~ps (60~ps) at the beginning (end) of the operation of HL-LHC, with which MTD will reduce the pileup to a similar level of the current LHC operations of 40-60. 
The MTD is a new CMS detector component under development for the Phase II upgrade at the HL-LHC. 
MTD will provide a track time resolution of 30~ps (60~ps) at the beginning (end) of the operation of HL-LHC, which will be exploited to mitigate pileup effects in the reconstruction of collision events.
The barrel and endcap sectors are designed with different technologies to cope with the diﬀerent radiation levels and additional constraints from integration
within the existing CMS detector, cost, and power budget. This section introduces the barrel part while the next will discuss the endcap.

\begin{figure}[h]
\centering
\includegraphics[width=0.9\linewidth]{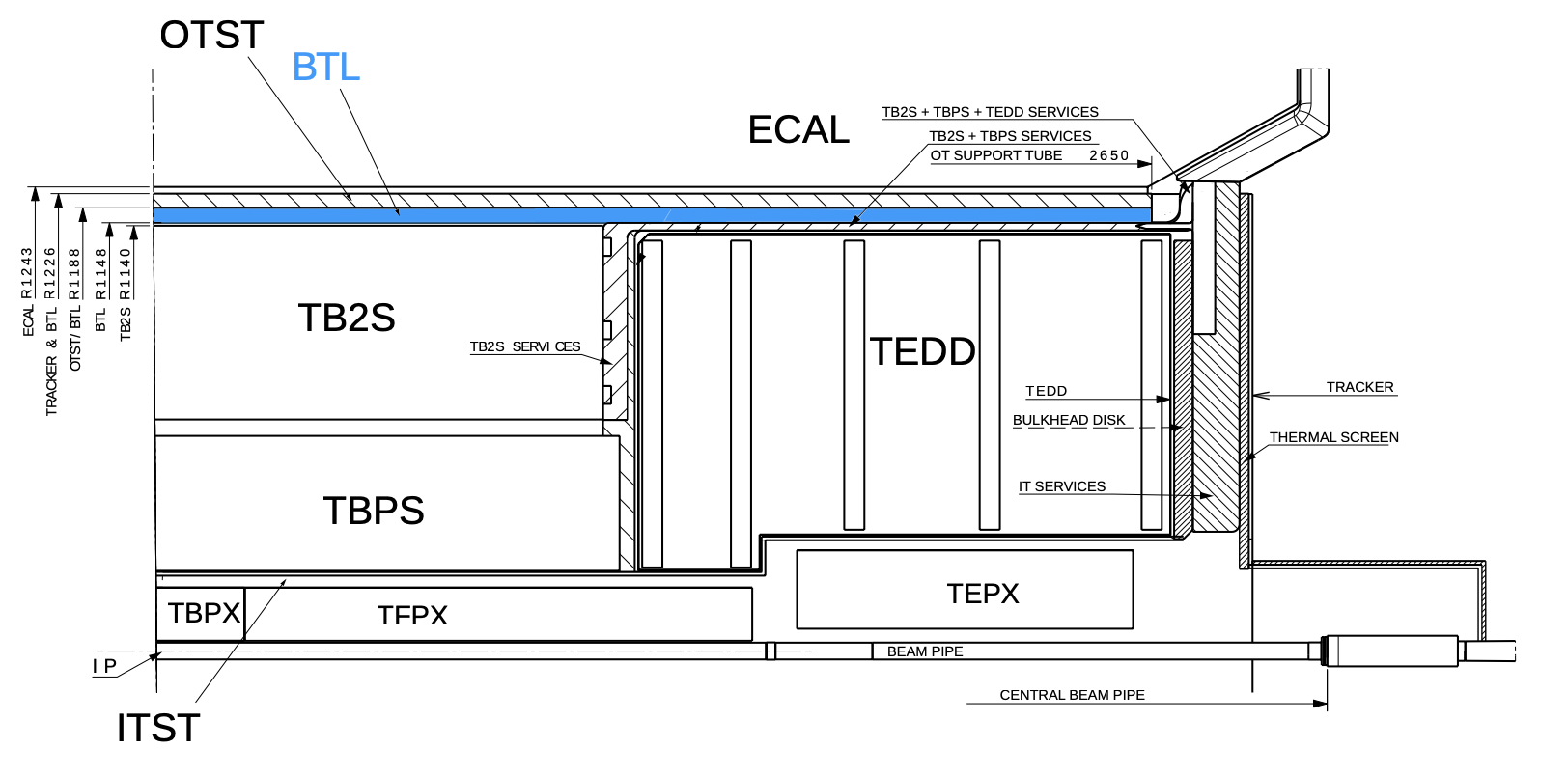}
\caption{The BTL integration in CMS}
\label{fig:btl_loc}
\end{figure}

The MTD Barrel Timing Layer (BTL) will be installed between the tracking system and the electromagnetic calorimeter (ECAL) with an inner radius of 1148 mm and outer of 1188 mm, as shown in Fig.~\ref{fig:btl_loc}. The sensors consist of Lutetium Yttrium Orthosilicate crystals doped with Cerium (LYSO:Ce) coupled to SiPMs at both ends. 16 LYSO bars with the dimension of 54.7 $\times$ 3.12 $\times$ 3.75 mm$^3$ wrapped by a thin layer of Enhanced Specular Reflector (3M ESR) are packed as an array. They are manufactured by Sichuan Tianle Photonics (STP). The LYSO array is coupled to two arrays of 16 SiPMs through a layer of RTV3145 glue about 100 $\mu m$-thick. Accordingly, the SiPM array contains 16 SiPMs carried by a PCB the opposite side of which mounts 4 Thermal Electric Coolers (TECs) to change the local temperature for operation and annealing, and an RTD for onboard temperature monitoring. The SiPMs have a dimension that matches the cross section of each LYSO bar. They are manufactured by Hamamatsu Photonics (HPK) with a cell-size of 25 $\mu m$ and a fill factor of 0.76. The photon detection efficiency is 57 (30)\% and a gain of 10$^6$ (3.6 $\times$ 10$^5$ ) for $V_{OV}$= 3.5 (1) V for the beginning (end) of the operation. 
%The lower $V_{OV}$ in the end of operation is to suppress the dark count rate (DCR) due to radiation damage and limit the power consumption.
One of the main challenges of BTL is the radiation damage (up to 2$\times$10$^{14}$~\neut after 3000~fb$^{-1}$) in the SiPMs. The approaches to mitigate the DCR, which will be O(10) GHz per SiPM at the end of operation, include operation at cold (T~=~-45$^{\circ}$C), SiPM annealing at high temperature (T~=~+60$^{\circ}$C) during technical stops of the accelerator and tuning of the operating voltage along the BTL lifetime.

\begin{figure}[h]
\centering
\includegraphics[width=0.75\linewidth]{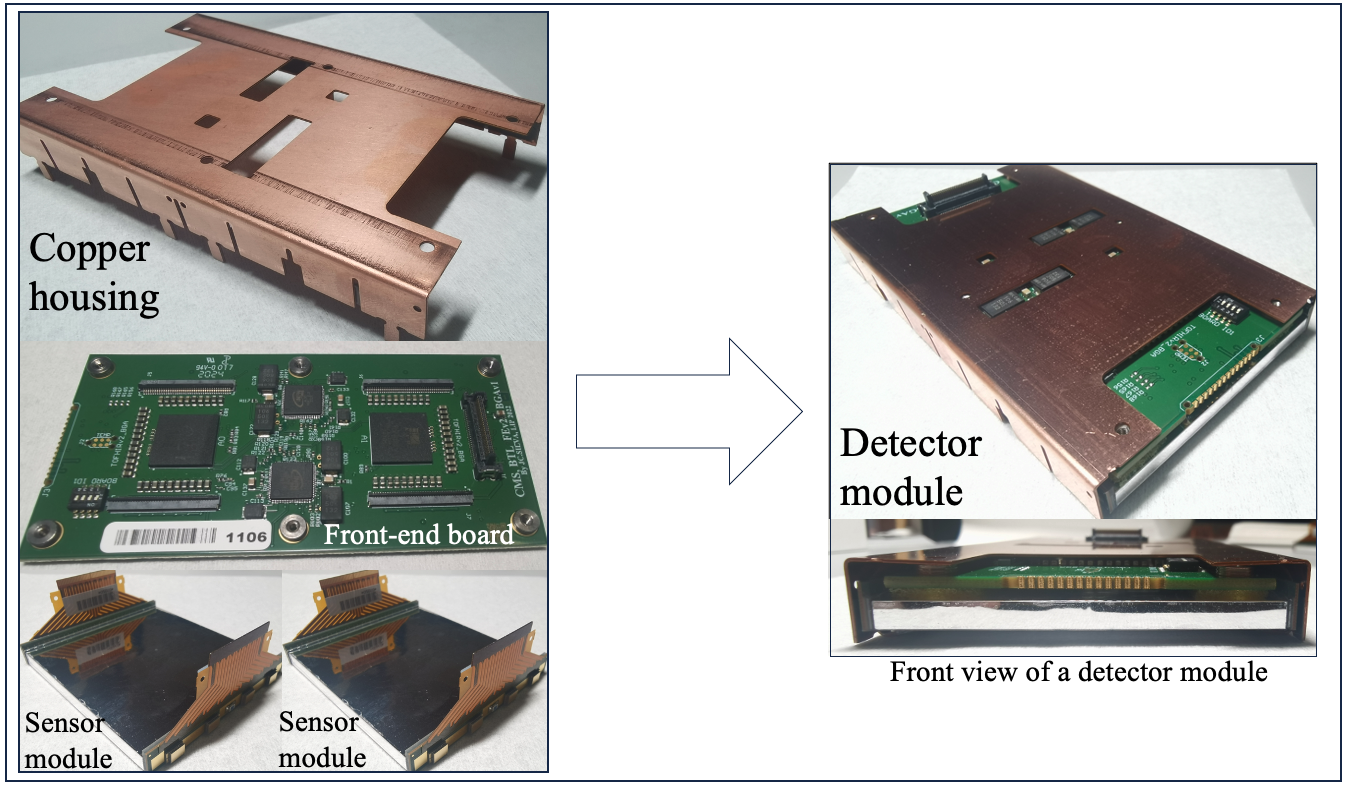}
\includegraphics[width=0.75\linewidth]{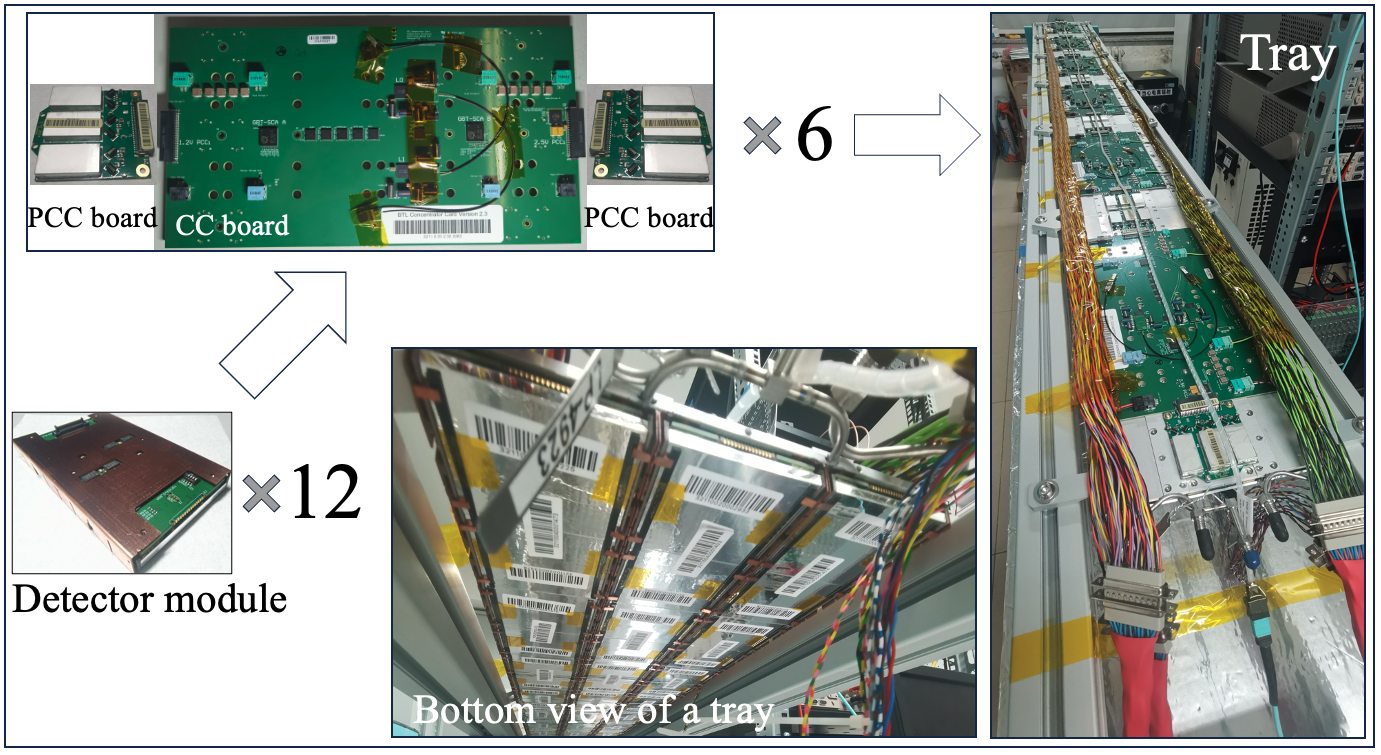}
\caption{The assembly scheme of CMS BTL.}
\label{fig:btl_assm}
\end{figure}

The sensor module (SM) consists of the wrapped 16 LYSO bars coupled to 16 SiPMs in each end covering a rectangular area of about 60 $\times$ 52 mm$^2$. The detector module (DM) then mounts 2 sensor modules to the front-end board that carries 2 front-end ASICs (TOFHIR2~\cite{Albuquerque_2024}) and 2 voltage regulator chips (ALDO~\cite{ALDO_2024}). 
TOFHIR2 provides measurements of the time of arrival of the MIP signals, using a TDC with 10~ps binning,
and of the amplitude of the signals through charge integration.
A matrix of 3 $\times$ 4 DMs and electronics (1 concentrator card, CC and 2 power converter cards, PCC) mounted the each face with a cooling plate make a readout unit (RU) with a rectangular area of about 182.5 $\times$ 413.3 mm$^2$. 6 of RUs connected in a line form a tray. The BTL barrel is split into two sectors forward and backward along the beam, each of which consists of 36 trays longitudinally. In total, 72 trays cover the entire surface of outer of the tracking system. The total number of channels include 331,776 SiPMs and 165,888 LYSO bars. Fig.~\ref{fig:btl_assm} provides a basic idea of the assembly scheme of CMS BTL.

The TOFHIR2 chip is designed by PETsys Electronics, using
CMOS 130~nm technology. The ASIC features 32 channels, and each channel integrates the analog front-end, time and charge digitizers and the channel digital control.
TOFHIR2 is engineered to withstand the total ionization dose TID (2.9 Mrad) and to the integrated particle fluence (2$\times$10$^{14}$ \neut) expected in the BTL.
A key feature of TOFHIR2 is the implementation, for the first time in a CMOS integrated circuit, of a dedicated DCR noise filtering based on the differential leading edge discriminator (DLED) concept~\cite{DLED_GolaPiemonte}. In the DLED technique, an inverted and delayed pulse is added to the original pulse, allowing baseline restoration while preserving the fast rising edge of the signals. In TOFHIR2, the delay line is approximated by a programmable RC network (yielding delays in the range 200-1800 ps). 

\begin{figure}[h]
\centering
\includegraphics[width=0.49\linewidth]{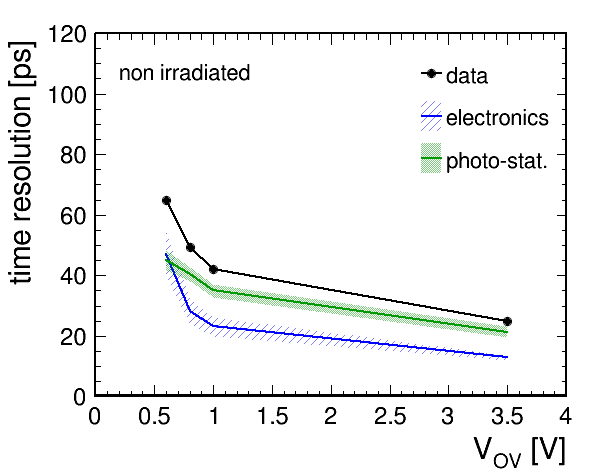}
\includegraphics[width=0.49\linewidth]{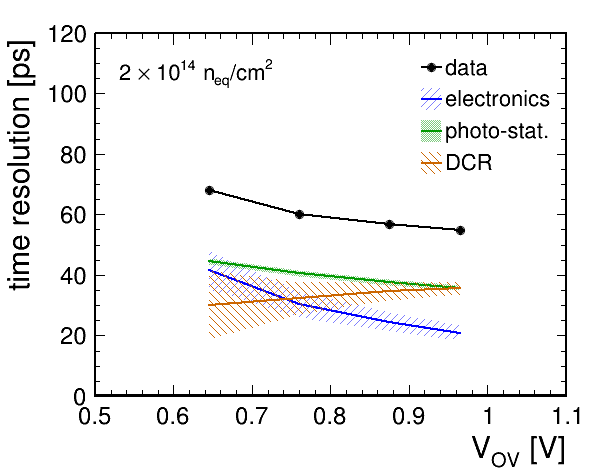}
\caption{Time resolution as a function of the SiPM over-voltage for modules with non-irradiated SiPMs (left) and SiPMs irradiated to 2~$\times$~10$^{14}$~\neut~(right). The time resolution measured with beam data is shown with black dots, while the colored lines represents the main individual contributions to the time resolution: electronics (blue), photo-statistics (green), and DCR (orange)~\cite{Addesa:2025kyl}.}
\label{fig:btl_timereso}
\end{figure}

By the time of writing, the BTL design has been completed and the assembly is undergoing. The designed time resolution of 30 ps for the condition of the beginning of the operation and 60 ps for the end has been reached with beam tests as shown in Fig.~\ref{fig:btl_timereso}.

\subsection{The CMS Endcap Timing Layer}
% - layout
% - sensors 
% - electronics
% - ...
The forward part of the CMS MTD, the Endcap Timing Layer (ETL)~\cite{CERN-LHCC-2019-003}, consists of a two-disc system positioned on either side of the interaction region and instrumented with LGAD sensors. Each pair of disks is located between the CMS endcap calorimeter and the end of the tracker, about 3~m from the nominal interaction point and covers an annular region with a radius between 315~mm and 1200~mm, corresponding to a pseudorapidity acceptance of 1.6 $< |\eta| <$ 3.0. Each disk has silicon devices on both faces, for a total active area of about 14~m$^2$. More than 85\% of the area of each disk is sensitive to (MIPs). The arrangement of two such disks per endcap, adjacent to each other with about 20~mm z-separation, allows for hermetic coverage and an average of about 1.7 hits per track. The target time resolution is 50~ps per single hit and 35~ps per track. The ETL will operate at a temperature of -30$^{\circ}$C with CO${_2}$ dual-phase cooling.

\begin{figure}[h]
\centering
\includegraphics[width=0.45\linewidth]{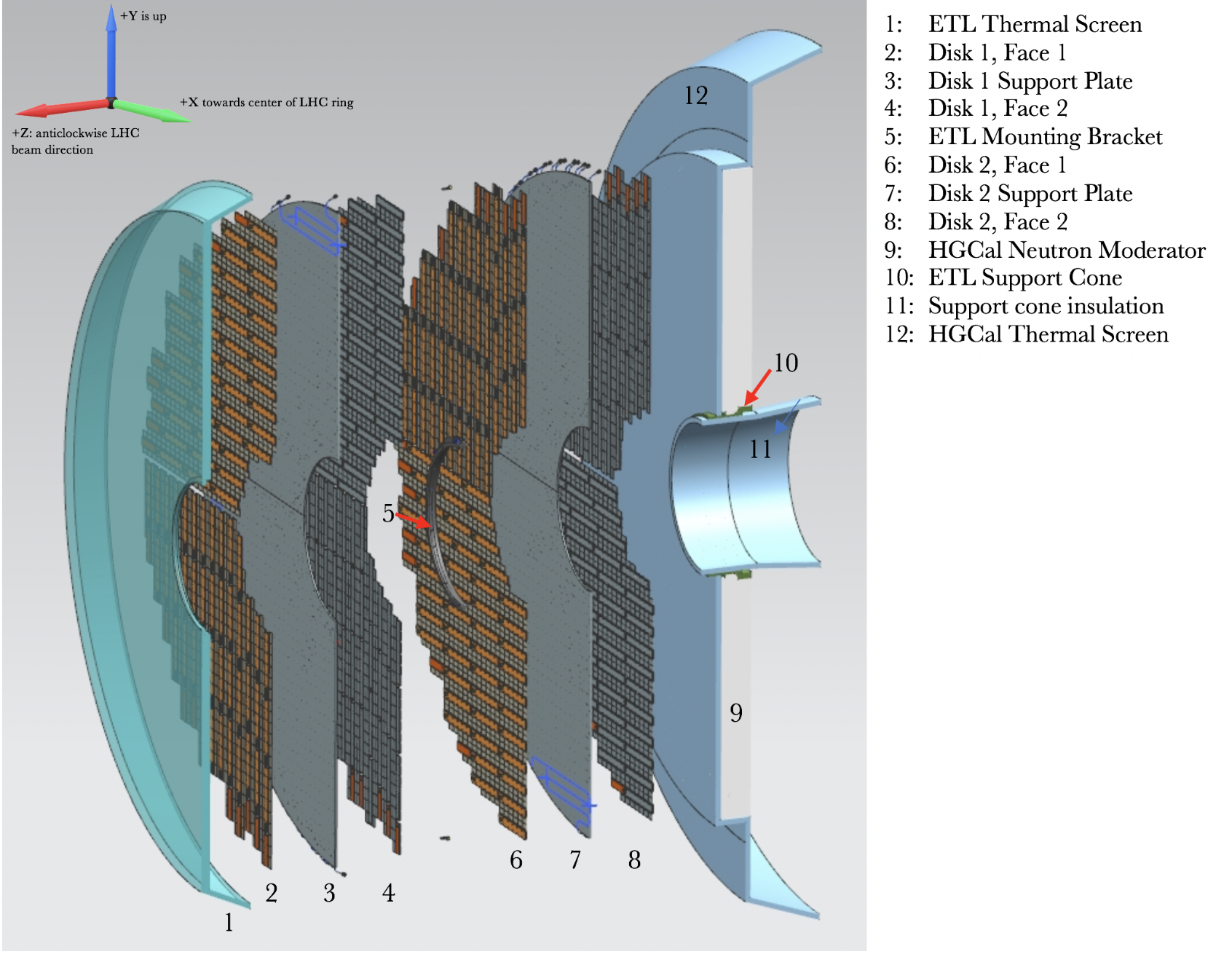}
\includegraphics[width=0.53\linewidth]{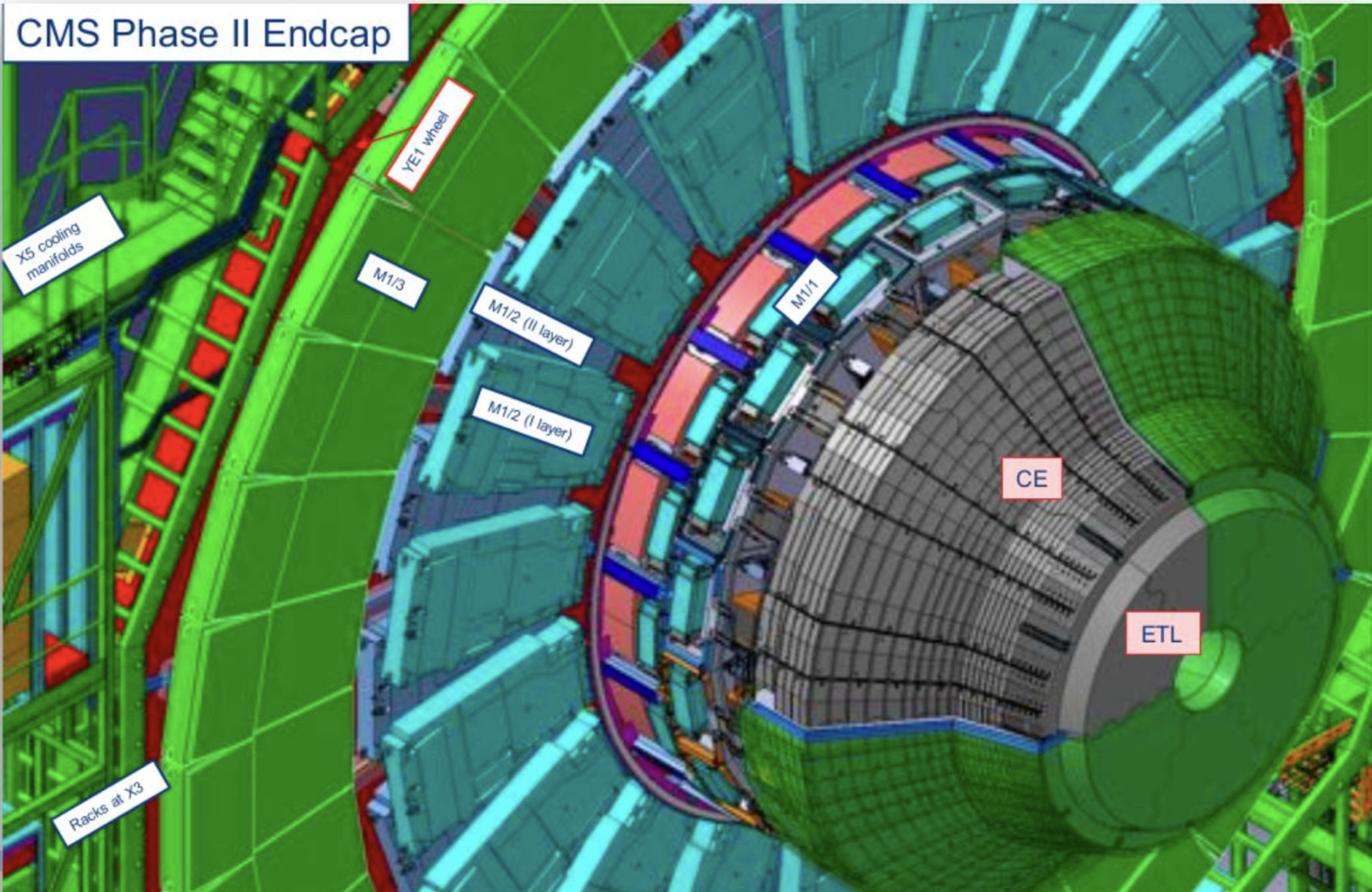}
\caption{ETL layout (left) and position of the ETL on the CMS endcap calorimeter nose (right)~\cite{CERN-LHCC-2019-003}.}
\label{fig:etl_layout}
\end{figure}

The radiation levels expected in ETL are the main driver for choosing silicon sensors over other technologies. The fluence varies with the distance from the beamline and is expected to reach 1.1$\times$10$^{14}$~\neut\ in the outermost detector region and up to 1.7$\times$10$^{15}$~\neut\ for $|\eta|$~=~3 at the end of operation. About 15\% of the detector will be exposed to fluences larger than 1$\times$10$^{15}$~\neut.

\begin{figure}[!h]
\centering
\includegraphics[width=0.37\linewidth]{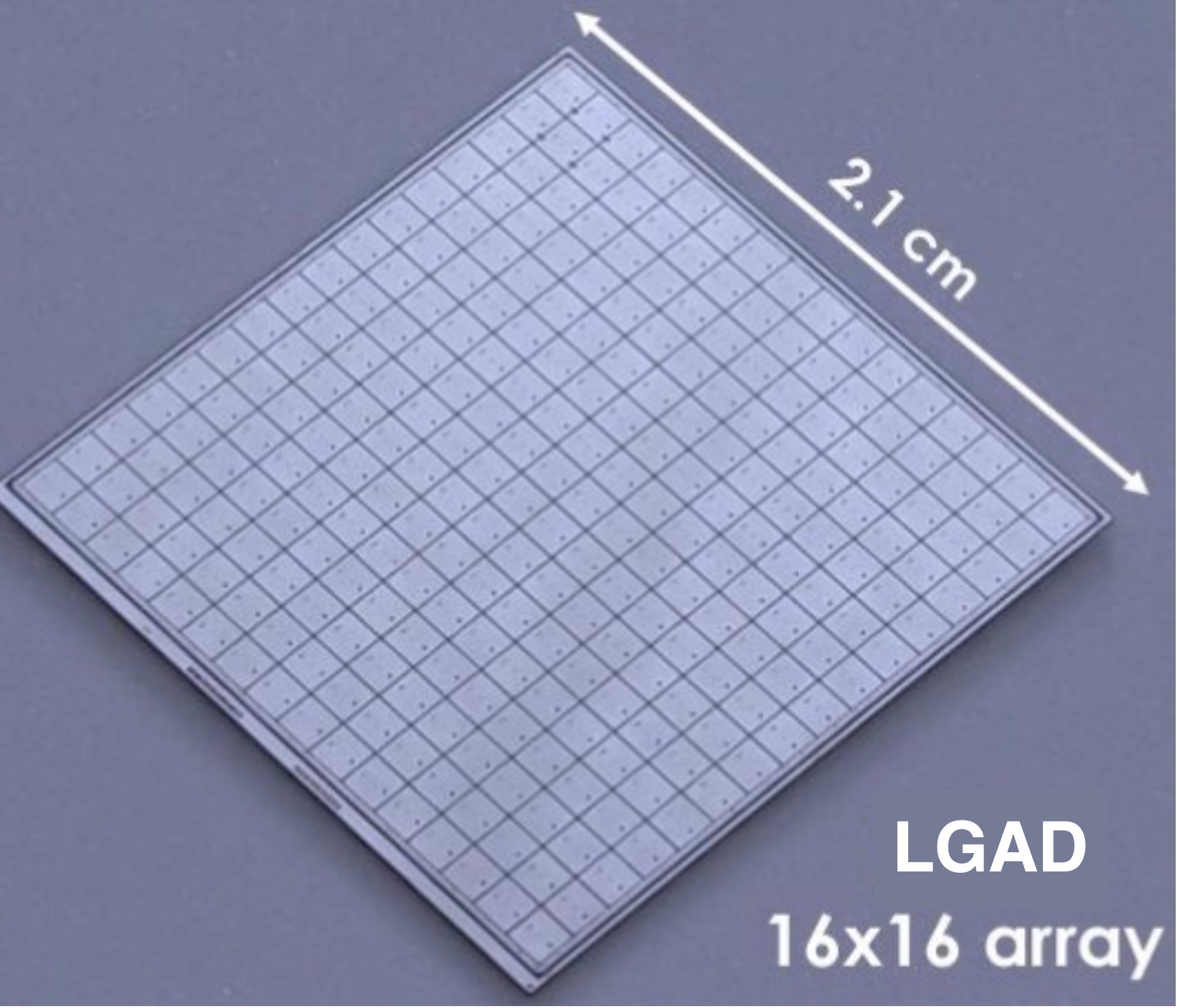}
\includegraphics[width=0.61\linewidth]{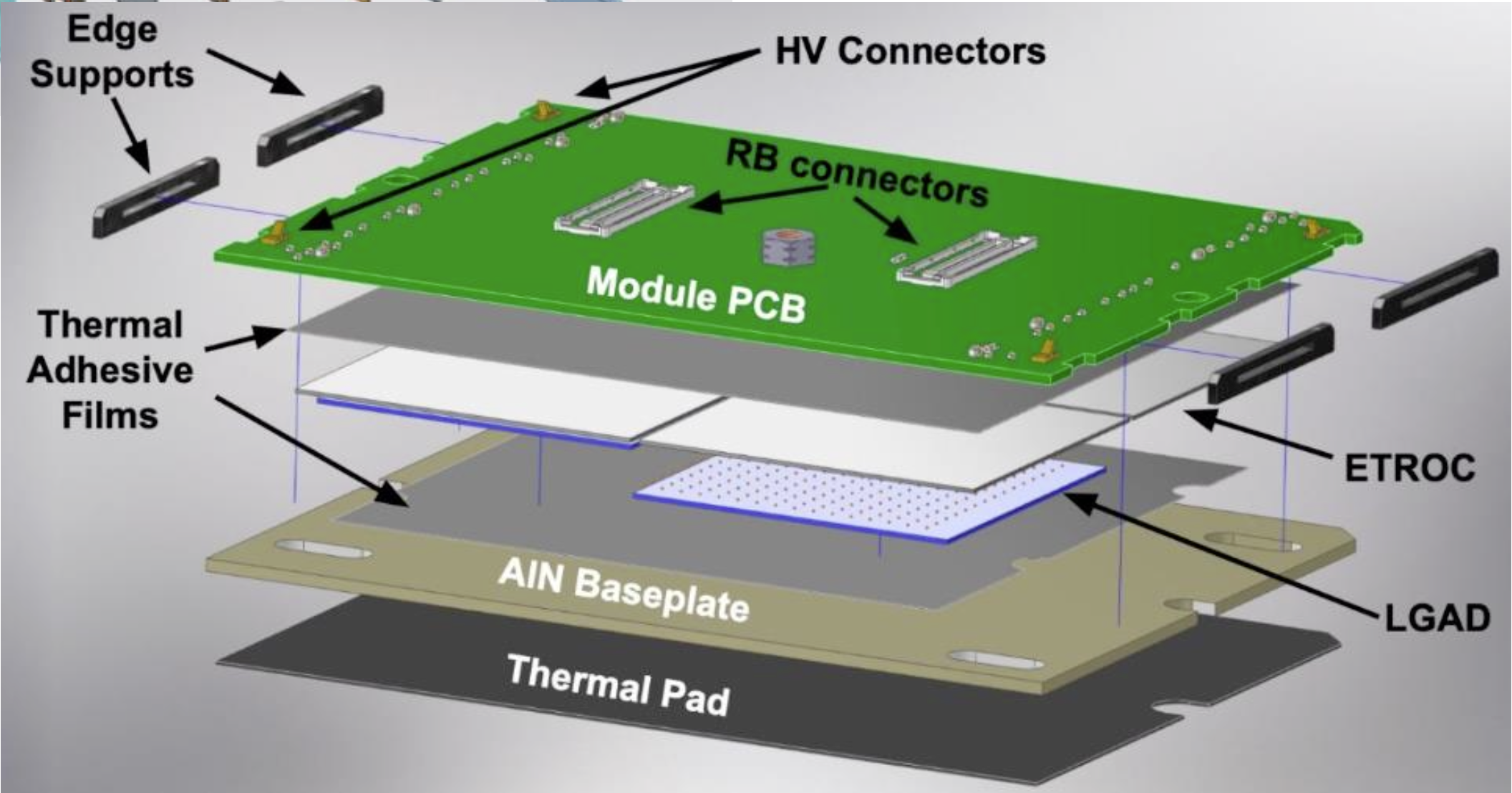}
\caption{Picture of a 16$\times$16 LGAD sensor for the CMS ETL (left) and layout of a module housing four sensors~\cite{CERN-LHCC-2019-003}.}
\label{fig:etl_sensors}
\end{figure}

The ETL sensor is a 16$\times$16 pad array (Fig.~\ref{fig:etl_sensors}-left)), with each pad having an active area of 1.3~$\times$~1.3~mm$^2$ area and 50~$\mu m$ active thickness. The collected charge from MIP signals is expected to be greater than 8~fC until the end of operation.

\begin{figure}[!h]
\centering
\includegraphics[width=0.41\linewidth]{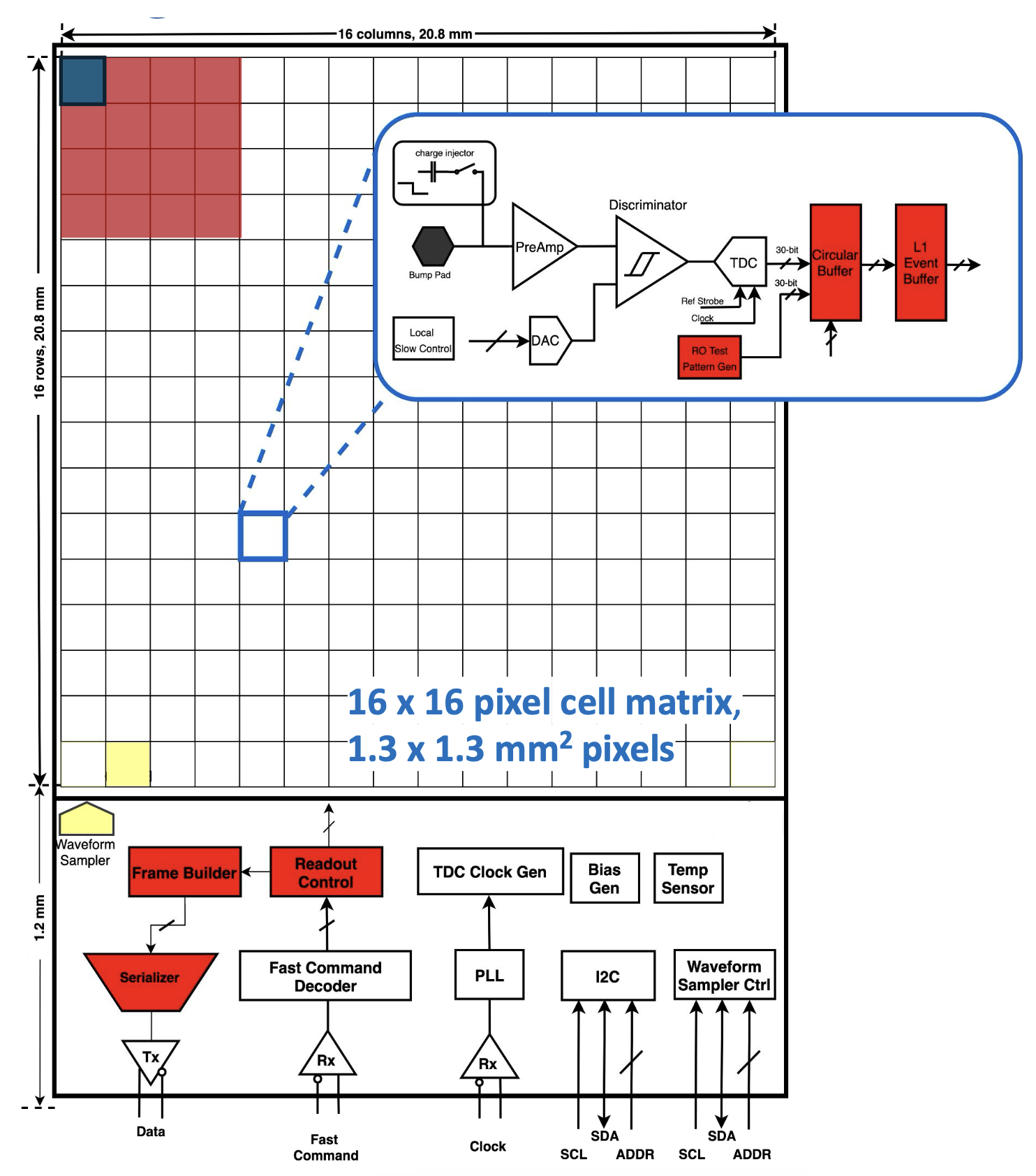}
\includegraphics[width=0.57\linewidth]{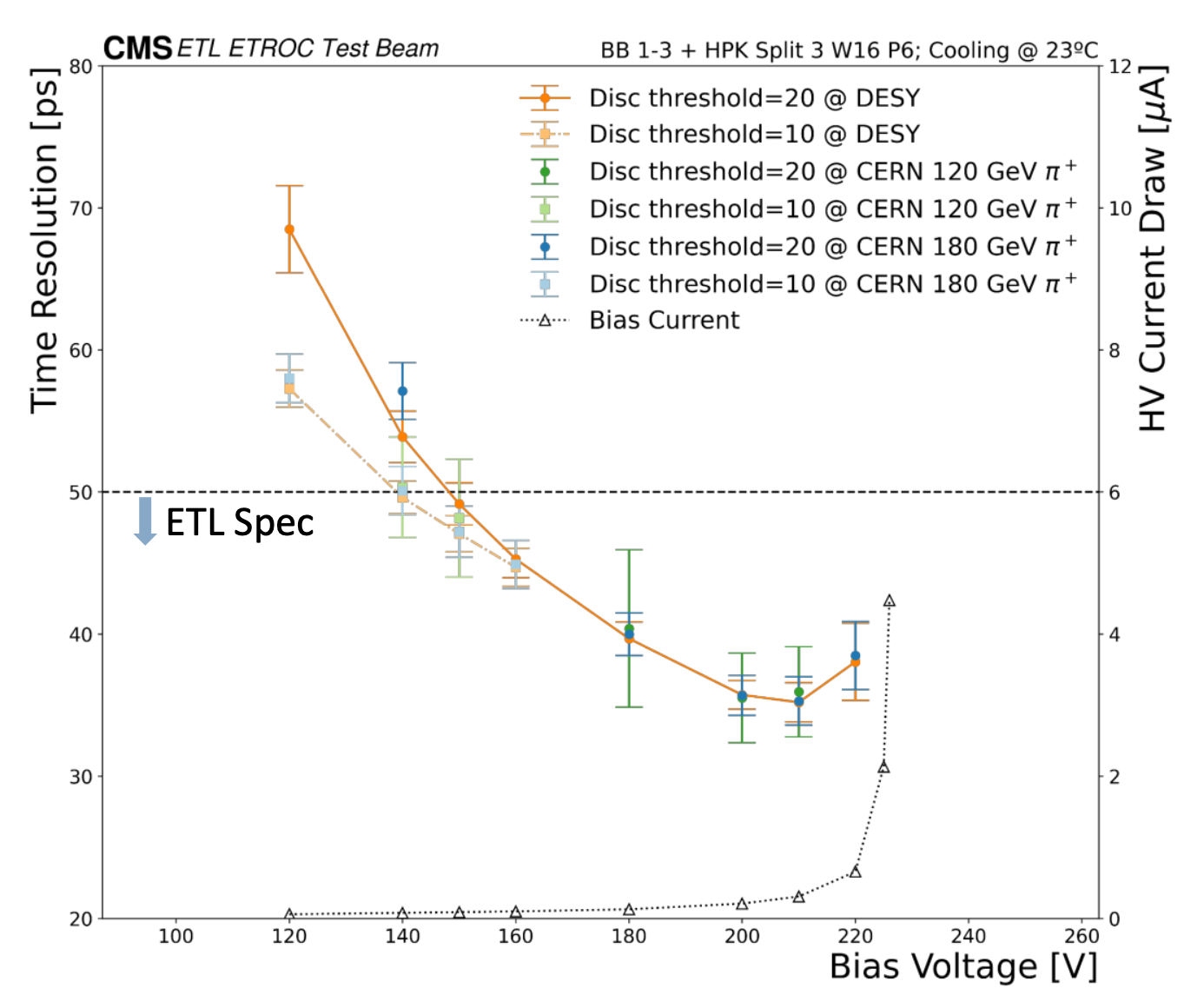}
\caption{ETROC block diagram (left) and time resolution as a function of the sensors bias voltage measured on beam with LGAD sensors read out with the CMS ETL readout chip ETROC2 (right)~\cite{TWEPP2024-TedLiu}.}
\label{fig:etroc2}
\end{figure}

A 16$\times$16 channels custom ASIC, the Endcap Timing Read Out Chip (ETROC), developed in CMOS-65~nm technology and bump bonded to the sensors, is used for read out.
The ETROC is designed to ensure the ASIC contribution to the total time resolution remains below 40~ps with limited power consumption ($\lesssim$1~W/chip or $\lesssim$4~mW/channel). 
%The pad cell size is a compromise between small capacitance (3.4~pF), which translates to a higher signal-to-noise ratio in the front-end, and the overall channel count.
Each channel comprises a preamplifier, a discriminator, a TDC used to digitize the time-of-arrival and time-over-threshold measurements, and a memory for data storage and readout (Fig.~\ref{fig:etroc2}-left).

The elemental unit of ETL is the module, a structure
housing 4 LGAD sensors bump bonded to one ETROC each, an AIN baseplate for thermal contact with the support disk, and a PCB interfacing to a multi-module read-out board (Fig.~\ref{fig:etl_sensors}-right).
The ETL will be made of 8000 modules, for a total of about 8 millions of read out channels.

Fig.~\ref{fig:etroc2} shows the time resolution measured on beam for prototype 16$\times$16 sensors bump bonded to the ETROC2 chip, demonstrating the target time resolution of $<$~50~ps per hit.

%ASIC contribution to time resolution < 40ps
%Targeted signal charge (1MIP): 6fC
%TDC range: up to 5ns TOA and 10ns TOT
%L1 buffer latency: 12.5 us
%with power consumption < 4 mW/channel
\subsection{The ATLAS High Granularity Timing Detector}
\begin{figure}[!h]
\centering
\includegraphics[width=0.65\linewidth]{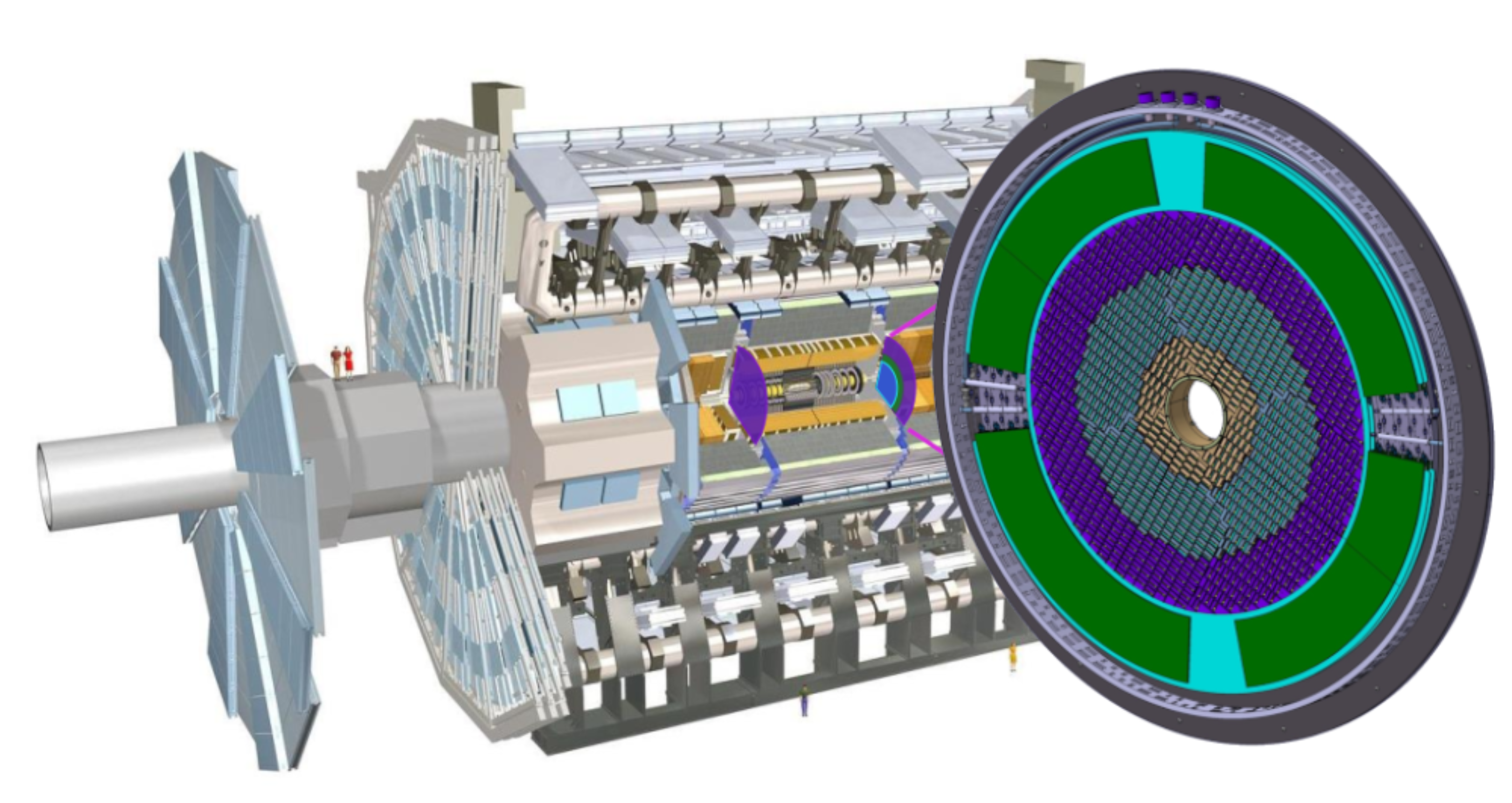} ~
\includegraphics[width=0.59\linewidth]{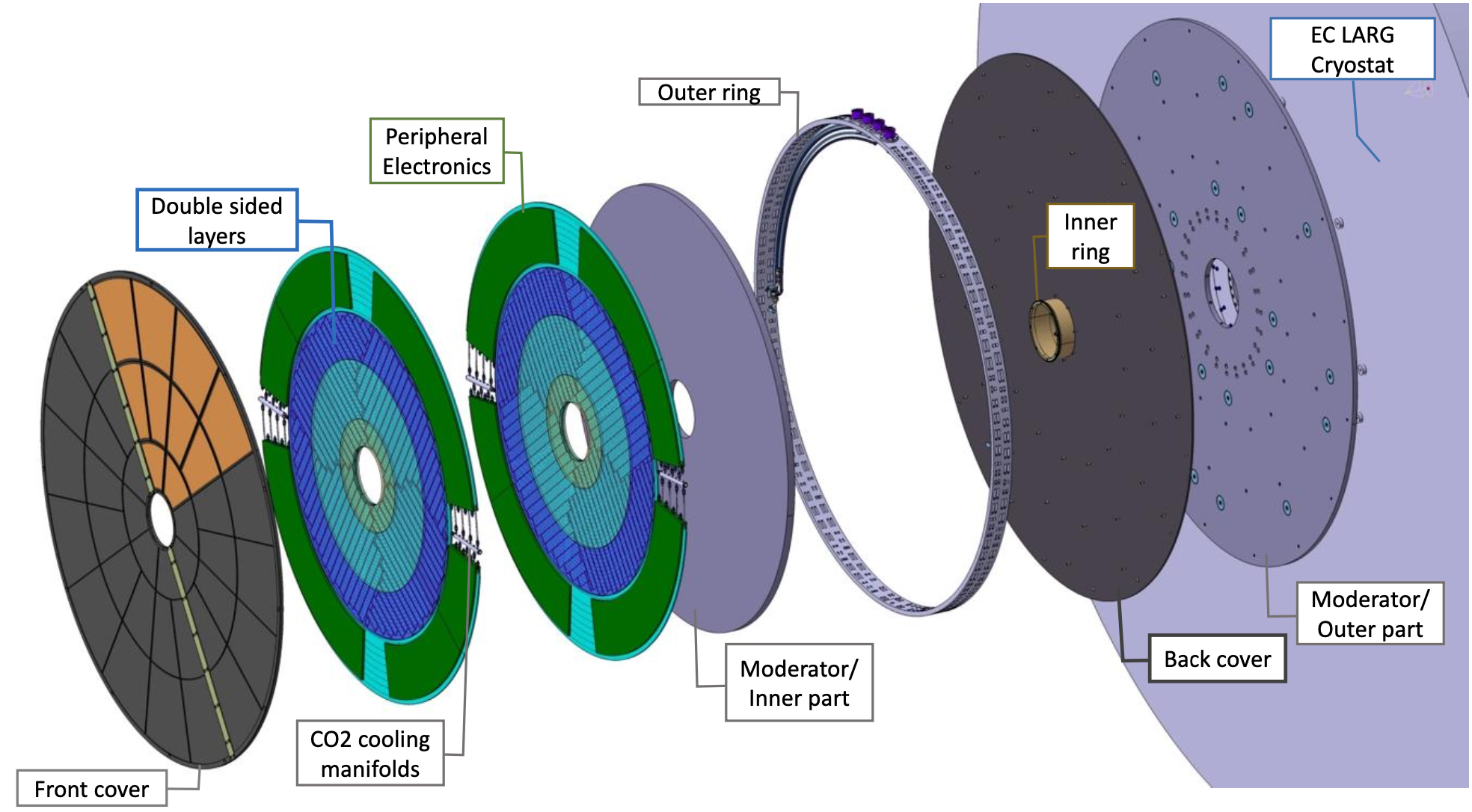}
\includegraphics[width=0.59\linewidth]{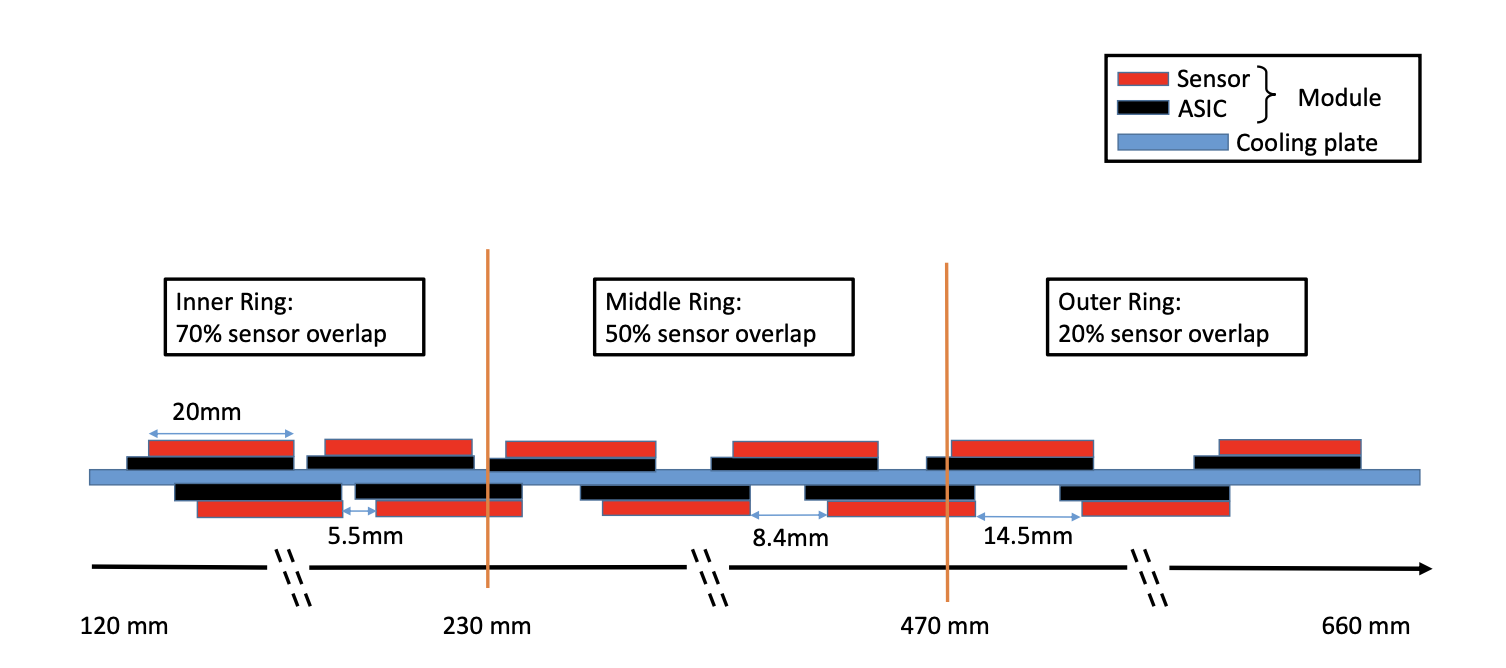}
\includegraphics[width=0.39\linewidth]{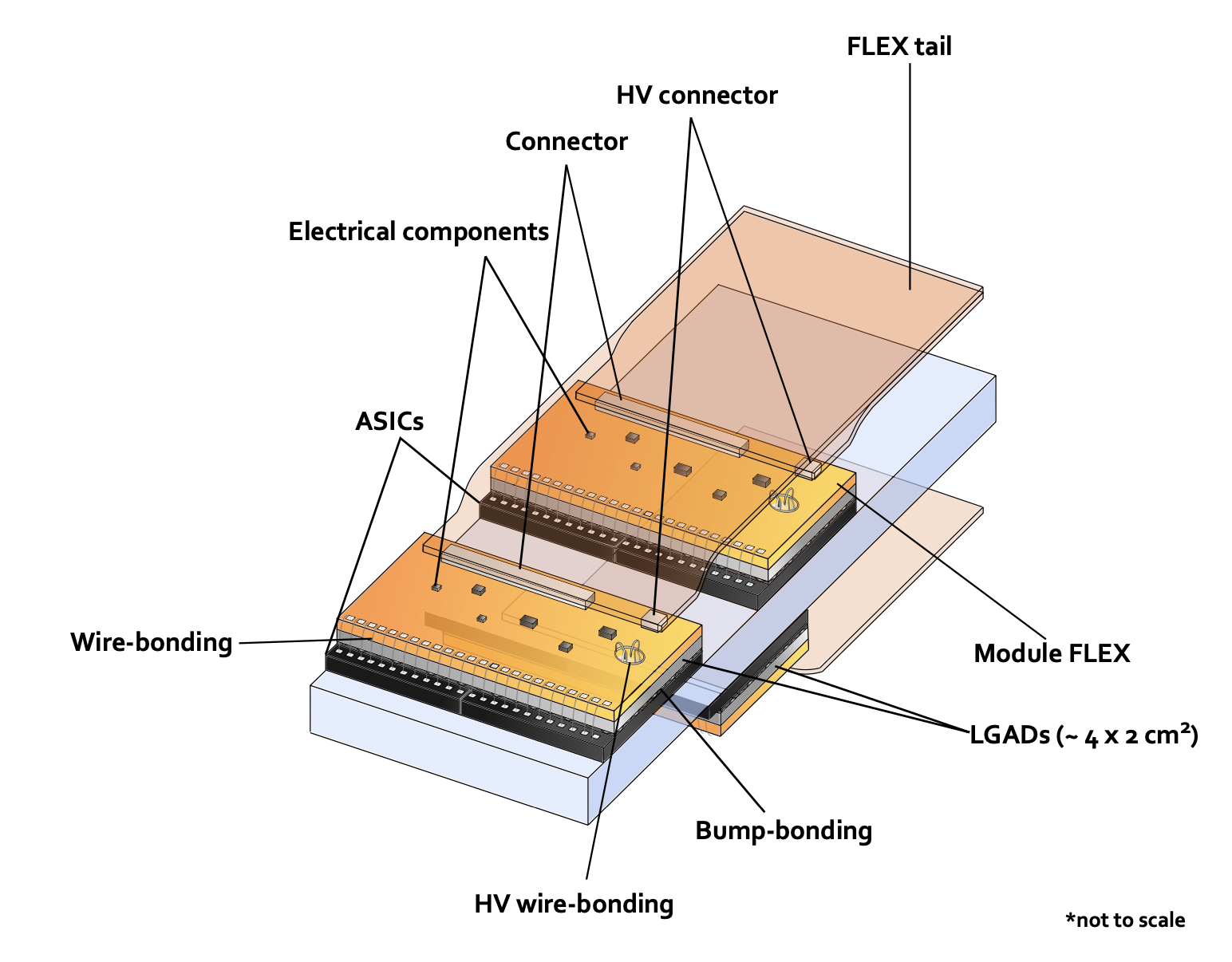}
\caption{Position of the HGTD within the ATLAS detector (top) and global view of the HGTD with its various components (middle). Bottom: the HGTD double-sided layer with the cooling plate (in blue) instrumented with LGAD sensors (in red) with the different overlap regions shown by the vertical lines and a schematic drawing of two adjacent modules on the top side and one on the bottom side of the cooling plate~\cite{CERN-LHCC-2020-007}.}
\label{fig:hgtd_layout1}
\end{figure}

% - layout
% - sensors 
% - electronics
% - ...
The High Granularity Timing Detector (HGTD)~\cite{CERN-LHCC-2020-007} is a new detector included in the ATLAS experiment upgrade and based on the LGAD technology similarly to the CMS ETL.
The HGTD consists of two endcaps located between the barrel and the forward calorimeters and positioned at a distance of $\pm$3.5~m from the nominal interaction point, as shown in Fig.~\ref{fig:hgtd_layout1}. 

The detector active area covers the pseudorapidity range 2.4~$<|\eta|<$~4.0, corresponding to a region with a radius between 120~mm and 640~mm. Each endcap is made of two instrumented double-sided layers arranged in a three-ring structure, where the detector units overlap on the front and back of the disk is optimized to give approximately uniform performance across the $\eta$ range (Fig. ~\ref{fig:hgtd_layout1}. The total active surface is 6.4~m$^2$. Each ring will be exposed to different radiation doses, depending on the radius. A replacement of two innermost rings is planned after a total fluence of 2.5~$\times$10$^{15}$~\neut, corresponding to an integrated luminosity of 1000 and 2000~fb$^{-1}$ for the innermost and middle ring, respectively. The detector targets at a time resolution of 30~ps per track at the beginning of the HL-LHC and 50~ps after a fluence of 2.5$\times$10$^{15}$~\neut.

\begin{figure}[!h]
\centering
\includegraphics[width=0.6\linewidth]{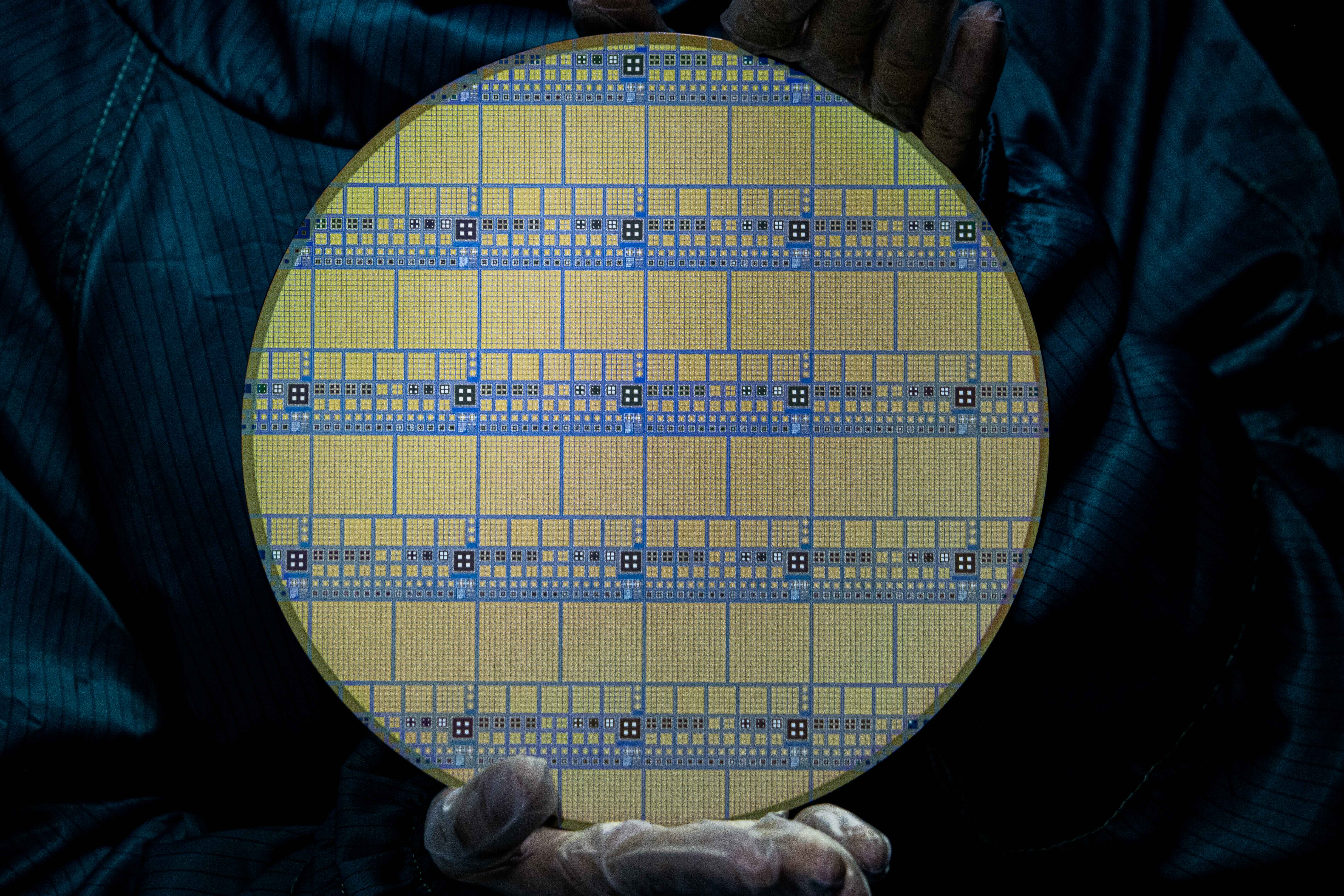}
\caption{An 8-inch prototype wafer of low-gain avalanche diodes for the High Granularity Timing Detector~\cite{Wu:2771199}.}
\label{fig:hgtd_sensors}
\end{figure}

%\begin{figure}[!h]
%\centering
%\includegraphics[width=0.59\linewidth]{hgtd_overlap.png}
%\includegraphics[width=0.39\linewidth]{hgtd_module.png}
%\caption{On the left, the HGTD double-sided layer with the cooling plate (in blue) instrumented with LGAD sensors (in red): the different overlap regions are shown by the vertical lines. On the right, schematic drawing of two adjacent modules on the top side and one on the bottom side of the cooling plate. Figures from~\cite{CERN-LHCC-2020-007}}
%\label{fig:hgtd_layout2}
%\end{figure}

\begin{figure}[!h]
\centering
\includegraphics[width=0.485\linewidth]{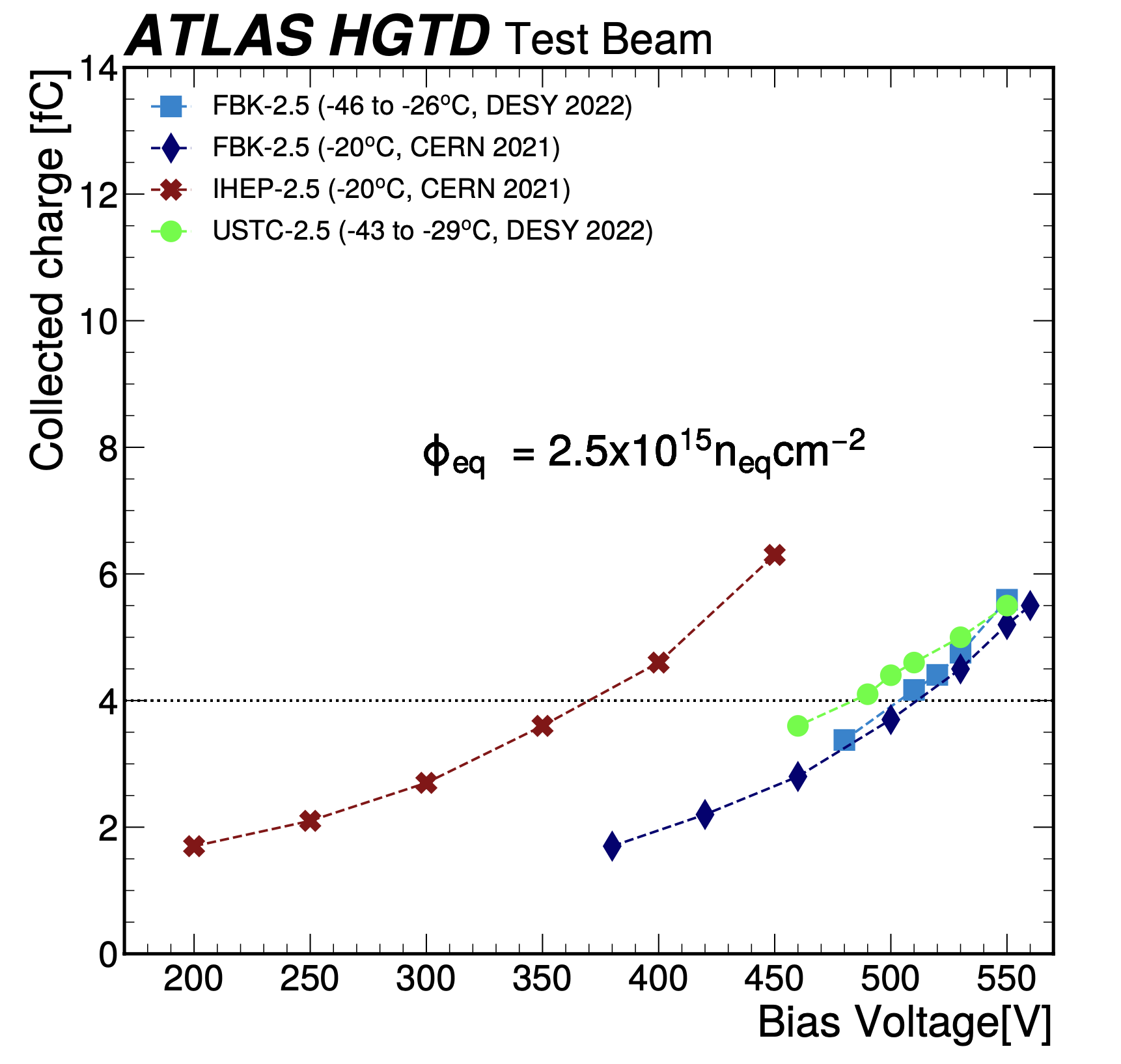}
\includegraphics[width=0.47\linewidth]{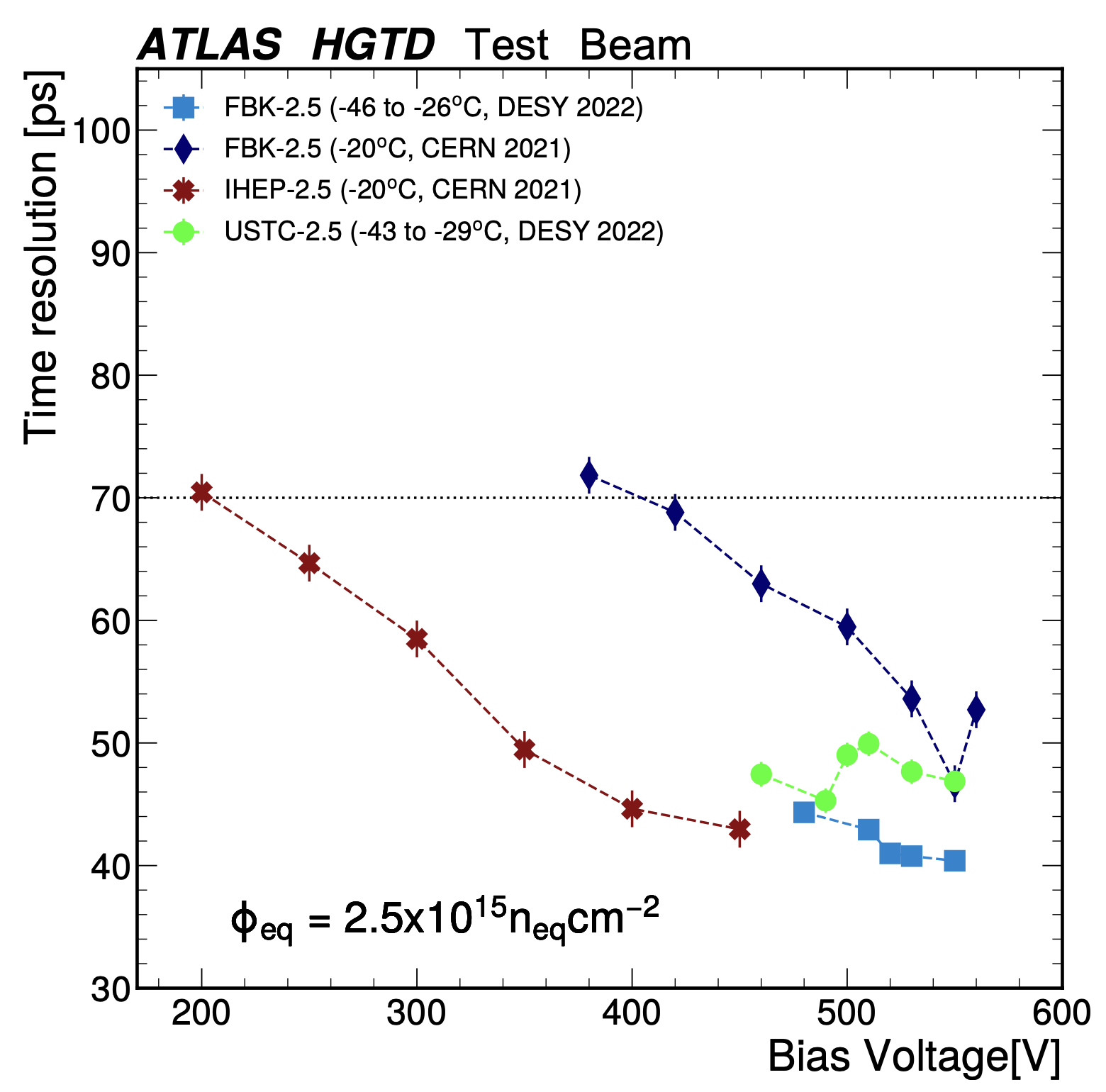}
\caption{Collected charge (left) and time resolution (right) as a function of the bias voltage for different single-pad prototype sensors for the ATLAS HGTD irradiated to a fluence of 2.5~$\times$10$^{15}$~\neut. The dashed lines represents the minimum required charge and maximum time resolution for the future HGTD after irradiation~\cite{Ali_2023}.}
\label{fig:hgtd_resolution}
\end{figure}

The HGTD LGAD sensor consists of an array of 15~$\times$~15 pixels (Fig.~\ref{fig:hgtd_sensors}), each with a size of 1.3~$\times$~1.3~mm$^2$ and an active thickness of 50~$\mu m$. Carbon-enriched LGADs are the chosen sensor technology for HGTD as they provide good performance after irradiation to the ultimate HGTD fluence, allowing safe operation up to 550~V (no SEB observed). Fig. ~\ref{fig:hgtd_resolution} reports the results obtained in beam tests measurements for prototype sensors irradiated to a fluence of 2.5~$\times$10$^{15}$~\neut, showing that the target of a collected charge larger than 4~fC and time resolution lower than 70~ps is achieved.

The fundamental unit of the HGTD is the module, an assembly of two LGAD sensors bump-bonded to two front-end ASICs and a flexible printed circuit board for connection to the peripheral electronics board. There are 8032 modules in the HGTD, for a total of 3.6 million channels.
The HGTD ASIC, named ALTIROC (ATLAS LGAD Timing Integrated Read Out Chip), is designed for 130~nm CMOS technology from TSMC and is a key component to achieve the target time resolution. It is required to have a jitter smaller than 25 (65) ps for signals exceeding 10~fC and below 70~ps for 10 (4)~fC charge collection.
The ALTIROC features 225 channels arranged in a 15$\times$15 matrix.
Each channel integrates a preamplifier, a high-speed
discriminator, and two Time to Digital Converters (TDC) in order to extract and digitize the Time of Arrival (TOA) and the Time Over Threshold (TOT) of each LGAD signal.

\subsection{ALICE Time-Of-Flight detector}

The Time-Of-Flight (TOF) detector of ALICE~\cite{CERN-LHCC-2000-012,Cortese:545834} at the LHC is installed at about 3.7 m away from the beam covering up to $|\eta|$=0.9 with a total area of about 141 m$^2$. The ALICE TOF is based on Multigap Resistive Plate Chambers (MRPCs) with a gas mixture of Freon (93\%) and SF$_6$ (7\%). There are in total 1593 MRPC strips mounted in 90 gas-tight modules, 5 of which make a SuperModules (SM). The whole TOF consists of 18 SMs installed every 20$^{\circ}$ in the azimuthal coordinate.

\begin{figure}[!h]
\centering
\includegraphics[width=0.75\linewidth]{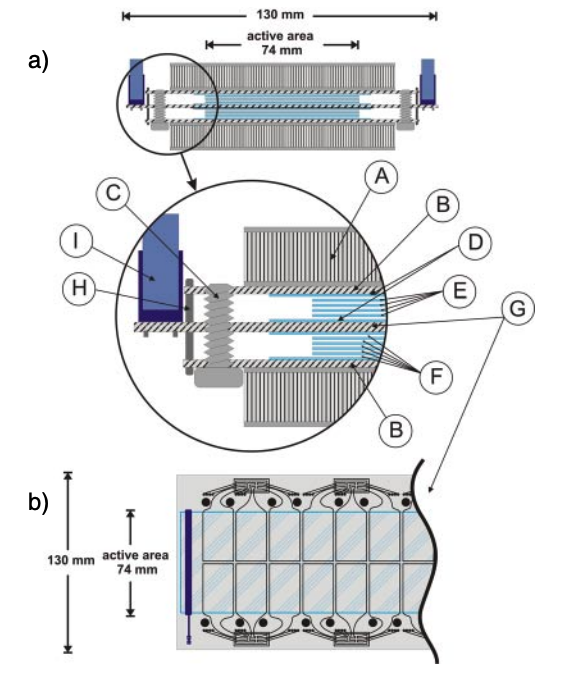}
\caption{a) Schematic diagram of a double-stack MRPC: A) phenolic honeycomb panel, B) cathode PCBs, C) nylon pin to stretch the fishing line, D) 550 $\mu$m glasses with resistive coating, E) 400 $\mu$m glasses, F) 250 $\mu$m gaps, G) anode PCBs, H) metallic pin soldered to cathode and anode PCB, I) 16 pin connector. b) Layout of the readout pads, from Ref.~\cite{AKINDINOV2004611,Akindinov:2009zze}.}
\label{fig:alice_sensor}
\end{figure}

One MRPC strip has an active area of 7.4 $\times$ 120 cm$^2$ and contains 2 rows of 48 pickup pads with an active area of 3.5 $\times$ 2.5 cm$^2$. A design of double stack MRPC is adopted as shown in Fig.~\ref{fig:alice_sensor}. This reduces the applied voltage by a factor of two while keeping the signals unchanged, and also suppresses the border effect between adjacent pickup pads given a shorter distance between the anode and cathode. The resistive plates are made of “soda-lime” float glass with a volume resistivity of 10$^{13} ~\Omega cm$ manufactured by Glaverbel S.A., Bruxelles. The outer glasses have a thickness of 550 $\mu m$, while the inner ones are 400 $\mu m$. The gaps thickness is 250 $\mu m$.

\begin{figure}[!h]
\centering
\includegraphics[width=0.65\linewidth]{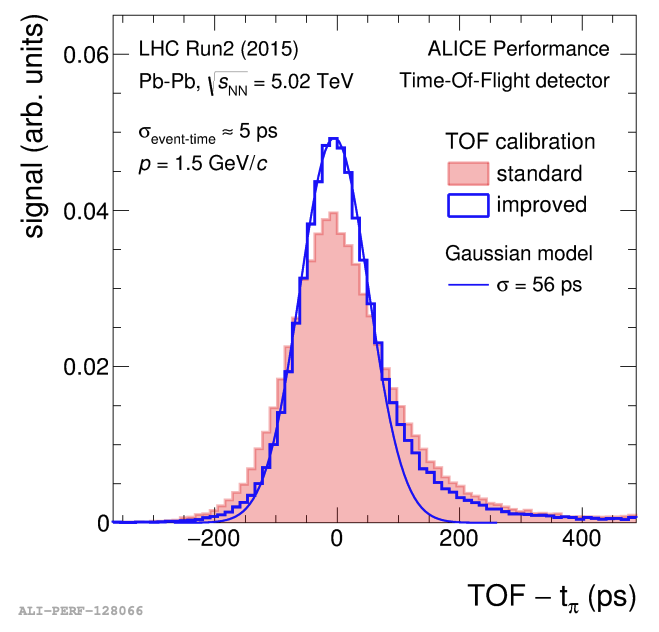}
\caption{TOF time resolution of 56 ps in Pb–Pb collisions at 5.02 TeV~\cite{Carnesecchi:2018oss}}
\label{fig:alice_timereso}
\end{figure}

The ALICE TOF underwent more than ten years in operation. The performance has been studied after Run 2, which showed no signs of degradation~\cite{Carnesecchi:2018oss}. The time resolution is 56 ps, as shown in Fig~\ref{fig:alice_timereso}. Currently, the TOF keeps serving the ALICE experiment.

\subsection{BESIII Time-Of-Flight detector}

\begin{figure}[!h]
\centering
\includegraphics[width=1.0\linewidth]{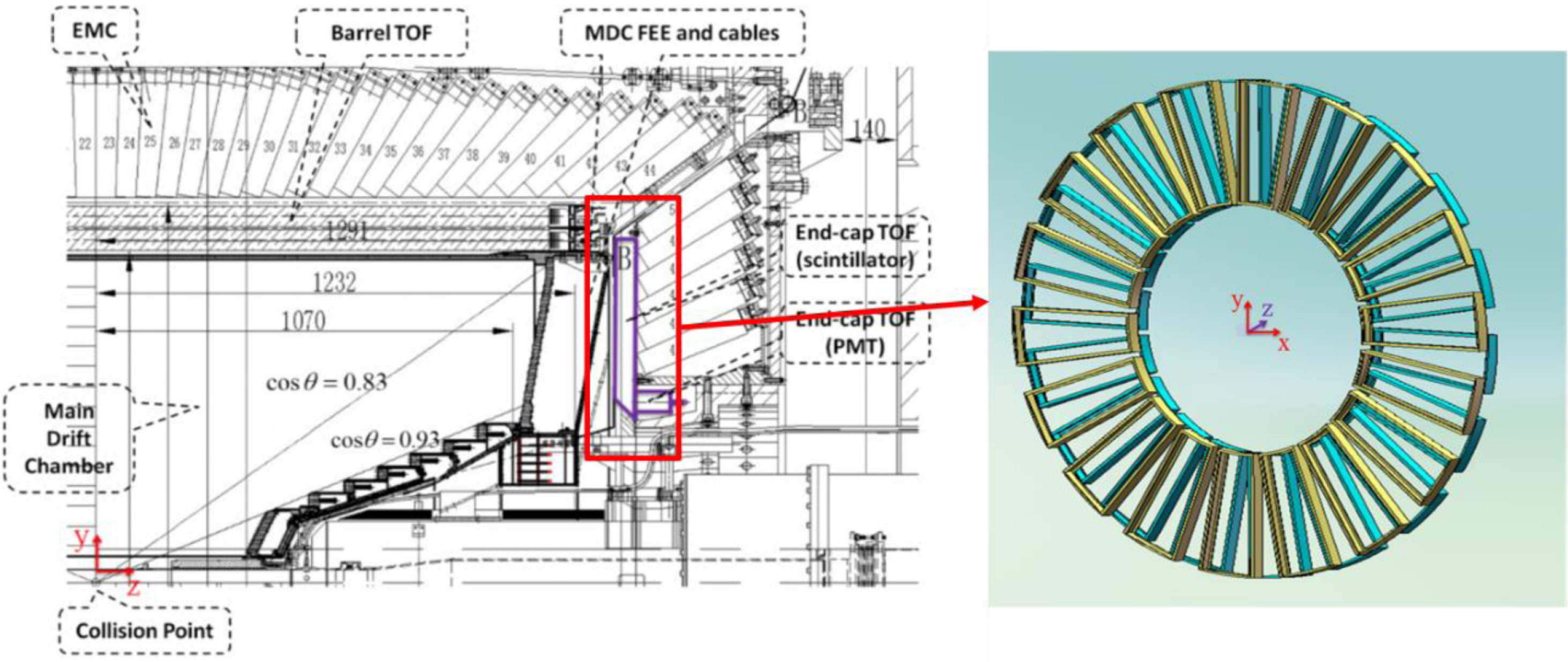}
\caption{Schematic drawing of the BESIII barrel and endcap TOF detectors~\cite{CAO2020163053}.}
\label{fig:bes3_layout}
\end{figure}

The TOF detector of the BESIII experiment is based on plastic scintillators read out by fine mesh photomultiplier tubes directly attached to the two end faces of the bars. It is installed out of  the main drift chamber (MDC) and inside of the CsI(Tl) electromagnetic calorimeter (EMC) as indicated in Fig.~\ref{fig:bes3_layout}. The TOF system includes a barrel sector covering $|cos(\theta)|<0.83$ and 2 endcap sectors in $0.85 < cos(\theta) < 0.95$. The barrel sector has 2 layers and each layer consists of 88 staggered scintillating bars (BC-408) that are 50 mm thick and 2300 mm long with a trapezoidal cross-section. The inner layer is 0.81 m away from the beam and the outer one is 0.86 m away. The total time resolution of the barrel sector is 70 ps with the dominant source of uncertainty coming from the intrinsic time resolution of the scintillators.

\begin{figure}[!h]
\centering
\includegraphics[width=1.0\linewidth]{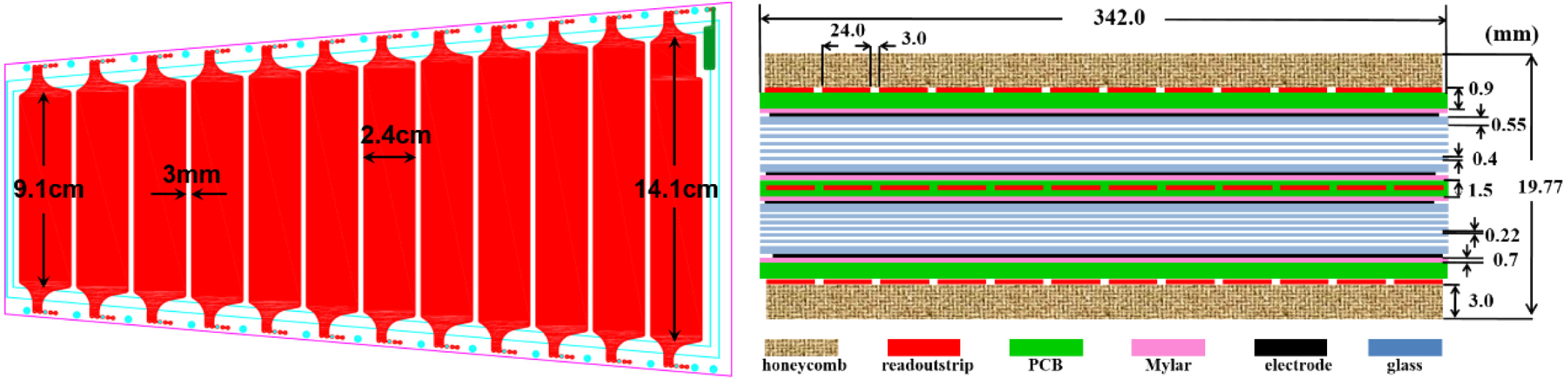}
\caption{The layout of the MRPC readout board (left), and the schematic drawing of the MRPC module cross-section (right) in BESIII.~\cite{CAO2020163053}.}
\label{fig:bes3_mrpc}
\end{figure}

The endcap sectors were originally built with 50 mm thick plastic scintillator pads (BC-404) and were later upgraded to MRPC~\cite{8824298, CAO2020163053} improving the time resolution from about 140 ps to 65 ps. In total, 2 $\times$ 36 MRPC modules are mounted to form 2 endcap sectors with an inner radius of 501 mm and an outer of 822 mm, installed out of the MDC about 1330 mm away from the collision point along the beam line. Each MRPC module contains 12 readout strips as shown in Fig.~\ref{fig:bes3_mrpc}, which increase the granularity by a factor of 12 compared to the old scintillator setup. Similar to ALICE, the double-stack (2 $\times$ 6) structure is adopted as shown in Fig.~\ref{fig:bes3_mrpc}.

%\subsection{Others detectors with timing capabilities in HEP}
%We may quickly mention:
%\begin{itemize}
%\item CMS Precision Proton Spectrometer ~\cite{RANTANEN2024169710}, timing with diamond sensors
%\item calorimeters with timing (CMS Phase 2 upgrade: ECAL, HGCAL)
%\item ....
%\end{itemize}

\subsection{The CMS PPS Timing Detector}

The CMS Precision Proton Spectrometer (PPS)~\cite{Albrow:1753795}, operating at the LHC, uses timing detectors based on planar single crystal CVD (Chemical Vapour Deposition) diamonds to measure the arrival time of protons scattered in the very forward region. The time information is used to reconstruct the longitudinal position of the proton interaction vertex and suppress pile-up background~\cite{BOSSINI2023167823}. Diamond is characterized by a large band gap (5.5~eV), very low leakage current, small dielectric constant and high carrier mobility. 
A detailed review of planar diamond sensors can be found in \cite{10.3389/fphy.2020.00248}.
The choice of diamond technology for the PPS was primarily motivated by its superior radiation hardness. PPS sensors have to sustain highly non-uniform irradiation, with a peak of about 5$\times$10$^{15}$ protons/cm$^2$ in the near beam region for an integrated
LHC luminosity of 100~fb$^{-1}$. The sensors consist of single-crystal CVD diamonds with a surface of 4.5$\times$4.5~mm$^2$ and a thickness of 500~$\mu$m. For the LHC Run 3 (2022–2025), each PPS timing station was equipped with four planes of sensors in double-diamond configuration~\cite{Bossini_2020, Berretti_2017}. A time resolution of approximately 60 ps has been achieved, with the target resolution of 30 ps expected to be reached through refined detector calibrations~\cite{RANTANEN2024169710}. 

\subsection{Timing in calorimetry}
The CMS upgrade for the High Luminosity phase of the LHC will introduce precision timing capabilities also in calorimeters to enhance event reconstruction and pile-up mitigation.

The central part of the CMS electromagnetic calorimeter is made of scintillating PbWO$_4$ crystals read out by avalanche photo-diodes (APDs). While the current ECAL timing performance is limited to approximately 150~ps~\cite{Hayrapetyan_2024}, the upgrade of the electronics for the High Luminosity phase of the LHC will enable precision timing measurements with 30~ps resolution for electrons and photons above 30~GeV. In particular, the new front-end electronics features two new radiation hard ASICs: a custom dual gain trans-impedance amplifier implemented in a 130~nm CMOS technology (CATIA, Calorimeter Trans-Impedance Amplifier) and a 12-bit ADC with 160~MHz sampling frequency~\cite{CERN-LHCC-2017-011}. The TIA effectively preserves the fast pulse shape of signals and is more resilient to noise caused by radiation-induced leakage current in the APDs.

In the forward region of the upgraded CMS detector, a new High Granularity Calorimeter (HGCal) will be installed to replace the current calorimeters~\cite{CERN-LHCC-2017-023}. The HGCAL is a high-precision sampling calorimeter composed of silicon and scintillator modules, arranged across 47 active layers with over 6 million readout channels. In the electromagnetic section, the silicon modules are interleaved with copper, copper–tungsten, and lead absorbers. The hadronic section, equipped with steel absorbers, uses silicon sensors in high-radiation areas where the total fluence is expected to be above $5\times$10$^{13}$ \neut, while plastic scintillator tiles are used in regions of lower radiation. The HGCAL aims to achieve a timing resolution of better than 30~ps for clusters resulting from particles with transverse momentum greater than 5~GeV. Test beam measurements of prototype elements have indicated a time resolution of 60~ps for single-channel measurements and better than 20~ps for full showers at the highest energies~\cite{Acar_2024}. 

%{\bf fixme: add a comment about 5D calorimetry in future experiments: precision timing in calorimetry provide depth segmentation in fiber calorimeters and improved jet energy resolution with 5D Particle Flow}
%\input{SystemAspects}
\section{Novel technologies and outlook}
\label{sec:noveltechnologies}

\subsection{Trends in fast scintillators}

Engineering the scintillator is an important direction for developing fast sensor materials~\cite{Lecoq,6303850}. Doping and co-doping can potentially help to improve the scintillation decay time~\cite{https://doi.org/10.1002/adom.201400571,doi:10.1021/cg501005s}. Heavy doping of PWO with La and Y at concentrations an order of magnitude higher than usual can reduce the fast component decay time to approximately 640 ps, opening an opportunity for the development of ultra-fast scintillators based on the family of tungstate crystals~\cite{KORZHIK2022166781}. 
Another promising approach involves magnesium (Mg) co-doping in GAGG:Ce crystals. In Refs~\cite{KAMADA201563,PhysRevApplied.2.044009,LUCCHINI2016176}, Mg co-doped GAGG:Ce crystals have been compared to standard GAGG:Ce, the former showing a decrease in light yield but a faster scintillation decay. This is explained by the fact that Mg$^{2+}$ tends to enrich the Ce$^{4+}$ centers, which can compete with electron traps from the conduction band with the shallow electron traps, providing fast radiative recombination. More studies are ongoing to optimize the doping concentration for fast scintillation in GAGG~\cite{D2MA00626J}, which is being evaluated~\cite{MARTINAZZOLI2021165231,AN2023167629} as a candidate toward a time resolution of 15 ps for fibres in spaghetti calorimeters (SPACAL)~\cite{Pauwels_2013} in the LHCb upgrade program of electromagnetic calorimeter~\cite{9006906}, discussed more in Sec.~\ref{sec:hl-lhc-more}.

Cherenkov light is usually faster than the scintillation light and can be explored for ultra-fast time measurements. It requires the crystal to have a high index of refraction $n$ given the Cherenkov threshold of $1/(\beta n)$. Various materials have been studied for this purpose, including PWO, LSO:Ce, LuAG:Ce, LuAP:Ce, Bismuth germanate BGO (or mixed with Bismuth silicate BSO), PbF$_2$, TlBr, TlCl and so on~\cite{Lecoq,6303850,6651684,AKCHURIN2008359,CALA2022166527,9222347,Kratochwil_2021,10.3389/fphy.2022.785627}. Accordingly, there are developments of photo-detectors such as SiPM enhanced for UV range~\cite{s19020308}. Efforts towards using directly the protection layer of the SiPMs as medium for Cherenkov light radiation are also ongoing, with promising time resolution results of about 20~ps~\cite{Carnesecchi2023_SiPMs}, discussed more in Sec.~\ref{sec:hl-lhc-more}.

%Furthermore, it is worth noting that an effort on the TOF timing layer of ALICE 3~\cite{arXiv:2211.02491} utilizes directly the protection layer of the SiPM as the medium for Cherenkov light and realizes a time resolution below 20 ps~\cite{Carnesecchi2023_SiPMs}.

\begin{figure}[h]
\centering
\includegraphics[width=0.6\linewidth]{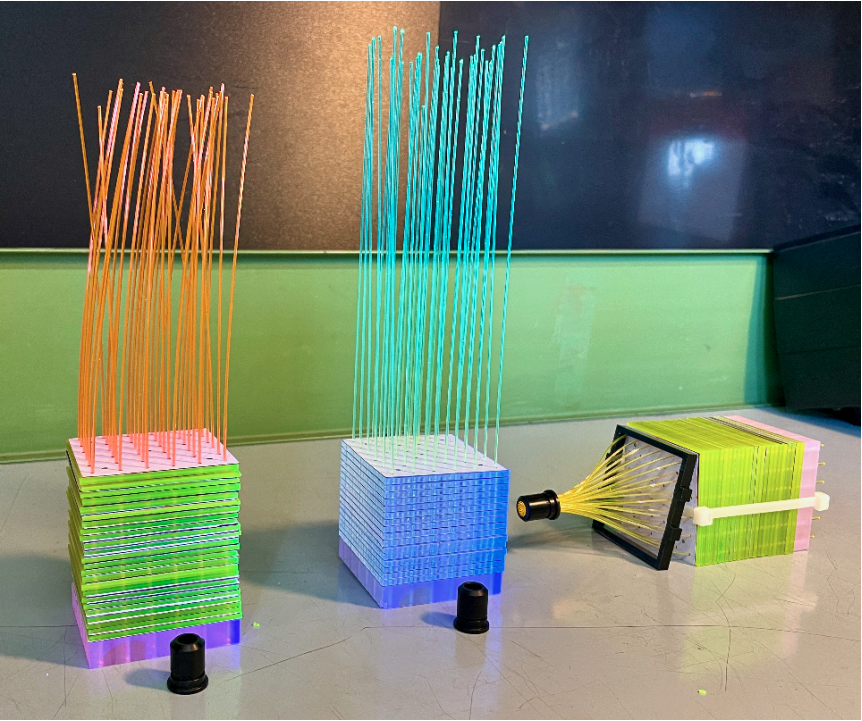}
\caption{Shashlik modules of NanoCal~\cite{refId0}.}
\label{fig:nanocrystal}
\end{figure}

New materials beyond crystals and plastics are also under development for fast timing applications. Scintillating glasses re-gained interest in recent years with new developments that brought better radiation hardness~\cite{NGUYEN2021164898,DORMENEV2021165762}. Studies in Ref.~\cite{LUCCHINI2023168214} show that cerium-doped Alkali Free Fluorophosphate scintillating glasses can reach about 10 ps for a MIP. Given their lower cost compared to crystals, glasses are now considered as potential candidates for the instrumentation of large-volume homogeneous hadron calorimeters, potentially spanning tens of cubic meters~\cite{6154597,Mao_2012}. Based on quantum confinement, quantum dots or nano-crystals embedded in meta-structured materials could bring the time for light emission down to sub-nanosecond~\cite{C1CC14687D,Grim2014,Turtos_2016,https://doi.org/10.1002/pssr.201600288,D2TC02060B,10.3389/fphy.2022.1021787}. First attempts to make calorimeter from nano-crystals kicked off. The AIDAinnova blue-sky project “NanoCal” laid the ground for a new generation of fine-sampling calorimeters using perovskite (e.g., caesium lead bromide - CsPbBr3) nano-crystals dissolved in polymethyl methacrylate (PMMA)~\cite{refId0}, as shown in Fig.~\ref{fig:nanocrystal}.

Photonic crystals represent another active field of research~\cite{cryst8020078}. Conventional crystal scintillators are generally characterized by a high refractive index for a shorter decay time and a higher stopping power. However, the mismatch in refractive index between the crystal and the coupling medium (air or optical grease) results in a reduced light collection efficiency ($LCE$). Photonic crystals can improve the time resolution by recovering $LCE$ at the exit surface through more efficient light extraction.

\subsection{Trends in gaseous detectors}

As the randomized location of the primary ionization along the trajectory of the incident particle dominates the time jitter, the reduction of gas gap thickness practically improves the time resolution. Refs.~\cite{AN200839,LIU2019396} report efforts in this direction, achieving a time resolution of 20 ps. The detector of Ref.~\cite{AN200839} is designed with 24 gas gaps with a width of 160~$\mu$m, for a total thickness of 22 mm between the two external pickup plates. Technologies to manufacture even narrower gaps exist, such as down to 50~$\mu$m or 25~$\mu$m as used in microbulk detectors~\cite{Gonzalez-Diaz_2017}, can be tried for MRPC to explore the frontier of time resolution, as pointed out in Ref.~\cite{microbulk}.

\begin{figure}[h]
\centering
\includegraphics[width=1.0\linewidth]{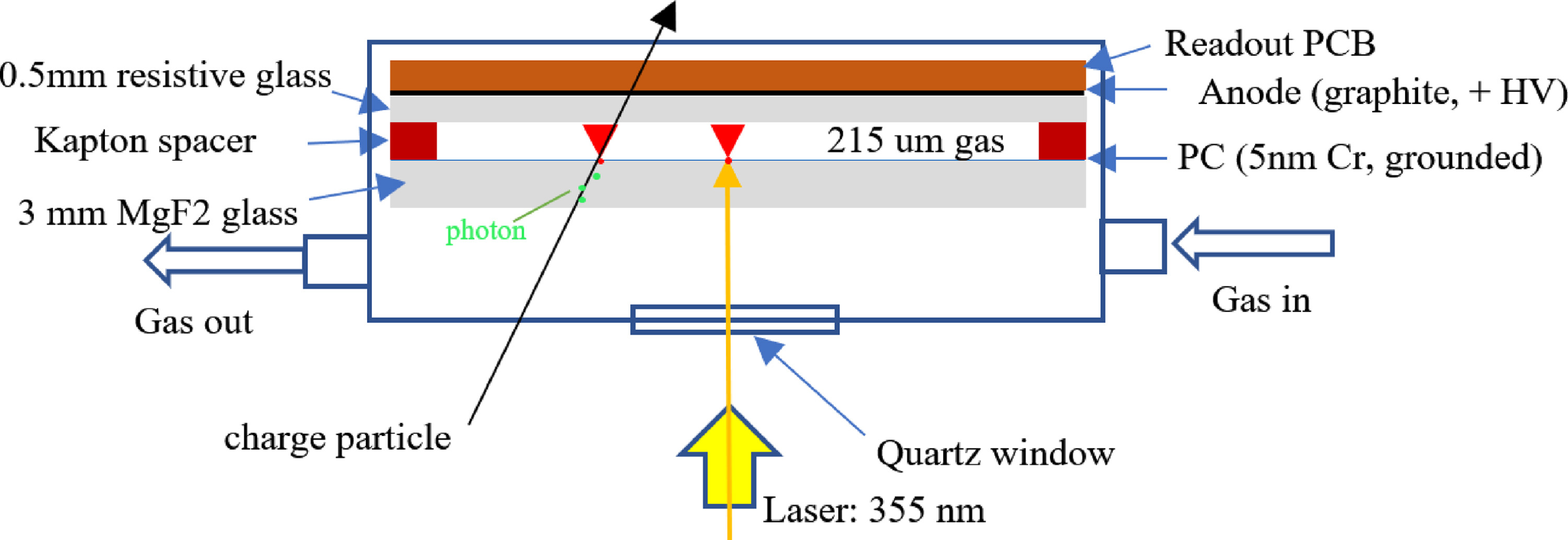}
\caption{Layout of a photoelectric detector with RPC~\cite{ZHAO2025170593}.}
\label{fig:rpc-cherenkov}
\end{figure}

An alternative approach to localizing the primary ionization involves using Cherenkov light produced in a medium and converting the emitted UV photons into electrons via a photocathode positioned at a fixed location within the gas gap. These photoelectrons then serve as the primary ionization source.
Such hybrid designs have been studied using parallel-plate avalanche chambers back in 1990s~\cite{CHARPAK199163}, and more recently adapted for  RPCs~\cite{FONTE200530,CARLSON2003189,FRANCKE2004163,MATSUOKA2023168378,ZHAO2025170593}, eventually pushing the time resolution down to 25 ps. Fig.~\ref{fig:rpc-cherenkov} shows the layout of a photoelectric detector with RPC designed in Ref.~\cite{ZHAO2025170593}. The photocathode is mounted at the entrance of the gas gap ensuring that all photo-electrons are produced in the same location. Relevant studies are ongoing towards practical applications.

\begin{figure}[h]
\centering
\includegraphics[width=1.0\linewidth]{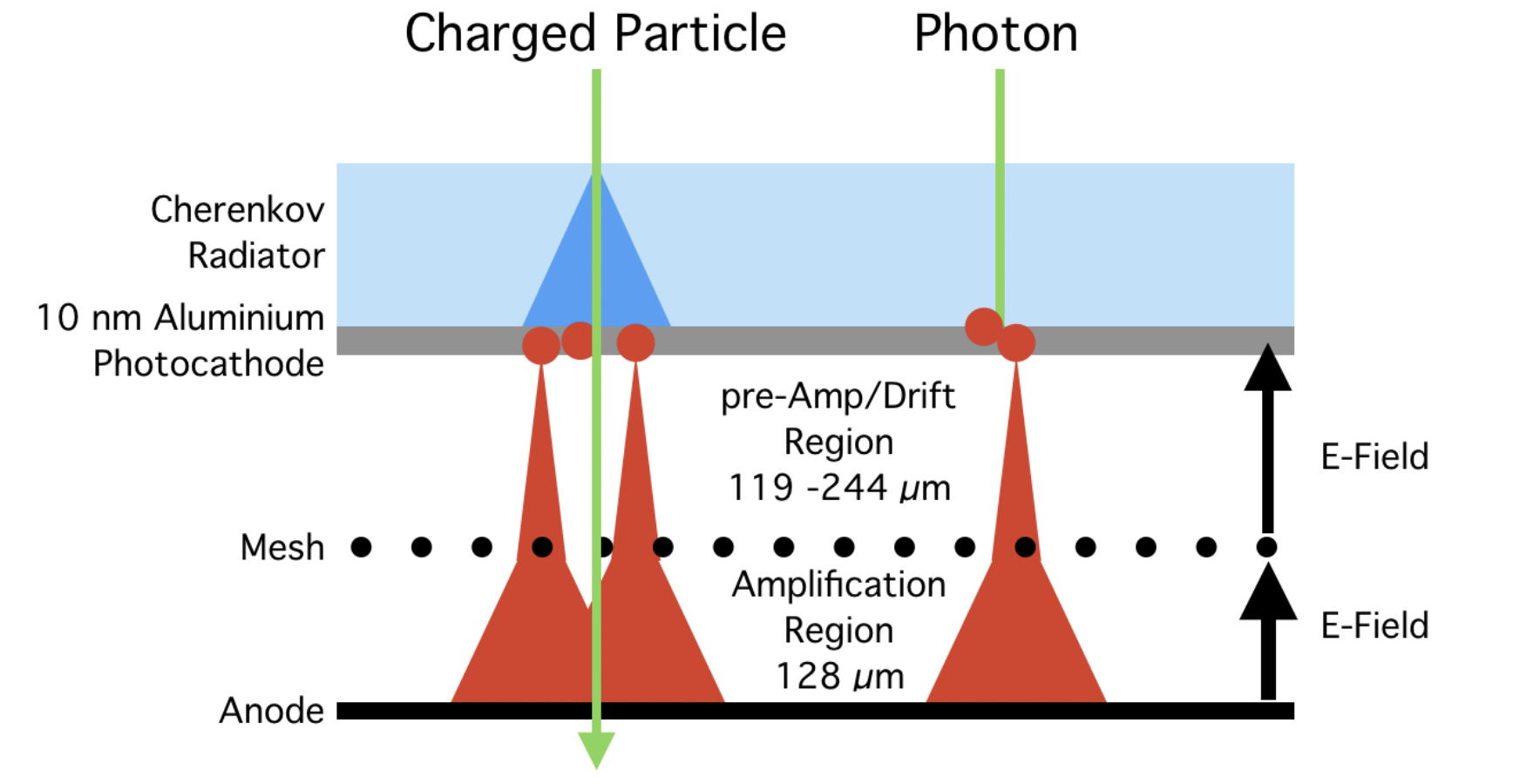}
\caption{Layout of PICOSEC-Micromegas~\cite{Sohl_2020}.}
\label{fig:picosec}
\end{figure}

In the field of Micromegas (Micro-Mesh Gaseous Structure) detectors, a similar idea of using Cherenkov light to localize the ionization is also under development. The design of the PICOSEC-Micromegas detector~\cite{BORTFELDT2018317,Sohl_2020} is shown in Fig.~\ref{fig:picosec}. With the fixed location of primary electrons, PICOSEC-Micromegas can also reach a time resolution of 25 ps.

\subsection{Trends in solid state detectors}
The evolution of technologies is mainly driven by the requirements for detectors in the next generations of high-energy physics experiments to cope with increasing luminosity and consequent event pile-up. 
4D tracking, i.e. the reconstruction of the trajectory of a charged particle in three spatial dimensions plus time as a fourth dimension, is one of the directions to deal with future collider experiment challenges. It requires sensors capable of providing precision measurements of both position and time with 10~$\mu$m and 10~ps resolution, respectively, along with increased radiation hardness and low power consumption. 
Several R\&D efforts are currently ongoing on several technologies. A comprehensive summary of the key research directions and challenges in 4D tracking is given in \cite{Detector:2784893}.

\begin{figure}[!h]
\centering
\includegraphics[width=0.8\linewidth]{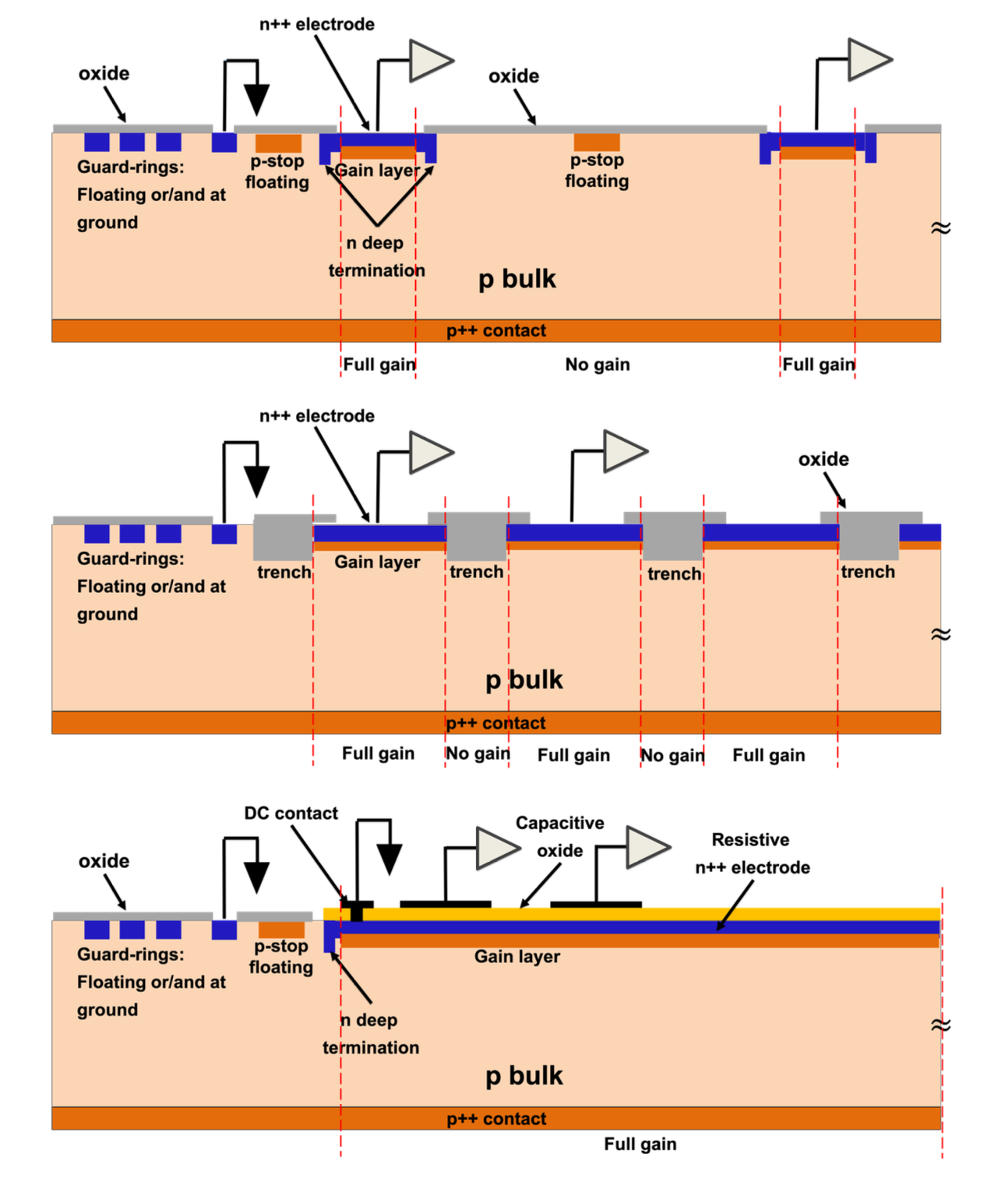}
\caption{Cross section, not to scale, of a standard LGAD segmentation based on JTE and p-stop (top), trench Isolated LGAD (middle) and resistive AC-coupled Silicon Detector (bottom). Vertical dashed red lines indicate a sharp separation between full-gain and no-gain region~\cite{Ferrero_2020}.}
\label{fig:newTech_lgads}
\end{figure}

\paragraph{Evolution of LGADs}
Structural variants of the LGAD design aim at reducing the no-gain region and hence increasing the sensor fill factor.
One of the drawbacks of standard LGADs is the presence of isolation structures (JTE, Junction Termination Extension) between adjacent pads needed to isolate the pixels and avoid premature breakdown, introducing an inter-pixel region 60–80~$\mu$m wide in which the gain is suppressed. 
One advanced design to overcome this limitation while maintaining high temporal resolution is the AC-LGAD (also referred to as resistive silicon detectors): at variance with standard LGADs, AC-LGADs feature a continuous gain layer, a resistive n+ implant, and a thin dielectric layer for AC-coupled read-out~\cite{Ferrero_2020}, as shown in Fig.~\ref{fig:newTech_lgads}.
Other novel technologies include trench-isolated LGADs (TI-LGAD)~\cite{9081916}, where segmentation with high fill-factor is achieved by substituting the structures used for the termination of the gain layer and channel isolation with narrow trenches (1~$\mu$m) filled with dielectric material, and deep-junction LGADs, which use a p-n junction buried several microns below the surface of the device~\cite{Zhao_2022}. 

\begin{figure}[!h]
\centering
\includegraphics[width=1.0\linewidth]{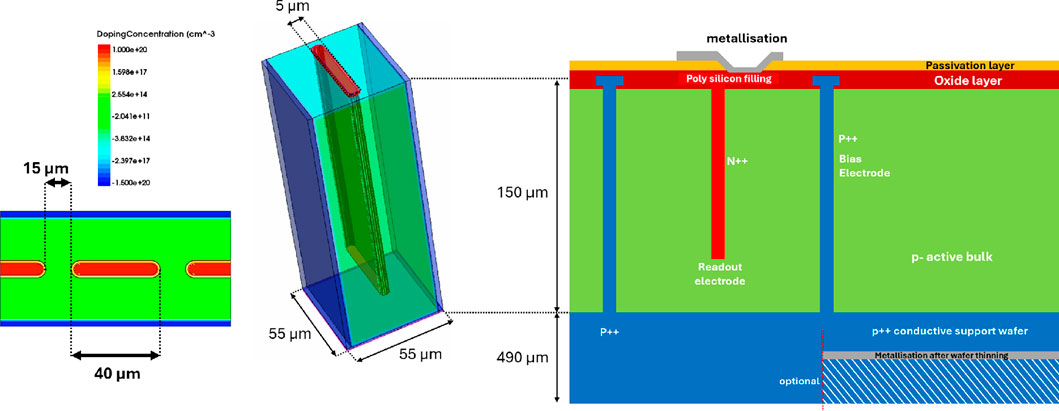}
\caption{ Right: structure and technological profile of the 3D trench sensor developed within the TimeSPOT project~\cite{10.3389/fphy.2024.1497267}.}
\label{fig:newTech_3Dsilicon}
\end{figure}

\paragraph{3D sensors}
3D silicon sensors are attractive technology featuring fast signals and robustness to bulk damages thanks to the short inter-electrode distance. Columnar 3D sensors have demonstrated a time resolution comparable to LGADs~\cite{KRAMBERGER201926}. The main limitation to the timing performance of columnar 3D sensors is due to electric field and weighting field spatial non-uniformities. These effects can be mitigated by employing trench-like geometries (Fig.~\ref{fig:newTech_3Dsilicon}): a time resolution of O(10)~ps has been reached in prototypes developed by the TimeSPOT project even after exposure to extreme radiation fluences 
up to 1$\times$10$^{17}$ \neut ~\cite{Lampis_2023,10.3389/fphy.2024.1497267}. Despite these remarkable results, the fabrication technology of 3D silicon sensors with timing capabilities requires still further developments for applications in large-scale HEP experiment. Novel technologies like the 3D diamond sensors have also been studied within the TimeSPOT initiative~\cite{instruments5040039}.

\paragraph{Monolithic sensors}
Monolithic sensors with embedded readout represent an appealing solution thanks to the advantage of a lower material budget, easier assembly procedure and lower production costs. Examples of ongoing developments with architectures optimized for precision timing include both monolithic sensors without internal gain (e.g. MiniCactus~\cite{Gan_2025} in CMOS technology, monolithic pixel sensors is SiGe BiCMOS technology~\cite{Zambito_2023}, MOST~\cite{selina2025}) and monolithic sensors with internal gain developed for example within the MONOLITH project~\cite{9620045} and ARCADIA project~\cite{9075426, Follo_2024}.

\subsection{Precision timing at the HL-LHC beyond Run4}\label{sec:hl-lhc-more}
%Precision timing is an emerging capability for the next generation of detectors at all future colliders. 
In addition to the CMS and ATLAS timing detectors which are set to be operational at the beginning of the HL-LHC in 2030, the other two LHC experiments LHCb and ALICE have proposed upgrades for the LHC Run 5 (currently scheduled to start in 2036, after the Long Shutdown 4) to add timing capabilities to the detectors at the tens of picoseconds level to meet challenges due to increased peak luminosity and pileup.  

For the LHCb Upgrade II~\cite{CERN-LHCC-2021-012}~\cite{LHCbCollaboration:2920835} the new VErtex LOcator (VELO), which is the tracking detector surrounding the interaction region, will feature pixels with time resolution of 50~ps or better, leading to 20~ps time resolution per track. This upgrade will enable full 4D tracking and allow for maintaining high vertex reconstruction efficiency in high pileup scenarios. Among the different technologies under investigation, the most promising one is represented by 3D silicon sensors, which have excellent radiation hardness potential and have been successfully used (without timing) in LHC experiments.
The key aspect of the LHCb physics program is its excellent performance in particle identification. 
An R$\&$D program is underway to enhance the Ring Imaging Cherenkov detectors with
sensors and readout electronics designed for time resolutions better than 100~ps. The sensor technologies under consideration are SiPMs and MCPs. A new Time Of internally Reflected CHerenkov light-based Time of Flight detector (TORCH) with quartz planes read by MCP-PMTs will provide charged hadron identification over a 2–20 GeV/c momentum range, on a 9.5~m flight distance from the LHC interaction point. To achieve this level of performance, a 15~ps timing resolution per track is required. Prototypes have already shown performances close to the target of 70~ps per photon~\cite{HARNEW2023167991}.
The LHCb Upgrade II detector will comprise also substantial modifications of the current LHCb electromagnetic calorimeter to cope with high radiation doses in the central region and increased particle densities, with advanced timing capabilities for pileup mitigation. The PicoCal will feature SpaCal and Shashlik modules with fine granularity and 20~ps time resolution. In particular, the innermost part close to the beam axis, where radiation doses up to 1 MGy and 6$\times$10$^{15}$ \neut are expected, will consist of SpaCal modules made from tungsten absorbers with radiation-hard scintillating crystal fibres (e.g. garnet crystals), while the intermediate region will have modules with lead absorbers and radiation-tolerant organic scintillators. An extensive R$\&$D campaign on absorber and radiation hard and fast scintillator materials is ongoing, and time resolutions of better than 20 ps at high energy were observed in test beam measurements of SpaCal and Shashlik prototype modules~\cite{KHOLODENKO2024169656}.

ALICE3~\cite{arXiv:2211.02491} is a compact, next-generation multipurpose detector at the LHC proposed for physics data-taking in the LHC Run 5 and beyond as a follow-up to the present ALICE experiment. 
The Time-of-Flight system is being redesigned to meet the demands of precision particle identification in the HL-LHC environment while maintaining full acceptance up to $|\eta| = 4$.
The TOF system will consist of an inner layer, an outer barrel, and two forward disks, for a total surface area of around 45~m$^2$. The system aims at $e/\pi$ separation up to $\sim$~500~MeV/c, $\pi/k$ separation up to 2 GeV and $k/p$ separation up to 4~GeV/c. Achieving this requires timing resolution below 20 ps while keeping a low material budget (1–3\% of a radiation length). Three sensor technologies are being evaluated: thin LGADs are suitable candidates, having demonstrated $\sim$ 30~ps resolution with 50~$\mu$m thickness and can potentially achieve even better time resolution with reduced thicknesses~\cite{Carnesecchi2023}; monolithic CMOS sensors are a cost-effective solution with integrated readout and low material, but their current time resolution is insufficient and R\&D is ongoing aiming at the integration of a gain layer to achieve the required time resolution of at least 20~ps~\cite{Follo_2024}; SiPM sensors 
for the detection of Cherenkov light generated in the protection layer by the passage of charged particles have shown promising time resolution of about 20~ps in test beam measurements for a sufficiently large number of fired SPADs~\cite{Carnesecchi2023_SiPMs}.

\subsection{Precision timing applications at next generation colliders}

At future colliders beyond the HL-LHC, precision timing applications include 4D tracking for pileup mitigation and background rejection, TOF-based particle identification, and heavy-flavour tagging.

\paragraph{FCC-ee}
FCC-ee is a proposed high-luminosity, high-energy electron-positron collider, part of the broader Future Circular Collider (FCC) program. Aimed to start operation in the late 2040s, its goal is to provide unprecedented precision measurements in the electroweak sector and search for indirect probes of new phenomena beyond the standard model.
Due to the cleaner environment of $e^+e^-$ colliders compared to hadron colliders, the applications of precision timing at FCC-ee include TOF-based PID, flavour-tagging and long-lived particles searches. Some of the detector concepts under study integrate timing layers around the main tracker, like a large radius ($\sim$2~m) timing layer in front of the calorimeter to provide time-of-flight measurements with about 100~ps resolution, complementing other PID sub-detectors (dE/dx or Cherenkov based) in the low momentum ($\sim$ 1 ~GeV) region~\cite{Abbrescia:2926782}. Moreover, TOF offers redundant $\pi$/k separation up to 5~GeV, vertex timing capabilities and will be important for long-lived particle searches.

\paragraph{FCC-hh}
The second stage of the FCC project envisions a highest-energy proton collider, FCC-hh, with a centre-of-mass energy of 100~TeV. The unprecedented pileup conditions, with O(1000) pileup events per beam crossing, represent a clear case for the use of 4D technology in tracking layers. To achieve a similar environment as HL-LHC, an O(5)~ps resolution per track is needed. Significant detector R\&D is required to develop sensors capable of operating at radiation doses above 10$^{17}$ \neut.

\paragraph{Muon collider}
A Muon Collider is another option under study for a next-generation collider facility post-High Luminosity LHC~\cite{Long2021}. On the detector side, the main challenge is the large beam-induced background (BIB) from muon decays and showering of electrons on the shielding. 4D tracking is critical for separating BIB hits from those deposited by the collision products, although the time resolution requirements in this case are less stringent (30-60~ps) compared to those of, for example, FCC-hh.

\section{Summary}
\label{sec:summary}

Precision timing is highly demanded in modern collider experiments for time-of-flight measurements used in particle identification, time tagging minimum ionizing particles to mitigate the severe pileup condition in future colliders, and searches for new particles that have exceptionally long lifetime and beyond. Advances in detector technologies now allow for time resolutions down to tens of picoseconds, pushing the frontier of precision measurements of the Standard Model and explorations for new physics at the colliders.

%Over the past decades, a wide range of fast timing technologies, ranging from gaseous to solid sensor materials, have been developed. As adopted in modern large-scale particle experiments, such as the gaseous based timing detector MPRC in BESIII and ALICE, the timing detector based on cystals coupled to SiPM in CMS MTD barrel, the timging detector based on LGAD in CMS MTD endcap and ATLAS, the timing technologies are proven to perform in execellent timing resolution that is down to 30 ps under various hash environments at colliders and push physics programs to the energy and luminosity frontier with outstanding performance on TOF and pileup mitigation.

Over the past decades, a wide range of fast timing technologies, spanning from gaseous to solid sensor materials, have been developed and are now implemented in large-scale particle physics experiments. Applications include the gas-based MRPC timing detectors used in ALICE and BESIII, crystal scintillators coupled to SiPMs in CMS MTD barrel timing layer, and LGAD-based timing detectors in the CMS MTD endcap timing layer and ATLAS HGTD. These technologies can provide time resolutions as low as 30 ps under harsh experimental conditions. Their performance in TOF applications and pileup mitigation plays a key role in enabling precision physics at the energy and luminosity frontiers.

Novel technologies in scintillator engineering, exploration of Cherenkov light, quantum confinement, hybrid gaseous chambers, resistive silicon sensors, columnar 3D sensors and monolithic sensors keep emerging. R\&D efforts targeting the sub-20 ps resolution frontier are actively ongoing, driven by precision timing applications at next-generation collider experiments.

%New technologies keep evolving for reaching new limits of timing precision. TODO

\section*{Acknowledgments}
%The authors are grateful to 
This review has been inspired by lectures and seminars given by Dr. N. Cartiglia, Prof. T. Tabarelli de Fatis and Prof. C. Tully.
The effort has been supported by funding from
%Big Data and Quantum Computing funded by the NextGenerationEU program (Italy),
MoST (China) under the National Key R\&D Program of China (No. 2022YFA1602100) and the NSFC (China) under global scientific research funding projects (No. W2443006).

\clearpage
\newpage

%\begin{thebibliography}{00}

%% For numbered reference style

%\bibitem{lamport94}
%  \textit{\LaTeX: a document preparation system},
%  Addison Wesley, Massachusetts,

%\bibitem{Bethe}
%H. Bethe and W. Heitler, \textit{On the stopping of fast particles and on the creation of positive electrons}, Proc. Royal Soc. London, 146(856):83, 1934.

%\bibitem{Lecoq}
%P. Lecoq et al,  \textit{Factors Influencing Time Resolution of Scintillators and Ways to Improve Them}, IEEE Trans. Nucl. Sci. 2010, 57, 2411–2416.

%\end{thebibliography}

\bibliographystyle{unsrt}
\bibliography{mybibfile}

@article{Ash1974,
author={Ash, W. and others},
title={{Experimental study of the new 3.1 GeV particle by e+e- collision at ADONE}},
journal={Lettere al Nuovo Cimento (1971-1985)},
year={1974},
month={Dec},
day={01},
volume={11},
number={17},
pages={705-710},
issn={1827-613X},
doi={10.1007/BF02762934},
}

@article{AFANASIEV1999210,
title = {The {NA49} large acceptance hadron detector},
journal = {Nuclear Instruments and Methods in Physics Research Section A: Accelerators, Spectrometers, Detectors and Associated Equipment},
volume = {430},
number = {2},
pages = {210-244},
year = {1999},
issn = {0168-9002},
doi = {https://doi.org/10.1016/S0168-9002(99)00239-9},
author = {S. Afanasiev and others}
}

@Article{Ambrosinietal.1999,
author={Ambrosini, G. and others},
title={Measurement of charged particle production from 450 GeV/c protons on beryllium},
journal={The European Physical Journal C - Particles and Fields},
year={1999},
month={Oct},
day={01},
volume={10},
number={4},
pages={605-627},
issn={1434-6052},
doi={10.1007/s100529900145},
}

@article{CABRERA2002416,
title = {The {CDF-II} time-of-flight detector},
journal = {Nuclear Instruments and Methods in Physics Research Section A: Accelerators, Spectrometers, Detectors and Associated Equipment},
volume = {494},
number = {1},
pages = {416-423},
year = {2002},
note = {Proceedings of the 8th International Conference on Instrumentatio n for Colliding Beam Physics},
issn = {0168-9002},
doi = {https://doi.org/10.1016/S0168-9002(02)01512-7},
author = {S Cabrera and others}
}

@article{CARLEN1999123,
title = {A large-acceptance spectrometer for tracking in a high-multiplicity environment, based on space point measurements and high-resolution time-of-flight},
journal = {Nuclear Instruments and Methods in Physics Research Section A: Accelerators, Spectrometers, Detectors and Associated Equipment},
volume = {431},
number = {1},
pages = {123-133},
year = {1999},
issn = {0168-9002},
doi = {https://doi.org/10.1016/S0168-9002(99)00261-2},
author = {L. Carlén and others}
}

@article{LLOPE2004252,
title = {The {TOFp/pVPD} time-of-flight system for {STAR}},
journal = {Nuclear Instruments and Methods in Physics Research Section A: Accelerators, Spectrometers, Detectors and Associated Equipment},
volume = {522},
number = {3},
pages = {252-273},
year = {2004},
issn = {0168-9002},
doi = {https://doi.org/10.1016/j.nima.2003.11.414},
author = {W.J. Llope and others},
}

@article{Apollinari:2120673,
      author        = "Apollinari, G. and Brüning, O. and Nakamoto, T. and
                       Rossi, Lucio",
      title         = "{Chapter 1: High Luminosity Large Hadron Collider HL-LHC.
                       High Luminosity Large Hadron Collider HL-LHC}",
      archivePrefix = "arXiv",
      eprint        = "1705.08830",
      reportNumber  = "FERMILAB-PUB-15-699-TD",
      journal       = "CERN Yellow Report",
      number        = "5",
      pages         = "1-19",
      year          = "2015",
      note          = "21 pages, chapter in High-Luminosity Large Hadron Collider
                       (HL-LHC) : Preliminary Design Report",
      doi           = "10.5170/CERN-2015-005.1",
}

@Article{FCC-ee,
author={Abada, A. and others},
title={{FCC-ee: The Lepton Collider}},
journal={The European Physical Journal Special Topics},
year={2019},
month={Jun},
day={01},
volume={228},
number={2},
pages={261-623},
issn={1951-6401},
doi={10.1140/epjst/e2019-900045-4},
}

@Article{FCC-hh,
author={Abada, A. and others},
title={{FCC-hh: The Hadron Collider}},
journal={The European Physical Journal Special Topics},
year={2019},
month={Jul},
day={01},
volume={228},
number={4},
pages={755-1107},
issn={1951-6401},
doi={10.1140/epjst/e2019-900087-0},
}

@unpublished{Collaboration:2231915,
      author        = "CMS Collaboration and Mc Cauley, Thomas",
      title         = "{Collisions recorded by the CMS detector on 14 Oct 2016
                       during the high pile-up fill}",
      year          = "2016",
      note          = "CMS Collection.",
}

@article{Bethe:1934za,
    author = "Bethe, H. and Heitler, W.",
    title = "{On the Stopping of fast particles and on the creation of positive electrons}",
    doi = "10.1098/rspa.1934.0140",
    journal = "Proc. Roy. Soc. Lond. A",
    volume = "146",
    pages = "83--112",
    year = "1934"
}

@article{GEDCKE1967377,
title = {A constant fraction of pulse height trigger for optimum time resolution},
journal = {Nuclear Instruments and Methods},
volume = {55},
pages = {377-380},
year = {1967},
issn = {0029-554X},
doi = {https://doi.org/10.1016/0029-554X(67)90145-0},
author = {D.A. Gedcke and W.J. McDonald}
}

@article{GEDCKE1968253,
title = {Design of the constant fraction of pulse height trigger for optimum time resolution},
journal = {Nuclear Instruments and Methods},
volume = {58},
number = {2},
pages = {253-260},
year = {1968},
issn = {0029-554X},
doi = {https://doi.org/10.1016/0029-554X(68)90473-4},
author = {D.A. Gedcke and W.J. McDonald}
}

@article{SBaron_2012,
doi = {10.1088/1748-0221/7/12/C12023},
year = {2012},
month = {dec},
publisher = {},
volume = {7},
number = {12},
pages = {C12023},
author = {S Baron and T Mastoridis and J Troska and P Baudrenghien},
title = {Jitter impact on clock distribution in LHC experiments},
journal = {Journal of Instrumentation},
}

@unpublished{Rivetti,
title= {Fast timing techniques},
author = {A. Rivetti},
year = {2016},
note= {{10th International Meeting on Front-End Electronics (FEE 2016)}},
}

@article{Lecoq,
    author  = "Paul Lecoq and others",
    title   = "Factors Influencing Time Resolution of Scintillators and Ways to Improve Them",
    year    = "2010",
    journal = "IEEE Transactions on Nuclear Science",
    volume  = "57",
    number  = "5",
    pages   = ""
}

@article{pdg:2024cfk,
    author = "Navas, S. and others",
    collaboration = "Particle Data Group",
    title = "{Review of particle physics}",
    doi = "10.1103/PhysRevD.110.030001",
    journal = "Phys. Rev. D",
    volume = "110",
    number = "3",
    pages = "030001",
    year = "2024"
}

@book{Rossi:99081,
      author        = "Rossi, Bruno Benedetto",
      title         = "{High-energy particles}",
      publisher     = "Prentice-Hall",
      address       = "New York, NY",
      series        = "Prentice-Hall physics series",
      year          = "1952",
}

@article{Fano:1963xu,
    author = "Fano, U.",
    title = "{Penetration of protons, alpha particles, and mesons}",
    doi = "10.1146/annurev.ns.13.120163.000245",
    journal = "Ann. Rev. Nucl. Part. Sci.",
    volume = "13",
    pages = "1--66",
    year = "1963"
}

@article{BIALKOWSKI1974221,
title = {Further study of timing properties of scintillation counters},
journal = {Nuclear Instruments and Methods},
volume = {117},
number = {1},
pages = {221-226},
year = {1974},
issn = {0029-554X},
doi = {https://doi.org/10.1016/0029-554X(74)90400-5},
author = {J. BiaŁkowski and Z. Moroz and M. Moszyński}
}

@article{BENGTSON1974227,
title = {Energy-transfer and light-collection characteristics for different types of plastic scintillators},
journal = {Nuclear Instruments and Methods},
volume = {117},
number = {1},
pages = {227-232},
year = {1974},
issn = {0029-554X},
doi = {https://doi.org/10.1016/0029-554X(74)90401-7},
author = {B. Bengtson and M. Moszyński},

}

@article{MOSZYNSKI19791,
title = {Status of timing with plastic scintillation detectors},
journal = {Nuclear Instruments and Methods},
volume = {158},
pages = {1-31},
year = {1979},
issn = {0029-554X},
doi = {https://doi.org/10.1016/S0029-554X(79)90170-8},
author = {M. Moszyński and B. Bengtson}
}

@article{MELCHER1992212,
title = {A promising new scintillator: cerium-doped lutetium oxyorthosilicate},
journal = {Nuclear Instruments and Methods in Physics Research Section A: Accelerators, Spectrometers, Detectors and Associated Equipment},
volume = {314},
number = {1},
pages = {212-214},
year = {1992},
issn = {0168-9002},
doi = {https://doi.org/10.1016/0168-9002(92)90517-8},
author = {C.L. Melcher and J.S. Schweitzer}
}

@article{10.1063/1.1328775,
    author = {Cooke, D. W. and McClellan, K. J. and Bennett, B. L. and Roper, J. M. and Whittaker, M. T. and Muenchausen, R. E. and Sze, R. C.},
    title = {Crystal growth and optical characterization of cerium-doped Lu1.8Y0.2SiO5},
    journal = {Journal of Applied Physics},
    volume = {88},
    number = {12},
    pages = {7360-7362},
    year = {2000},
    month = {12},
    issn = {0021-8979},
    doi = {10.1063/1.1328775},
    eprint = {https://pubs.aip.org/aip/jap/article-pdf/88/12/7360/19134505/7360\_1\_online.pdf},
}

@INPROCEEDINGS{1462080,
  author={Jiaming Chen and Liyuan Zhang and Ren-yuan Zhu},
  booktitle={IEEE Symposium Conference Record Nuclear Science 2004.}, 
  title={Large size LYSO crystals for future high energy physics experiments}, 
  year={2004},
  volume={1},
  number={},
  pages={117-125 Vol. 1},
  doi={10.1109/NSSMIC.2004.1462080}
}

@ARTICLE{4291695,
  author={Chen, Jianming and Mao, Rihua and Zhang, Liyuan and Zhu, Ren-yuan},
  journal={IEEE Transactions on Nuclear Science}, 
  title={Gamma-Ray Induced Radiation Damage in Large Size LSO and LYSO Crystal Samples}, 
  year={2007},
  volume={54},
  number={4},
  pages={1319-1326},
  doi={10.1109/TNS.2007.902370}}

@ARTICLE{8720279,
  author={Hu, Chen and Xu, Chao and Zhang, Liyuan and Zhang, Qinghui and Zhu, Ren-Yuan},
  journal={IEEE Transactions on Nuclear Science}, 
  title={Development of Yttrium-Doped BaF2 Crystals for Future HEP Experiments}, 
  year={2019},
  volume={66},
  number={7},
  pages={1854-1860},
  doi={10.1109/TNS.2019.2918305}}

@article{Zhu:2018jpp,
    author = "Zhu, Ren-Yuan",
    editor = "Liu, Zhen-An",
    title = "{Applications of Very Fast Inorganic Crystal Scintillators in Future HEP Experiments}",
    doi = "10.1007/978-981-13-1316-5$\_$13",
    journal = "Springer Proc. Phys.",
    volume = "213",
    pages = "70--75",
    year = "2018"
}

@book{Fleck:2021njs,
    author = "Titov, Maxim and Grupen, Claus and Buvat, Ir\`ene",
    editor = "Fleck, Ivor and Titov, Maxim and Grupen, Claus and Buvat, Ir\`ene",
    title = "{Handbook of Particle Detection and Imaging}",
    doi = "10.1007/978-3-319-47999-6",
    isbn = "978-3-319-47999-6, 978-3-319-93784-7, 978-3-319-93785-4",
    publisher = "Springer",
    year = "2022"
}

@book{Fabjan:2020wnt,
    editor = "Fabjan, Christian Wolfgang and Schopper, Herwig",
    title = "{Particle Physics Reference Library. Volume 2: Detectors for Particles and Radiation}",
    doi = "10.1007/978-3-030-35318-6",
    isbn = "978-3-030-35317-9, 978-3-030-35320-9, 978-3-030-35318-6",
    publisher = "Springer",
    year = "2020"
}

@book{Lecoq:2006tzy,
    author = "Lecoq, Paul and Annenkov, Alexander and Gektin, Alexander and Korzhik, Mikhail and Pedrini, Christian",
    title = "{Inorganic Scintillators for Detector Systems. Physical Principles and Crystal Engineering}",
    doi = "10.1007/3-540-27768-4",
    isbn = "978-3-540-27766-8, 978-3-540-27768-2",
    publisher = "Springer",
    address = "Berlin",
    series = "Particle Acceleration and Detection",
    year = "2006"
}

@inproceedings{vasiliev2000,
author = {Vasil'ev, Andrey},
year = {2000},
month = {01},
pages = {43-52},
title = {Relaxation of hot electronic excitations in scintillators: Account for scattering, track effects, complicated electronic structure},
journal = {Proc. Fifth Int. Conf. on Inorganic Scintillators and Their Applications SCINT'99}
}

@article{Rodnyi95,
author = {Rodnyi, P. A. and Dorenbos, P. and van Eijk, C. W. E.},
title = {Energy Loss in Inorganic Scintillators},
journal = {physica status solidi (b)},
volume = {187},
number = {1},
pages = {15-29},
doi = {https://doi.org/10.1002/pssb.2221870102},
eprint = {https://onlinelibrary.wiley.com/doi/pdf/10.1002/pssb.2221870102},
year = {1995}
}

@article{BARTRAM1996225,
title = {Efficiency of electron-hole pair production in scintillators},
journal = {Journal of Luminescence},
volume = {68},
number = {5},
pages = {225-240},
year = {1996},
issn = {0022-2313},
doi = {https://doi.org/10.1016/0022-2313(96)00026-9},
author = {R.H. Bartram and A. Lempicki}
}

@article{VASILEV1996165,
title = {Polarization approximation for electron cascade in insulators after high-energy excitation},
journal = {Nuclear Instruments and Methods in Physics Research Section B: Beam Interactions with Materials and Atoms},
volume = {107},
number = {1},
pages = {165-171},
year = {1996},
issn = {0168-583X},
doi = {https://doi.org/10.1016/0168-583X(95)01023-8},
author = {A.N. Vasil'ev}
}

@article{rodnyi1992core,
  title={Core-valence band transitions in wide-gap ionic crystals},
  author={Rodnyi, PA},
  journal={Soviet physics. Solid state},
  volume={34},
  number={7},
  pages={1053--1066},
  year={1992}
}

@article{VANEIJK1994936,
title = {Cross-luminescence},
journal = {Journal of Luminescence},
volume = {60-61},
pages = {936-941},
year = {1994},
issn = {0022-2313},
doi = {https://doi.org/10.1016/0022-2313(94)90316-6},
author = {C.W.E. {van Eijk}}
}

@article{itoh1997temperature,
  title={Temperature dependence of Auger-free luminescence in alkali and alkaline-earth halides},
  author={Itoh, Minoru and Kamada, Masao and Ohno, Nobuhito},
  journal={Journal of the Physical Society of Japan},
  volume={66},
  number={8},
  pages={2502--2512},
  year={1997},
  publisher={The Physical Society of Japan}
}

@article{KHANIN2023113399,
title = {Recent advances in the study of core-valence luminescence (cross luminescence). Review},
journal = {Optical Materials},
volume = {136},
pages = {113399},
year = {2023},
issn = {0925-3467},
doi = {https://doi.org/10.1016/j.optmat.2022.113399},
author = {Vasilii Khanin and Ivan Venevtsev and Piotr Rodnyi}
}

@book{Sauli:1992bu,
    editor = "Sauli, F.",
    title = "{Instrumentation in high-energy physics}",
    doi = "10.1142/1356",
    year = "1992"
}

@article{MOSER199331,
title = {Principles and practice of plastic scintillator design},
journal = {Radiation Physics and Chemistry},
volume = {41},
number = {1},
pages = {31-36},
year = {1993},
issn = {0969-806X},
doi = {https://doi.org/10.1016/0969-806X(93)90039-W}
}

@book{Henderson06,
    author = {Henderson, B and Imbusch, G F},
    title = {Optical Spectroscopy of Inorganic Solids},
    publisher = {Oxford University Press},
    year = {2006},
    month = {05},
    isbn = {9780199298624},
    doi = {10.1093/oso/9780199298624.001.0001},
}

@article{10.1063/1.1718829,
    author = {Hyman, L. G. and Schwarcz, R. M. and Schluter, R. A.},
    title = {Study of High Speed Photomultiplier Systems},
    journal = {Review of Scientific Instruments},
    volume = {35},
    number = {3},
    pages = {393-406},
    year = {1964},
    month = {03},
    issn = {0034-6748},
    doi = {10.1063/1.1718829},
    eprint = {https://pubs.aip.org/aip/rsi/article-pdf/35/3/393/19242891/393\_1\_online.pdf},
}

@article{10.1063/1.1719516,
    author = {Hyman, L. G.},
    title = {Time Resolution of Photomultiplier Systems},
    journal = {Review of Scientific Instruments},
    volume = {36},
    number = {2},
    pages = {193-196},
    year = {1965},
    month = {02},
    issn = {0034-6748},
    doi = {10.1063/1.1719516},
    eprint = {https://pubs.aip.org/aip/rsi/article-pdf/36/2/193/19213492/193\_1\_online.pdf},
}

@INPROCEEDINGS{842466,
  author={Derenzo, S.E. and Weber, J.J. and Moses, W.W. and Dujardin, C.},
  booktitle={1999 IEEE Nuclear Science Symposium. Conference Record. 1999 Nuclear Science Symposium and Medical Imaging Conference (Cat. No.99CH37019)}, 
  title={Measurements of the intrinsic rise times of common inorganic scintillators}, 
  year={1999},
  volume={1},
  number={},
  pages={152-156 vol.1},
  doi={10.1109/NSSMIC.1999.842466}}

@ARTICLE{6303850,
  author={Lecoq, P.},
  journal={IEEE Transactions on Nuclear Science}, 
  title={New Approaches to Improve Timing Resolution in Scintillators}, 
  year={2012},
  volume={59},
  number={5},
  pages={2313-2318},
  doi={10.1109/TNS.2012.2212283}}

@article{Robbins_1980,
doi = {10.1149/1.2129574},
year = {1980},
month = {dec},
publisher = {The Electrochemical Society, Inc.},
volume = {127},
number = {12},
pages = {2694},
author = {Robbins, D. J.},
title = {On Predicting the Maximum Efficiency of Phosphor Systems Excited by Ionizing Radiation},
journal = {Journal of The Electrochemical Society}
}

@article{PhysRev.139.A1702,
  title = {Theory of the Yield and Fano Factor of Electron-Hole Pairs Generated in Semiconductors by High-Energy Particles},
  author = {van Roosbroeck, W.},
  journal = {Phys. Rev.},
  volume = {139},
  issue = {5A},
  pages = {A1702--A1716},
  numpages = {0},
  year = {1965},
  month = {Aug},
  publisher = {American Physical Society},
  doi = {10.1103/PhysRev.139.A1702},
}

@article{LEMPICKI1993304,
title = {Fundamental limits of scintillator performance},
journal = {Nuclear Instruments and Methods in Physics Research Section A: Accelerators, Spectrometers, Detectors and Associated Equipment},
volume = {333},
number = {2},
pages = {304-311},
year = {1993},
issn = {0168-9002},
doi = {https://doi.org/10.1016/0168-9002(93)91170-R},
author = {A. Lempicki and A.J. Wojtowicz and E. Berman}
}

@ARTICLE{6519337,
  author={Pro, Tiziana and Ferri, Alessandro and Gola, Alberto and Serra, Nicola and Tarolli, Alessandro and Zorzi, Nicola and Piemonte, Claudio},
  journal={IEEE Transactions on Nuclear Science}, 
  title={New Developments of Near-UV SiPMs at FBK}, 
  year={2013},
  volume={60},
  number={3},
  pages={2247-2253},
  doi={10.1109/TNS.2013.2259505}}

@article{Zappalà_2016,
doi = {10.1088/1748-0221/11/11/P11010},
year = {2016},
month = {nov},
publisher = {},
volume = {11},
number = {11},
pages = {P11010},
author = {Zappalà, G. and Acerbi, F. and Ferri, A. and Gola, A. and Paternoster, G. and Regazzoni, V. and Zorzi, N. and Piemonte, C.},
title = {Study of the photo-detection efficiency of FBK High-Density silicon photomultipliers},
journal = {Journal of Instrumentation}
}

@article{OTTE2017106,
title = {Characterization of three high efficiency and blue sensitive silicon photomultipliers},
journal = {Nuclear Instruments and Methods in Physics Research Section A: Accelerators, Spectrometers, Detectors and Associated Equipment},
volume = {846},
pages = {106-125},
year = {2017},
issn = {0168-9002},
doi = {https://doi.org/10.1016/j.nima.2016.09.053},
author = {Adam Nepomuk Otte and Distefano Garcia and Thanh Nguyen and Dhruv Purushotham}
}

@article{Yang:2016six,
    author = "Yang, Fan and Zhang, Liyuan and Zhu, Ren-Yuan",
    editor = "Dorenbos, Pieter and Auffray, Ettienette and Cherepy, Nerine and Fasoli, Mauro and Glodo, Jarek and Kim, Hongjoo and Melcher, Charles L. and Payne, Steve",
    title = "{Gamma-Ray Induced Radiation Damage Up to 340~Mrad in Various Scintillation Crystals}",
    doi = "10.1109/TNS.2015.2505721",
    journal = "IEEE Trans. Nucl. Sci.",
    volume = "63",
    number = "2",
    pages = "612--619",
    year = "2016"
}

@ARTICLE{7762196,
  author={Yang, Fan and Zhang, Liyuan and Zhu, Ren-Yuan and Kapustinsky, Jon and Nelson, Ron and Wang, Zhehui},
  journal={IEEE Transactions on Nuclear Science}, 
  title={Proton-Induced Radiation Damage in Fast Crystal Scintillators}, 
  year={2017},
  volume={64},
  number={1},
  pages={665-672},
  doi={10.1109/TNS.2016.2633427}}

@ARTICLE{9075296,
  author={Hu, Chen and Yang, Fan and Zhang, Liyuan and Zhu, Ren-Yuan and Kapustinsky, Jon and Mocko, Michael and Nelson, Ron and Wang, Zhehui},
  journal={IEEE Transactions on Nuclear Science}, 
  title={Neutron-Induced Radiation Damage in LYSO, BaF2, and PWO Crystals}, 
  year={2020},
  volume={67},
  number={6},
  pages={1086-1092},
  doi={10.1109/TNS.2020.2989116}}

@article{DISSERTORI20141,
title = {Results on damage induced by high-energy protons in LYSO calorimeter crystals},
journal = {Nuclear Instruments and Methods in Physics Research Section A: Accelerators, Spectrometers, Detectors and Associated Equipment},
volume = {745},
pages = {1-6},
year = {2014},
issn = {0168-9002},
doi = {https://doi.org/10.1016/j.nima.2014.02.003},
author = {G. Dissertori and D. Luckey and F. Nessi-Tedaldi and F. Pauss and M. Quittnat and R. Wallny and M. Glaser}
}

@article{Kharzheev:2019tfk,
    author = "Kharzheev, Yu. N.",
    title = "{Radiation Hardness of Scintillation Detectors Based on Organic Plastic Scintillators and Optical Fibers}",
    doi = "10.1134/S1063779619010027",
    journal = "Phys. Part. Nucl.",
    volume = "50",
    number = "1",
    pages = "42--76",
    year = "2019"
}

@article{Sirunyan_2020,
doi = {10.1088/1748-0221/15/06/P06009},
year = {2020},
month = {jun},
publisher = {},
volume = {15},
number = {06},
pages = {P06009},
author = {Sirunyan, A.M. et al.},
title = {Measurements with silicon photomultipliers of dose-rate effects in the radiation damage of plastic scintillator tiles in the CMS hadron endcap calorimeter},
journal = {Journal of Instrumentation}
}

@article{GILLEN19924358,
title = {Rigorous experimental confirmation of a theoretical model for diffusion-limited oxidation},
journal = {Polymer},
volume = {33},
number = {20},
pages = {4358-4365},
year = {1992},
issn = {0032-3861},
doi = {https://doi.org/10.1016/0032-3861(92)90280-A},
author = {Kenneth T Gillen and Roger L Clough}
}

@article{SEGUCHI1981195,
title = {Radiation induced oxidative degradation of polymers—I: Oxidation region in polymer films irradiated in oxygen under pressure},
journal = {Radiation Physics and Chemistry (1977)},
volume = {17},
number = {4},
pages = {195-201},
year = {1981},
issn = {0146-5724},
doi = {https://doi.org/10.1016/0146-5724(81)90331-9},
author = {Tadao Seguchi and Shoji Hashimoto and Kazuo Arakawa and Naohiro Hayakawa and Waichiro Kawakami and Isamu Kuriyama}
}

@phdthesis{Moll:1999kv,
    author = "Moll, Michael",
    title = "{Radiation damage in silicon particle detectors: Microscopic defects and macroscopic properties}",
    reportNumber = "DESY-THESIS-1999-040",
    school = "Hamburg U.",
    year = "1999"
}

@article{Gundacker:2020cnv,
    author = "Gundacker, Stefan and Heering, Arjan",
    title = "{The silicon-photomultiplier: fundamentals and applications of a modern solid-state photon detector}",
    doi = "10.1088/1361-6560/ab7b2d",
    journal = "Phys. Med. Biol.",
    volume = "65",
    number = "17",
    pages = "17TR01",
    year = "2020"
}

@article{FONTE2025170401,
title = {{Timing RPCs: 25 years}},
journal = {Nuclear Instruments and Methods in Physics Research Section A: Accelerators, Spectrometers, Detectors and Associated Equipment},
volume = {1075},
pages = {170401},
year = {2025},
issn = {0168-9002},
doi = {https://doi.org/10.1016/j.nima.2025.170401},
author = {Paulo Fonte}
}

@article{Gundacker_2020,
doi = {10.1088/1361-6560/ab7b2d},
year = {2020},
month = {aug},
publisher = {IOP Publishing},
volume = {65},
number = {17},
pages = {17TR01},
author = {Gundacker, Stefan and Heering, Arjan},
title = {The silicon photomultiplier: fundamentals and applications of a modern solid-state photon detector},
journal = {Physics in Medicine \& Biology},
}

@article{10.1063/1.1710367,
    author = {Shockley, W.},
    title = {Currents to Conductors Induced by a Moving Point Charge},
    journal = {Journal of Applied Physics},
    volume = {9},
    number = {10},
    pages = {635-636},
    year = {1938},
    month = {10},
    issn = {0021-8979},
    doi = {10.1063/1.1710367},
    eprint = {https://pubs.aip.org/aip/jap/article-pdf/9/10/635/18304047/635\_1\_online.pdf},
}

@article{1686997,
  author={Ramo, S.},
  journal={Proceedings of the IRE}, 
  title={Currents Induced by Electron Motion}, 
  year={1939},
  volume={27},
  number={9},
  pages={584-585},
  doi={10.1109/JRPROC.1939.228757}}

@article{PELLEGRINI201412,
title = {{Technology developments and first measurements of Low Gain Avalanche Detectors (LGAD) for high energy physics applications}},
journal = {Nuclear Instruments and Methods in Physics Research Section A: Accelerators, Spectrometers, Detectors and Associated Equipment},
volume = {765},
pages = {12-16},
year = {2014},
note = {HSTD-9 2013 - Proceedings of the 9th International "Hiroshima" Symposium on Development and Application of Semiconductor Tracking Detectors},
issn = {0168-9002},
doi = {https://doi.org/10.1016/j.nima.2014.06.008},
author = {G. Pellegrini and others},
}

@article{CARTIGLIA2015141,
title = {Design optimization of ultra-fast silicon detectors},
journal = {Nuclear Instruments and Methods in Physics Research Section A: Accelerators, Spectrometers, Detectors and Associated Equipment},
volume = {796},
pages = {141-148},
year = {2015},
note = {Proceedings of the 10th International Conference on Radiation Effects on Semiconductor Materials Detectors and Devices},
issn = {0168-9002},
doi = {https://doi.org/10.1016/j.nima.2015.04.025},
author = {Cartiglia, N and others},
}

@article{Cartiglia:2017O8,
  author = "Cartiglia, N and others",
  title = "{Tracking in 4 dimensions}",
  doi = "10.22323/1.314.0489",
  journal = "PoS",
  year = 2017,
  volume = "EPS-HEP2017",
  pages = "489"
}

@article{CARTIGLIA2019350,
title = {Timing layers, 4- and 5-dimension tracking},
journal = {Nuclear Instruments and Methods in Physics Research Section A: Accelerators, Spectrometers, Detectors and Associated Equipment},
volume = {924},
pages = {350-354},
year = {2019},
note = {11th International Hiroshima Symposium on Development and Application of Semiconductor Tracking Detectors},
issn = {0168-9002},
doi = {https://doi.org/10.1016/j.nima.2018.09.157},
author = {N. Cartiglia and others},
}

@article{CortinaGil:20245q,
  author = "Cortina Gil, E.  and  others",
  title = "{Operation and performance of the NA62 Gigatracker}",
  doi = "10.22323/1.448.0008",
  journal = "PoS",
  year = 2024,
  volume = "VERTEX2023",
  pages = "008"
}

@article{KRAMBERGER201926,
title = {Timing performance of small cell 3D silicon detectors},
journal = {Nuclear Instruments and Methods in Physics Research Section A: Accelerators, Spectrometers, Detectors and Associated Equipment},
volume = {934},
pages = {26-32},
year = {2019},
issn = {0168-9002},
doi = {https://doi.org/10.1016/j.nima.2019.04.088},
author = {Kramberger, G. and others},
}

@article{Lampis_2023,
doi = {10.1088/1748-0221/18/01/C01051},
year = {2023},
month = {jan},
publisher = {IOP Publishing},
volume = {18},
number = {01},
pages = {C01051},
author = {Lampis, A. and others},
title = {{10 ps timing with highly irradiated 3D trench silicon pixel sensors}},
journal = {Journal of Instrumentation},
}

@article{PARKER1997328,
title = {3D — A proposed new architecture for solid-state radiation detectors},
journal = {Nuclear Instruments and Methods in Physics Research Section A: Accelerators, Spectrometers, Detectors and Associated Equipment},
volume = {395},
number = {3},
pages = {328-343},
year = {1997},
note = {Proceedings of the Third International Workshop on Semiconductor Pixel Detectors for Particles and X-rays},
issn = {0168-9002},
doi = {https://doi.org/10.1016/S0168-9002(97)00694-3},
author = {S.I. Parker and C.J. Kenney and J. Segal},
}

@techreport{Capeans:1291633,
      author        = "{ATLAS Collaboration}",
      title         = "{ATLAS Insertable B-Layer Technical Design Report}",
      reportNumber  = "CERN-LHCC-2010-013, ATLAS-TDR-19",
      institution   = "CERN",
      year          = "2010",
}

@book{UltrafastSiliconDetectors,
author = {Ferrero, Marco and Arcidiacono, Roberta and Mandurrino, Marco and Sola, Valentina and Cartiglia, Nicolò},
year = {2021},
month = {06},
pages = {},
title = {An Introduction to Ultra-Fast Silicon Detectors: Design, Tests, and Performances},
isbn = {9781003131946},
doi = {10.1201/9781003131946},
publisher = {Taylor \& Francis}
}

@article{Sadrozinski_2018,
doi = {10.1088/1361-6633/aa94d3},
year = {2017},
month = {dec},
publisher = {IOP Publishing},
volume = {81},
number = {2},
pages = {026101},
author = {Sadrozinski, Hartmut F-W and Seiden, Abraham and Cartiglia, Nicolò},
title = {4D tracking with ultra-fast silicon detectors},
journal = {Reports on Progress in Physics}
}

@ARTICLE{MMoll,
  author={Moll, Michael},
  journal={IEEE Transactions on Nuclear Science}, 
  title={Displacement Damage in Silicon Detectors for High Energy Physics}, 
  year={2018},
  volume={65},
  number={8},
  pages={1561-1582},
  doi={10.1109/TNS.2018.2819506}
}

@article{FERRERO201916,
title = {Radiation resistant LGAD design},
journal = {Nuclear Instruments and Methods in Physics Research Section A: Accelerators, Spectrometers, Detectors and Associated Equipment},
volume = {919},
pages = {16-26},
year = {2019},
issn = {0168-9002},
doi = {https://doi.org/10.1016/j.nima.2018.11.121},
author = {M. Ferrero and others}
}

@article{Lastoviska-Medin_2023,
doi = {10.1088/1748-0221/18/02/C02059},
year = {2023},
month = {feb},
publisher = {IOP Publishing},
volume = {18},
number = {02},
pages = {C02059},
author = {Laštovička-Medin, G. and others},
title = {New insight into gain suppression and single event Burnout effects in LGAD},
journal = {Journal of Instrumentation}
}

@article{Beresford_2023,
doi = {10.1088/1748-0221/18/07/P07030},
year = {2023},
month = {jul},
publisher = {IOP Publishing},
volume = {18},
number = {07},
pages = {P07030},
author = {Beresford, L.A. and others},
title = {Destructive breakdown studies of irradiated LGADs at beam tests for the ATLAS HGTD},
journal = {Journal of Instrumentation}
}

@article{SOLA2024169453,
title = {The first batch of compensated LGAD sensors},
journal = {Nuclear Instruments and Methods in Physics Research Section A: Accelerators, Spectrometers, Detectors and Associated Equipment},
volume = {1064},
pages = {169453},
year = {2024},
issn = {0168-9002},
doi = {https://doi.org/10.1016/j.nima.2024.169453},
author = {Sola, V. and others}
}

@article{FISCHLE1991202,
title = {Experimental determination of ionization cluster size distributions in counting gases},
journal = {Nuclear Instruments and Methods in Physics Research Section A: Accelerators, Spectrometers, Detectors and Associated Equipment},
volume = {301},
number = {2},
pages = {202-214},
year = {1991},
issn = {0168-9002},
doi = {https://doi.org/10.1016/0168-9002(91)90460-8},
author = {Hansjörg Fischle and Joachim Heintze and Bernhard Schmidt}
}

@inbook{Sauli_2014, place={Cambridge}, series={Cambridge Monographs on Particle Physics, Nuclear Physics and Cosmology}, title={Drift and diffusion of charges in gases}, booktitle={Gaseous Radiation Detectors: Fundamentals and Applications}, publisher={Cambridge University Press}, author={Sauli, Fabio}, year={2014}, pages={76–128}, collection={Cambridge Monographs on Particle Physics, Nuclear Physics and Cosmology}}

@article{SAHIN2014104,
title = {High-precision gas gain and energy transfer measurements in Ar–CO2 mixtures},
journal = {Nuclear Instruments and Methods in Physics Research Section A: Accelerators, Spectrometers, Detectors and Associated Equipment},
volume = {768},
pages = {104-111},
year = {2014},
issn = {0168-9002},
doi = {https://doi.org/10.1016/j.nima.2014.09.061},
author = {Özkan Şahin and Tadeusz Z. Kowalski and Rob Veenhof}
}

@article{ALICE:1999osy,
    author = "Arnaldi, R. and others",
    editor = "Borchi, E. and Huston, J. and Majewski, S. and Penzo, A. and Rancoita, P. G.",
    collaboration = "ALICE",
    title = "{Study of resistive plate chambers for the Alice dimuon spectrometer}",
    doi = "10.1016/S0920-5632(99)00527-7",
    journal = "Nucl. Phys. B Proc. Suppl.",
    volume = "78",
    pages = "84--89",
    year = "1999"
}

@article{BACCI2000342,
title = {High altitude test of RPCs for the Argo YBJ experiment},
journal = {Nuclear Instruments and Methods in Physics Research Section A: Accelerators, Spectrometers, Detectors and Associated Equipment},
volume = {443},
number = {2},
pages = {342-350},
year = {2000},
issn = {0168-9002},
doi = {https://doi.org/10.1016/S0168-9002(99)01079-7},
author = {C. Bacci et al.}
}

@phdthesis{Lippmann:2003uaa,
    author = "Lippmann, Christian",
    title = "{Detector Physics of Resistive Plate Chambers}",
    reportNumber = "CERN-THESIS-2003-035",
    school = "Frankfurt U.",
    year = "2003"
}

@article{PARKHOMCHUCK1971269,
title = {A spark counter with large area},
journal = {Nuclear Instruments and Methods},
volume = {93},
number = {2},
pages = {269-270},
year = {1971},
issn = {0029-554X},
doi = {https://doi.org/10.1016/0029-554X(71)90475-7},
author = {V.V. Parkhomchuck and Yu.N. Pestov and N.V. Petrovykh}
}

@article{COOPER200941,
title = {The D0 silicon tracker},
journal = {Nuclear Instruments and Methods in Physics Research Section A: Accelerators, Spectrometers, Detectors and Associated Equipment},
volume = {598},
number = {1},
pages = {41-45},
year = {2009},
note = {Instrumentation for Collding Beam Physics},
issn = {0168-9002},
doi = {https://doi.org/10.1016/j.nima.2008.08.065},
author = {W.E. Cooper}
}

@article{SANTONICO1981377,
title = {Development of resistive plate counters},
journal = {Nuclear Instruments and Methods in Physics Research},
volume = {187},
number = {2},
pages = {377-380},
year = {1981},
issn = {0167-5087},
doi = {https://doi.org/10.1016/0029-554X(81)90363-3},
author = {R. Santonico and R. Cardarelli}
}

@article{CERRONZEBALLOS1996132,
title = {A new type of resistive plate chamber: The multigap RPC},
journal = {Nuclear Instruments and Methods in Physics Research Section A: Accelerators, Spectrometers, Detectors and Associated Equipment},
volume = {374},
number = {1},
pages = {132-135},
year = {1996},
issn = {0168-9002},
doi = {https://doi.org/10.1016/0168-9002(96)00158-1},
author = {E {Cerron Zeballos} and I Crotty and D Hatzifotiadou and J {Lamas Valverde} and S Neupane and M.C.S Williams and A Zichichi}
}

@inbook{RPCbook,
author = {M. Abbrescia, V. Peskov, P. Fonte},
publisher = {John Wiley \& Sons, Ltd},
isbn = {9783527698691},
title = {Further Developments in Resistive Plate Chambers},
booktitle = {Resistive Gaseous Detectors},
year = {2018},
chapter = {4},
pages = {111-159},
doi = {https://doi.org/10.1002/9783527698691.ch4}
}

@article{Gonzalez-Diaz_2017,
doi = {10.1088/1748-0221/12/03/C03029},
year = {2017},
month = {mar},
publisher = {},
volume = {12},
number = {03},
pages = {C03029},
author = {González-Díaz, D. and Palomo, F.R. and González, J. and Chen, H.},
title = {Detectors and Concepts for sub-100 ps timing with gaseous detectors},
journal = {Journal of Instrumentation}
}

@article{SAKAI2002153,
title = {Study of the effect of water vapor on a resistive plate chamber with glass electrodes},
journal = {Nuclear Instruments and Methods in Physics Research Section A: Accelerators, Spectrometers, Detectors and Associated Equipment},
volume = {484},
number = {1},
pages = {153-161},
year = {2002},
issn = {0168-9002},
doi = {https://doi.org/10.1016/S0168-9002(01)02032-0},
author = {H Sakai and H Sakaue and Y Teramoto and E Nakano and T Takahashi}
}

@article{AIELLI200486,
title = {Electrical conduction properties of phenolic–melaminic laminates},
journal = {Nuclear Instruments and Methods in Physics Research Section A: Accelerators, Spectrometers, Detectors and Associated Equipment},
volume = {533},
number = {1},
pages = {86-92},
year = {2004},
note = {Proceedings of the Seventh International Workshop on Resistive Plate Chambers and Related Detectors},
issn = {0168-9002},
doi = {https://doi.org/10.1016/j.nima.2004.07.006},
author = {G. Aielli and P. Camarri and R. Cardarelli and A. {Di Ciaccio} and A. {Di Simone} and B. Liberti and R. Santonico}
}

@techreport{CERN-LHCC-2019-003,
      author        = "{CMS Collaboration}",
      title         = "{A MIP Timing Detector for the CMS Phase-2 Upgrade}",
      institution   = "CERN",
      reportNumber  = "CERN-LHCC-2019-003, CMS-TDR-020",
      address       = "Geneva",
      year          = "2019",
}

@article{Albuquerque_2024,
doi = {10.1088/1748-0221/19/05/P05048},
year = {2024},
month = {may},
publisher = {IOP Publishing},
volume = {19},
number = {05},
pages = {P05048},
author = {Albuquerque, E. and others},
title = {{TOFHIR2: the readout ASIC of the CMS barrel MIP Timing Detector}},
journal = {Journal of Instrumentation}
}

@ARTICLE{ALDO_2024,
  author={Carniti, Paolo and Gotti, Claudio and Pessina, Gianluigi},
  journal={IEEE Transactions on Nuclear Science}, 
  title={A Multifunction Radiation-Hardened HV and LV Linear Regulator for SiPM-Based HEP Detectors}, 
  year={2024},
  volume={71},
  number={7},
  pages={1399-1408},
  doi={10.1109/TNS.2024.3409820}}

@ARTICLE{DLED_GolaPiemonte,
  author={Gola, Alberto and Piemonte, Claudio and Tarolli, Alessandro},
  journal={IEEE Transactions on Nuclear Science}, 
  title={{The DLED Algorithm for Timing Measurements on Large Area SiPMs Coupled to Scintillators}}, 
  year={2012},
  volume={59},
  number={2},
  pages={358-365},
  doi={10.1109/TNS.2012.2187927}
}

@misc{Addesa:2025kyl,
    author = "Addesa, F. and others",
    title = "{The CMS Barrel Timing Layer: test beam confirmation of module timing performance}",
    eprint = "2504.11209",
    archivePrefix = "arXiv",
    primaryClass = "physics.ins-det",
    month = "4",
    year = "2025"
}

@misc{TWEPP2024-TedLiu,
  author = {T. Liu},
  title = {The testing and performance of the ETROC2 for CMS MTD Endcap Timing Layer (ETL) upgrade}, 
  year = {2024},
  note = {in:  Topical Workshop on Electronics for Particle Physics},
}

@techreport{CERN-LHCC-2020-007,
    author = "{ATLAS Collaboration}",
    title         = "{Technical Design Report: A High-Granularity Timing
                       Detector for the ATLAS Phase-II Upgrade}",
    institution   = "CERN",
    reportNumber  = "CERN-LHCC-2020-007, ATLAS-TDR-031",
    address       = "Geneva",
    year          = "2020",
}

@article{Ali_2023,
    doi = {10.1088/1748-0221/18/05/P05005},
    year = {2023},
    month = {may},
    publisher = {IOP Publishing},
    volume = {18},
    number = {05},
    pages = {P05005},
    author = {Ali, S. and others},
    title = {{Performance in beam tests of carbon-enriched irradiated Low Gain Avalanche Detectors for the ATLAS High Granularity Timing Detector}},
    journal = {Journal of Instrumentation}
}

@unpublished{Wu:2771199,
      author        = "Wu, Kewei",
      title         = "{HGTD sensors | ATLAS Phase-2 Upgrade}",
      year          = "2021",
      note          = "General Photo",
}

@book{CERN-LHCC-2000-012,
      author = "{ALICE Collaboration}",
      title         = "{ALICE Time-Of-Flight system (TOF): Technical Design
                       Report}",
      publisher     = "CERN",
      address       = "Geneva",
      series        = "Technical design report. ALICE",
      year          = "2000",
}

@book{Cortese:545834,
      author        = "Cortese, P",
      collaboration = "ALICE",
      title         = "{ALICE Time-Of Flight system (TOF): addendum to the
                       Technical Design Report}",
      publisher     = "CERN",
      address       = "Geneva",
      series        = "Technical design report. ALICE",
      year          = "2002",
}

@article{AKINDINOV2004611,
title = {Results from a large sample of MRPC-strip prototypes for the ALICE TOF detector},
journal = {Nuclear Instruments and Methods in Physics Research Section A: Accelerators, Spectrometers, Detectors and Associated Equipment},
volume = {532},
number = {3},
pages = {611-621},
year = {2004},
issn = {0168-9002},
doi = {https://doi.org/10.1016/j.nima.2004.05.125},
author = {A.V. Akindinov and others},
}

@article{Akindinov:2009zze,
    author = "Akindinov, A. and others",
    title = "{The ALICE Time-Of-Flight system: Construction, assembly and quality tests}",
    doi = "10.1393/ncb/i2009-10761-3",
    journal = "Nuovo Cim. B",
    volume = "124",
    pages = "235--253",
    year = "2009"
}

@article{Carnesecchi:2018oss,
    author = "Carnesecchi, Francesca",
    collaboration = "ALICE",
    title = "{Performance of the ALICE Time-Of-Flight detector at the LHC}",
    eprint = "1806.03825",
    archivePrefix = "arXiv",
    primaryClass = "physics.ins-det",
    doi = "10.1088/1748-0221/14/06/C06023",
    journal = "JINST",
    volume = "14",
    number = "06",
    pages = "C06023",
    year = "2019"
}

@article{CAO2020163053,
title = {{Design and construction of the new BESIII endcap Time-of-Flight system with MRPC Technology}},
journal = {Nuclear Instruments and Methods in Physics Research Section A: Accelerators, Spectrometers, Detectors and Associated Equipment},
volume = {953},
pages = {163053},
year = {2020},
issn = {0168-9002},
doi = {https://doi.org/10.1016/j.nima.2019.163053},
author = {P. Cao and others},
}

@INPROCEEDINGS{8824298,
  author={Li, X. and others},
  booktitle={2018 IEEE Nuclear Science Symposium and Medical Imaging Conference Proceedings (NSS/MIC)}, 
  title={Design and Test of the BESIII ETOF with MRPC Technology}, 
  year={2018},
  volume={},
  number={},
  pages={1-7},
  doi={10.1109/NSSMIC.2018.8824298}
}

@techreport{Albrow:1753795,
    author = "Albrow, M and others",
    title         = "{CMS-TOTEM Precision Proton Spectrometer}",
    institution = "CERN",     
    reportNumber  = "CERN-LHCC-2014-021, TOTEM-TDR-003, CMS-TDR-013",
    year          = "2014",
}

@article{BOSSINI2023167823,
title = {{The CMS Precision Proton Spectrometer: Precision timing with scCVD diamond crystals}},
journal = {Nuclear Instruments and Methods in Physics Research Section A: Accelerators, Spectrometers, Detectors and Associated Equipment},
volume = {1047},
pages = {167823},
year = {2023},
issn = {0168-9002},
doi = {https://doi.org/10.1016/j.nima.2022.167823},
author = {E. Bossini}
}

@article{Bossini_2020,
doi = {10.1088/1748-0221/15/05/C05054},
year = {2020},
month = {may},
publisher = {},
volume = {15},
number = {05},
pages = {C05054},
author = {Bossini, E.},
title = {{The CMS Precision Proton Spectrometer timing system: performance in Run 2, future upgrades and sensor radiation hardness studies}},
journal = {Journal of Instrumentation}
}

@article{Berretti_2017,
doi = {10.1088/1748-0221/12/03/P03026},
year = {2017},
month = {mar},
publisher = {},
volume = {12},
number = {03},
pages = {P03026},
author = {Berretti, M. and others},
title = {Timing performance of a double layer diamond detector},
journal = {Journal of Instrumentation}
}

@ARTICLE{10.3389/fphy.2020.00248,
AUTHOR={Bossini, Edoardo  and Minafra, Nicola },       
TITLE={Diamond Detectors for Timing Measurements in High Energy Physics},
JOURNAL={Frontiers in Physics},
VOLUME={Volume 8 - 2020},
YEAR={2020},
DOI={10.3389/fphy.2020.00248},
ISSN={2296-424X}
}

@article{RANTANEN2024169710,
title = {{The Run 3 timing detector of the CMS Precision Proton Spectrometer: Status and performance}},
journal = {Nuclear Instruments and Methods in Physics Research Section A: Accelerators, Spectrometers, Detectors and Associated Equipment},
volume = {1068},
pages = {169710},
year = {2024},
issn = {0168-9002},
doi = {https://doi.org/10.1016/j.nima.2024.169710},
author = {Milla-Maarit Rantanen}
}

@article{Hayrapetyan_2024,
doi = {10.1088/1748-0221/19/09/P09004},
year = {2024},
month = {sep},
publisher = {IOP Publishing},
volume = {19},
number = {09},
pages = {P09004},
author = {{CMS Collaboration}},
title = {Performance of the CMS electromagnetic calorimeter in pp collisions at $\sqrt{s}$ = 13 TeV},
journal = {Journal of Instrumentation},
}

@techreport{CERN-LHCC-2017-011,
      author        = "{CMS Collaboration}",
      title         = "{The Phase-2 Upgrade of the CMS Barrel Calorimeters}",
      institution   = "CERN",
      reportNumber  = "CERN-LHCC-2017-011, CMS-TDR-015",
      address       = "Geneva",
      year          = "2017",
      note          = "This is the final version, approved by the LHCC",
}

@techreport{CERN-LHCC-2017-023,
      author = "{CMS Collaboration}",
      title         = "{The Phase-2 Upgrade of the CMS Endcap Calorimeter}",
      institution   = "CERN",
      reportNumber  = "CERN-LHCC-2017-023, CMS-TDR-019",
      address       = "Geneva",
      year          = "2017",
      doi           = "10.17181/CERN.IV8M.1JY2",
}

@article{Acar_2024,
doi = {10.1088/1748-0221/19/04/P04015},
year = {2024},
month = {apr},
publisher = {IOP Publishing},
volume = {19},
number = {04},
pages = {P04015},
author = {Acar, B and others},
title = {{Timing performance of the CMS High Granularity Calorimeter prototype}},
journal = {Journal of Instrumentation}
}

@article{Ferrero_2020,
doi = {10.1088/1748-0221/15/04/C04027},
year = {2020},
month = {apr},
publisher = {},
volume = {15},
number = {04},
pages = {C04027},
author = {Ferrero, M. and others},
title = {Evolution of the design of ultra fast silicon detector to cope with high irradiation fluences and fine segmentation},
journal = {Journal of Instrumentation}
}

@ARTICLE{9081916,
  author={Paternoster, G. and others},
  journal={IEEE Electron Device Letters}, 
  title={{Trench-Isolated Low Gain Avalanche Diodes (TI-LGADs)}}, 
  year={2020},
  volume={41},
  number={6},
  pages={884-887},
  doi={10.1109/LED.2020.2991351}}

@article{Zhao_2022,
doi = {10.1088/1742-6596/2374/1/012171},
year = {2022},
month = {nov},
publisher = {IOP Publishing},
volume = {2374},
number = {1},
pages = {012171},
author = {Zhao, Y. and others},
title = {A new approach to achieving high granularity for silicon diode detectors with impact ionization gain},
journal = {Journal of Physics: Conference Series}
}

@article{Gan_2025,
doi = {10.1088/1748-0221/20/04/C04013},
year = {2025},
month = {apr},
publisher = {IOP Publishing},
volume = {20},
number = {04},
pages = {C04013},
author = {Gan, Y. and others},
title = {{The first test-beam results of MiniCACTUS-V2: an ASIC prototype with 60 ps time resolution and a fast recovery time}},
journal = {Journal of Instrumentation}
}

@article{Zambito_2023,
doi = {10.1088/1748-0221/18/03/P03047},
year = {2023},
month = {mar},
publisher = {IOP Publishing},
volume = {18},
number = {03},
pages = {P03047},
author = {Zambito, S. and others},
title = {20 ps time resolution with a fully-efficient monolithic silicon pixel detector without internal gain layer},
journal = {Journal of Instrumentation}
}

@misc{selina2025,
title={{Exploring unique design features of the Monolithic Stitched Sensor with Timing (MOST): yield, powering, timing, and sensor reverse bias}}, 
author={Mariia Selina and others},
year={2025},
eprint={2504.13696},
archivePrefix={arXiv},
primaryClass={physics.ins-det},
}

@ARTICLE{9620045,
  author={Iacobucci, Giuseppe and Paolozzi, Lorenzo and Valerio, Pierpaolo},
  journal={IEEE Instrumentation \& Measurement Magazine}, 
  title={{Monolithic Picosecond Silicon Pixel Sensors for Future Physics: Experiments and Applications}}, 
  year={2021},
  volume={24},
  number={9},
  pages={5-11},
  doi={10.1109/MIM.2021.9620045}}

@ARTICLE{9075426,
  author={Pancheri, Lucio and Giampaolo, Raffaele A. and Salvo, Andrea Di and Mattiazzo, Serena and Corradino, Thomas and Giubilato, Piero and Santoro, Romualdo and Caccia, Massimo and Margutti, Giovanni and Olave, Jonhatan E. and Rolo, Manuel and Rivetti, Angelo},
  journal={IEEE Transactions on Electron Devices}, 
  title={{Fully Depleted MAPS in 110-nm CMOS Process With 100–300~$\mu$m Active Substrate}}, 
  year={2020},
  volume={67},
  number={6},
  pages={2393-2399},
  doi={10.1109/TED.2020.2985639}}

@Article{instruments5040039,
AUTHOR = {Anderlini, Lucio and others},
TITLE = {{Fabrication and First Full Characterisation of Timing Properties of 3D Diamond Detectors}},
JOURNAL = {Instruments},
VOLUME = {5},
YEAR = {2021},
NUMBER = {4},
ARTICLE-NUMBER = {39},
ISSN = {2410-390X},
DOI = {10.3390/instruments5040039}
}

@ARTICLE{10.3389/fphy.2024.1497267,
AUTHOR={Addison, M.  and others},
TITLE={{Characterisation of 3D trench silicon pixel sensors irradiated at 1$\cdot$ 10$^{17}$ 1 MeV n$_{eq}$cm$^{-2}$}},
JOURNAL={Frontiers in Physics},
VOLUME={Volume 12 - 2024},
YEAR={2024},
DOI={10.3389/fphy.2024.1497267},
ISSN={2296-424X},
}

@techreport{Detector:2784893,
      author        = "{ECFA Detector R\&D Roadmap Process Group}",
      title         = "{The 2021 ECFA detector research and development roadmap}",
      institution   = "CERN",
      reportNumber  = "CERN-ESU-017",
      address       = "Geneva",
      year          = "2020",
      doi           = "10.17181/CERN.XDPL.W2EX",
}

@techreport{CERN-LHCC-2021-012,
      author = "{LHCb Collaboration}",
      title         = "{Framework TDR for the LHCb Upgrade II: Opportunities in
                       flavour physics, and beyond, in the HL-LHC era}",
      institution   = "CERN",
      reportNumber  = "CERN-LHCC-2021-012, LHCB-TDR-023",
      address       = "Geneva",
      year          = "2021",
      doi           = "10.17181/CERN.NTVH.Q21W",
}

@techreport{LHCbCollaboration:2920835,
      author        = "{LHCb Collaboration}",
      title         = "{Technology developments for LHCb Upgrade II}",
      institution   = "CERN",
      reportNumber  = "LHCb-PUB-2025-002, CERN-LHCb-PUB-2025-002",
      address       = "Geneva",
      year          = "2025",
}

@article{HARNEW2023167991,
title = {{The TORCH time-of-flight detector}},
journal = {Nuclear Instruments and Methods in Physics Research Section A: Accelerators, Spectrometers, Detectors and Associated Equipment},
volume = {1048},
pages = {167991},
year = {2023},
issn = {0168-9002},
doi = {https://doi.org/10.1016/j.nima.2022.167991},
author = {Harnew, N. and others}
}

@article{KHOLODENKO2024169656,
title = {{Scintillating sampling ECAL technology for the LHCb PicoCal}},
journal = {Nuclear Instruments and Methods in Physics Research Section A: Accelerators, Spectrometers, Detectors and Associated Equipment},
volume = {1066},
pages = {169656},
year = {2024},
issn = {0168-9002},
doi = {https://doi.org/10.1016/j.nima.2024.169656},
author = {Sergei Kholodenko}
}

@techreport{arXiv:2211.02491,
      author        = "{ALICE Collaboration}",
      title         = "{Letter of intent for ALICE 3: A next generation heavy-ion
                       experiment at the LHC}",
      institution   = "CERN",
      archivePrefix = "arXiv",
      eprint        = "2211.02491",
      reportNumber  = "CERN-LHCC-2022-009, LHCC-I-038, LHCC-I-038",
      address       = "Geneva",
      year          = "2022",
}

@Article{Carnesecchi2023,
author={Carnesecchi, F. and others},
title={Beam test results of 25 and 35 $\mu$m thick FBK ultra-fast silicon detectors},
journal={The European Physical Journal Plus},
year={2023},
month={Jan},
day={30},
volume={138},
number={1},
pages={99},
issn={2190-5444},
doi={10.1140/epjp/s13360-022-03619-1},
}

@article{Follo_2024,
doi = {10.1088/1748-0221/19/07/P07033},
year = {2024},
month = {jul},
publisher = {IOP Publishing},
volume = {19},
number = {07},
pages = {P07033},
author = {Follo, U. and others},
title = {First results on monolithic CMOS detector with internal gain},
journal = {Journal of Instrumentation}
}

@Article{Carnesecchi2023_SiPMs,
author={Carnesecchi, F. and others}, 
title={Measurements of the Cherenkov effect in direct detection of charged particles with SiPMs},
journal={The European Physical Journal Plus},
year={2023},
month={Sep},
day={05},
volume={138},
number={9},
pages={788},
issn={2190-5444},
doi={10.1140/epjp/s13360-023-04397-0},
}

@misc{Abbrescia:2926782,
      author        = "Abbrescia, M. and others", 
      title         = "{The IDEA detector concept for FCC-ee}",
      archivePrefix = "arXiv",
      eprint        = "2502.21223",
      reportNumber  = "FERMILAB-PUB-25-0189-PPD",
      year          = "2025",
}

@Article{Long2021,
author={Long, K. R.
and others},
title={Muon colliders to expand frontiers of particle physics},
journal={Nature Physics},
year={2021},
month={Mar},
day={01},
volume={17},
number={3},
pages={289-292},
issn={1745-2481},
doi={10.1038/s41567-020-01130-x},
}

@article{AN200839,
title = {{A 20 ps timing device—A Multigap Resistive Plate Chamber with 24 gas gaps}},
journal = {Nuclear Instruments and Methods in Physics Research Section A: Accelerators, Spectrometers, Detectors and Associated Equipment},
volume = {594},
number = {1},
pages = {39-43},
year = {2008},
issn = {0168-9002},
doi = {https://doi.org/10.1016/j.nima.2008.06.013},
author = {S. An and Y.K. Jo and J.S. Kim and M.M. Kim and D. Hatzifotiadou and M.C.S. Williams and A. Zichichi and R. Zuyeuski}
}

@article{microbulk,
doi = {10.1088/1748-0221/7/04/P04007},
year = {2012},
month = {apr},
publisher = {},
volume = {7},
number = {04},
pages = {P04007},
author = {F J Iguaz and E Ferrer-Ribas and A Giganon and I Giomataris},
title = {Characterization of microbulk detectors in argon- and neon-based mixtures},
journal = {Journal of Instrumentation}
}

@article{LIU2019396,
title = {Timing performance study of Multigap Resistive Plate Chamber with different gap size},
journal = {Nuclear Instruments and Methods in Physics Research Section A: Accelerators, Spectrometers, Detectors and Associated Equipment},
volume = {927},
pages = {396-400},
year = {2019},
issn = {0168-9002},
doi = {https://doi.org/10.1016/j.nima.2019.02.068},
author = {Z. Liu and F. Carnesecchi and M.C.S. Williams and A. Zichichi and R. Zuyeuski}
}

@article{CHARPAK199163,
title = {Investigation of operation of a parallel-plate avalanche chamber with a CsI photocathode under high gain conditions},
journal = {Nuclear Instruments and Methods in Physics Research Section A: Accelerators, Spectrometers, Detectors and Associated Equipment},
volume = {307},
number = {1},
pages = {63-68},
year = {1991},
issn = {0168-9002},
doi = {https://doi.org/10.1016/0168-9002(91)90131-9},
author = {G. Charpak and P. Fonte and V. Peskov and F. Sauli and D. Scigocki and D. Stuart}
}

@article{FONTE200530,
title = {Novel single photon detectors for UV imaging},
journal = {Nuclear Instruments and Methods in Physics Research Section A: Accelerators, Spectrometers, Detectors and Associated Equipment},
volume = {553},
number = {1},
pages = {30-34},
year = {2005},
note = {Proceedings of the fifth International Workshop on Ring Imaging Detectors},
issn = {0168-9002},
doi = {https://doi.org/10.1016/j.nima.2005.08.002},
author = {P. Fonte and T. Francke and N. Pavlopoulos and V. Peskov and I. Rodionov}
}

@article{CARLSON2003189,
title = {Beyond the RICH: innovative photosensitive gaseous detectors for new fields of applications},
journal = {Nuclear Instruments and Methods in Physics Research Section A: Accelerators, Spectrometers, Detectors and Associated Equipment},
volume = {502},
number = {1},
pages = {189-194},
year = {2003},
note = {Experimental Techniques of Cherenkov Light Imaging. Proceedings of the Fourth International Workshop on Ring Imaging Cherenkov Detectors},
issn = {0168-9002},
doi = {https://doi.org/10.1016/S0168-9002(03)00272-9},
author = {P. Carlson and C. Iacobeaus and T. Francke and B. Lund-Jensen and L. Periale and V. Peskov and I. Rodionov}
}

@article{FRANCKE2004163,
title = {High rate (up to 105Hz/cm2), high position resolution (30$\mu$m) photosensitive RPCs},
journal = {Nuclear Instruments and Methods in Physics Research Section A: Accelerators, Spectrometers, Detectors and Associated Equipment},
volume = {533},
number = {1},
pages = {163-168},
year = {2004},
note = {Proceedings of the Seventh International Workshop on Resistive Plate Chambers and Related Detectors},
issn = {0168-9002},
doi = {https://doi.org/10.1016/j.nima.2004.07.021},
author = {T. Francke and V. Peskov and I. Rodionov and P. Fonte}
}

@article{MATSUOKA2023168378,
title = {Demonstration of a 25-picosecond single-photon time resolution with gaseous photomultiplication},
journal = {Nuclear Instruments and Methods in Physics Research Section A: Accelerators, Spectrometers, Detectors and Associated Equipment},
volume = {1053},
pages = {168378},
year = {2023},
issn = {0168-9002},
doi = {https://doi.org/10.1016/j.nima.2023.168378},
author = {K. Matsuoka and R. Okubo and Y. Adachi}
}

@article{ZHAO2025170593,
title = {A high rate and high timing photoelectric detector prototype with RPC structure},
journal = {Nuclear Instruments and Methods in Physics Research Section A: Accelerators, Spectrometers, Detectors and Associated Equipment},
volume = {1078},
pages = {170593},
year = {2025},
issn = {0168-9002},
doi = {https://doi.org/10.1016/j.nima.2025.170593},
author = {Yiding Zhao and others}
}

@article{BORTFELDT2018317,
title = {PICOSEC: Charged particle timing at sub-25 picosecond precision with a Micromegas based detector},
journal = {Nuclear Instruments and Methods in Physics Research Section A: Accelerators, Spectrometers, Detectors and Associated Equipment},
volume = {903},
pages = {317-325},
year = {2018},
issn = {0168-9002},
doi = {https://doi.org/10.1016/j.nima.2018.04.033},
author = {J. Bortfeldt and others}
}

@article{Sohl_2020,
doi = {10.1088/1748-0221/15/04/C04053},
year = {2020},
month = {apr},
publisher = {},
volume = {15},
number = {04},
pages = {C04053},
author = {Sohl, L. and others},
title = {Single photoelectron time resolution studies of the PICOSEC-Micromegas detector},
journal = {Journal of Instrumentation}
}

@article{KAMADA201563,
title = {Alkali earth co-doping effects on luminescence and scintillation properties of Ce doped Gd3Al2Ga3O12 scintillator},
journal = {Optical Materials},
volume = {41},
pages = {63-66},
year = {2015},
note = {5th International Workshop on Photoluminescence in Rare Earths (PRE'14): Photonic Materials and Devices May 13-16, 2014, San Sebastian, Spain},
issn = {0925-3467},
doi = {https://doi.org/10.1016/j.optmat.2014.10.008},
author = {Kei Kamada and others}
}

@article{PhysRevApplied.2.044009,
  title = {Role of ${\mathrm{Ce}}^{4+}$ in the Scintillation Mechanism of Codoped ${\mathrm{Gd}}_{3}{\mathrm{Ga}}_{3}{\mathrm{Al}}_{2}{\mathrm{O}}_{12}\ensuremath{\mathbin:}\mathrm{Ce}$},
  author = {Wu, Yuntao and Meng, Fang and Li, Qi and Koschan, Merry and Melcher, Charles L.},
  journal = {Phys. Rev. Appl.},
  volume = {2},
  issue = {4},
  pages = {044009},
  numpages = {13},
  year = {2014},
  month = {Oct},
  publisher = {American Physical Society},
  doi = {10.1103/PhysRevApplied.2.044009},
}

@article{LUCCHINI2016176,
title = {{Effect of Mg2+ ions co-doping on timing performance and radiation tolerance of Cerium doped Gd3Al2Ga3O12 crystals}},
journal = {Nuclear Instruments and Methods in Physics Research Section A: Accelerators, Spectrometers, Detectors and Associated Equipment},
volume = {816},
pages = {176-183},
year = {2016},
issn = {0168-9002},
doi = {https://doi.org/10.1016/j.nima.2016.02.004},
author = {M.T. Lucchini and others}
}

@article{https://doi.org/10.1002/adom.201400571,
author = {Nikl, Martin and Yoshikawa, Akira},
title = {{Recent R\&D Trends in Inorganic Single-Crystal Scintillator Materials for Radiation Detection}},
journal = {Advanced Optical Materials},
volume = {3},
number = {4},
pages = {463-481},
doi = {https://doi.org/10.1002/adom.201400571},
year = {2015}
}

@article{doi:10.1021/cg501005s,
author = {Nikl, Martin and others},
title = {Defect Engineering in Ce-Doped Aluminum Garnet Single Crystal Scintillators},
journal = {Crystal Growth \& Design},
volume = {14},
number = {9},
pages = {4827-4833},
year = {2014},
doi = {10.1021/cg501005s},
}

@article{AN2023167629,
title = {Performance of a spaghetti calorimeter prototype with tungsten absorber and garnet crystal fibres},
journal = {Nuclear Instruments and Methods in Physics Research Section A: Accelerators, Spectrometers, Detectors and Associated Equipment},
volume = {1045},
pages = {167629},
year = {2023},
issn = {0168-9002},
doi = {https://doi.org/10.1016/j.nima.2022.167629},
author = {L. An and others}
}

@article{MARTINAZZOLI2021165231,
title = {Scintillation properties and timing performance of state-of-the-art Gd3Al2Ga3O12 single crystals},
journal = {Nuclear Instruments and Methods in Physics Research Section A: Accelerators, Spectrometers, Detectors and Associated Equipment},
volume = {1000},
pages = {165231},
year = {2021},
issn = {0168-9002},
doi = {https://doi.org/10.1016/j.nima.2021.165231},
author = {Loris Martinazzoli and Nicolaus Kratochwil and Stefan Gundacker and Etiennette Auffray}
}

@ARTICLE{9006906,
  author={Martinazzoli, Loris},
  journal={IEEE Transactions on Nuclear Science}, 
  title={Crystal Fibers for the LHCb Calorimeter Upgrade}, 
  year={2020},
  volume={67},
  number={6},
  pages={1003-1008},
  doi={10.1109/TNS.2020.2975570}}

@article{Pauwels_2013,
doi = {10.1088/1748-0221/8/09/P09019},
year = {2013},
month = {sep},
publisher = {},
volume = {8},
number = {09},
pages = {P09019},
author = {K Pauwels and others},
title = {{Single crystalline LuAG fibers for homogeneous dual-readout calorimeters}},
journal = {Journal of Instrumentation}
}

@Article{D2MA00626J,
author ="Martinazzoli, Loris and others",
title  ="Compositional engineering of multicomponent garnet scintillators: towards an ultra-accelerated scintillation response",
journal  ="Mater. Adv.",
year  ="2022",
volume  ="3",
issue  ="17",
pages  ="6842-6852",
publisher  ="RSC",
doi  ="10.1039/D2MA00626J"}

@article{KORZHIK2022166781,
title = {{Ultrafast PWO scintillator for future high energy physics instrumentation}},
journal = {Nuclear Instruments and Methods in Physics Research Section A: Accelerators, Spectrometers, Detectors and Associated Equipment},
volume = {1034},
pages = {166781},
year = {2022},
issn = {0168-9002},
doi = {https://doi.org/10.1016/j.nima.2022.166781},
author = {M. Korzhik and others}
}

@article{Kratochwil_2021,
doi = {10.1088/1361-6560/ac212a},
year = {2021},
month = {sep},
publisher = {IOP Publishing},
volume = {66},
number = {19},
pages = {195001},
author = {Kratochwil, Nicolaus and Gundacker, Stefan and Auffray, Etiennette},
title = {{A roadmap for sole Cherenkov radiators with SiPMs in TOF-PET}},
journal = {Physics in Medicine \& Biology}
}

@ARTICLE{9222347,
  author={Kratochwil, Nicolaus and Auffray, Etiennette and Gundacker, Stefan},
  journal={IEEE Transactions on Radiation and Plasma Medical Sciences}, 
  title={Exploring Cherenkov Emission of BGO for TOF-PET}, 
  year={2021},
  volume={5},
  number={5},
  pages={619-629},
  doi={10.1109/TRPMS.2020.3030483}}

@article{AKCHURIN2008359,
title = {Separation of crystal signals into scintillation and Cherenkov components},
journal = {Nuclear Instruments and Methods in Physics Research Section A: Accelerators, Spectrometers, Detectors and Associated Equipment},
volume = {595},
number = {2},
pages = {359-374},
year = {2008},
issn = {0168-9002},
doi = {https://doi.org/10.1016/j.nima.2008.07.136},
author = {N. Akchurin and others}
}

@Article{s19020308,
AUTHOR = {Gola, Alberto and Acerbi, Fabio and Capasso, Massimo and Marcante, Marco and Mazzi, Alberto and Paternoster, Giovanni and Piemonte, Claudio and Regazzoni, Veronica and Zorzi, Nicola},
TITLE = {NUV-Sensitive Silicon Photomultiplier Technologies Developed at Fondazione Bruno Kessler},
JOURNAL = {Sensors},
VOLUME = {19},
YEAR = {2019},
NUMBER = {2},
ARTICLE-NUMBER = {308},
PubMedID = {30646553},
ISSN = {1424-8220},
DOI = {10.3390/s19020308}
}

@ARTICLE{6651684,
  author={Brunner, S. E. and Gruber, L. and Marton, J. and Suzuki, K. and Hirtl, A.},
  journal={IEEE Transactions on Nuclear Science}, 
  title={Studies on the Cherenkov Effect for Improved Time Resolution of TOF-PET}, 
  year={2014},
  volume={61},
  number={1},
  pages={443-447},
  doi={10.1109/TNS.2013.2281667}}

@article{CALA2022166527,
title = {{Characterization of mixed $Bi_4(Ge_xSi_{1-x})_3O_12$ for crystal calorimetry at future colliders}},
journal = {Nuclear Instruments and Methods in Physics Research Section A: Accelerators, Spectrometers, Detectors and Associated Equipment},
volume = {1032},
pages = {166527},
year = {2022},
issn = {0168-9002},
doi = {https://doi.org/10.1016/j.nima.2022.166527},
author = {R. Cala’ and others}
}

@ARTICLE{10.3389/fphy.2022.785627,
  
AUTHOR={Terragni, Giulia  and others},
         
TITLE={Time Resolution Studies of Thallium Based Cherenkov Semiconductors},
        
JOURNAL={Frontiers in Physics},
        
VOLUME={Volume 10 - 2022},

YEAR={2022},


DOI={10.3389/fphy.2022.785627},

ISSN={2296-424X}}

@article{LUCCHINI2023168214,
title = {{Sub-10 ps time tagging of electromagnetic showers with scintillating glasses and SiPMs}},
journal = {Nuclear Instruments and Methods in Physics Research Section A: Accelerators, Spectrometers, Detectors and Associated Equipment},
volume = {1051},
pages = {168214},
year = {2023},
issn = {0168-9002},
doi = {https://doi.org/10.1016/j.nima.2023.168214},
author = {Marco T. Lucchini and others}
}

@article{NGUYEN2021164898,
title = {Boron-loaded organic glass scintillators},
journal = {Nuclear Instruments and Methods in Physics Research Section A: Accelerators, Spectrometers, Detectors and Associated Equipment},
volume = {988},
pages = {164898},
year = {2021},
issn = {0168-9002},
doi = {https://doi.org/10.1016/j.nima.2020.164898},
author = {Lucas Q. Nguyen and others}
}

@article{DORMENEV2021165762,
title = {{Multipurpose Ce-doped Ba-Gd silica glass scintillator for radiation measurements}},
journal = {Nuclear Instruments and Methods in Physics Research Section A: Accelerators, Spectrometers, Detectors and Associated Equipment},
volume = {1015},
pages = {165762},
year = {2021},
issn = {0168-9002},
doi = {https://doi.org/10.1016/j.nima.2021.165762},
author = {V. Dormenev and others}
}

@INPROCEEDINGS{6154597,
  author={Genser, Krzysztof and Para, Adam and Wenzel, Hans},
  booktitle={2011 IEEE Nuclear Science Symposium Conference Record}, 
  title={Very high resolution hadron calorimetry}, 
  year={2011},
  volume={},
  number={},
  pages={1177-1182},
  doi={10.1109/NSSMIC.2011.6154597}}

@article{Mao_2012,
doi = {10.1088/1742-6596/404/1/012029},
year = {2012},
month = {dec},
publisher = {},
volume = {404},
number = {1},
pages = {012029},
author = {Mao, Rihua and Zhang, Liyuan and Zhu, Ren-Yuan},
title = {Crystals for the HHCAL Detector Concept},
journal = {Journal of Physics: Conference Series}
}

@ARTICLE{10.3389/fphy.2022.1021787,
  
AUTHOR={Pagano, Fiammetta  and others },
         
TITLE={A new method to characterize low stopping power and ultra-fast scintillators using pulsed X-rays},
        
JOURNAL={Frontiers in Physics},
        
VOLUME={Volume 10 - 2022},

YEAR={2022},


DOI={10.3389/fphy.2022.1021787},

ISSN={2296-424X}}

@Article{D2TC02060B,
author ="Děcká, Kateřina and others",
title  ="Timing performance of lead halide perovskite nanoscintillators embedded in a polystyrene matrix",
journal  ="J. Mater. Chem. C",
year  ="2022",
volume  ="10",
issue  ="35",
pages  ="12836-12843",
publisher  ="The Royal Society of Chemistry",
doi  ="10.1039/D2TC02060B"}

@article{https://doi.org/10.1002/pssr.201600288,
author = {Turtos, Rosana M. and others},
title = {T{iming performance of ZnO:Ga nanopowder composite scintillators}},
journal = {physica status solidi (RRL) – Rapid Research Letters},
volume = {10},
number = {11},
pages = {843-847},
doi = {https://doi.org/10.1002/pssr.201600288},
eprint = {https://onlinelibrary.wiley.com/doi/pdf/10.1002/pssr.201600288},
year = {2016}
}

@article{Turtos_2016,
doi = {10.1088/1748-0221/11/10/P10015},
year = {2016},
month = {oct},
publisher = {},
volume = {11},
number = {10},
pages = {P10015},
author = {Turtos, R.M. and Gundacker, S. and Polovitsyn, A. and Christodoulou, S. and Salomoni, M. and Auffray, E. and Moreels, I. and Lecoq, P. and Grim, J.Q.},
title = {Ultrafast emission from colloidal nanocrystals under pulsed X-ray excitation},
journal = {Journal of Instrumentation}
}

@Article{Grim2014,
author={Grim, Joel Q.
and Christodoulou, Sotirios
and Di Stasio, Francesco
and Krahne, Roman
and Cingolani, Roberto
and Manna, Liberato
and Moreels, Iwan},
title={Continuous-wave biexciton lasing at room temperature using solution-processed quantum wells},
journal={Nature Nanotechnology},
year={2014},
month={Nov},
day={01},
volume={9},
number={11},
pages={891-895},
issn={1748-3395},
doi={10.1038/nnano.2014.213},
}

@Article{C1CC14687D,
author ="Khare, Ankur and Wills, Andrew W. and Ammerman, Lauren M. and Norris, David J. and Aydil, Eray S.",
title  ="{Size control and quantum confinement in Cu2ZnSnS4 nanocrystals}",
journal  ="Chem. Commun.",
year  ="2011",
volume  ="47",
issue  ="42",
pages  ="11721-11723",
publisher  ="The Royal Society of Chemistry",
doi  ="10.1039/C1CC14687D"}

@article{ refId0,
	author = {{Soldani, M.} and others},
	title = {Innovative nanocrystal-based scintillators for next-generation sampling calorimeters},
	DOI= "10.1051/epjconf/202532000020",
	journal = {EPJ Web Conf.},
	year = 2025,
	volume = 320,
	pages = "00020",
}

@Article{cryst8020078,
AUTHOR = {Salomoni, Matteo and Pots, Rosalinde and Auffray, Etiennette and Lecoq, Paul},
TITLE = {Enhancing Light Extraction of Inorganic Scintillators Using Photonic Crystals},
JOURNAL = {Crystals},
VOLUME = {8},
YEAR = {2018},
NUMBER = {2},
ARTICLE-NUMBER = {78},
ISSN = {2073-4352},
DOI = {10.3390/cryst8020078}
}

\end{document}